\pdfoutput=1

\documentclass{ws-ijmpe}
\usepackage[super,compress]{cite}
\usepackage{hyperref}
\usepackage{xspace}
\usepackage{array}
\usepackage{srcs/commands}
\usepackage{amsmath}
\begin{document}

\markboth{Kevin Dusling, Wei Li, Bj\"orn Schenke}{Novel collective phenomena}

\catchline{}{}{}{}{}
%%%%%%%%%%%%%%%%%%%%%%%%%%%%%%%%%%%%%%%%%%%%%%%%%%%%%%%%%%%%%%%%%%%%

\title{Novel Collective Phenomena in High-Energy\\Proton-Proton and Proton-Nucleus Collisions}

\author{Kevin Dusling}

\address{American Physical Society, 1 Research Road\\
Ridge, NY 11961, USA\\
kdusling@mailaps.org}

\author{Wei Li}
\address{Department of Physics and Astronomy, 6100 Main St., MS-315\\
Houston, TX 77005, USA\\
davidlw@rice.edu}

\author{Bj\"orn Schenke}
\address{Physics Department, Bldg. 510A, Brookhaven National Laboratory, \\
Upton, NY 11973, USA\\
bschenke@bnl.gov}

\maketitle

\begin{history}
\received{Day Month Year}
\revised{Day Month Year}
%\accepted{Day Month Year}
%\comby{(xxxxxxxxxx)}
\end{history}

\begin{abstract}
The observation of long-range rapidity correlations among particles in
high-multiplicity p-p and p-Pb collisions has created new opportunities for
investigating novel high-density QCD phenomena in small colliding systems. We
review experimental results related to the study of collective phenomena in
small systems at RHIC and the LHC along with the related developments in theory
and phenomenology. Perspectives on possible future directions for research are
discussed with the aim of exploring emergent QCD phenomena.
\end{abstract}


\ccode{PACS numbers: 25.75.−q, 25.75.Gz, 25.75.Ld, 12.38.−t, 12.38.Aw, 12.38.Mh}

\tableofcontents

\section{Introduction}

Collective phenomena are ubiquitous across physics and observed in systems
having disparate temporal and spatial scales ranging from atomic nuclei,
ultra-cold atomic gases and biological systems, up to the cosmological scales
involved in galaxy formation. In a broad sense, the concept of collectivity 
refers to a common behavior exhibited by a group of entities 
(e.g., particles moving with a common velocity or along a preferred direction).
Collectivity is often associated with an emergent phenomenon of a
complex, many-body system, for which the basic interactions (for example, at
the two-body level) may be well understood.  An important goal of studying
collective phenomena is to unravel how the macroscopic behavior of a many-body
system emerges from its fundamental degrees of freedom.

One of the most striking displays of collectivity is the strong expansion of
the medium produced in ultra-relativistic heavy-ion collisions.  A careful
analysis of identified particle spectra, elliptic flow, Handbury-Brown-Twiss (HBT) interferometry and the
suppression of particles at high transverse momenta has indicated that a
hot and dense QCD medium, the ``Quark-Gluon plasma (QGP)'' undergoes a nearly
ideal hydrodynamic expansion when created~\cite{BRAHMS,PHENIX,PHOBOS,STAR}.

\begin{figure}[t]
\center
\includegraphics[width=\linewidth]{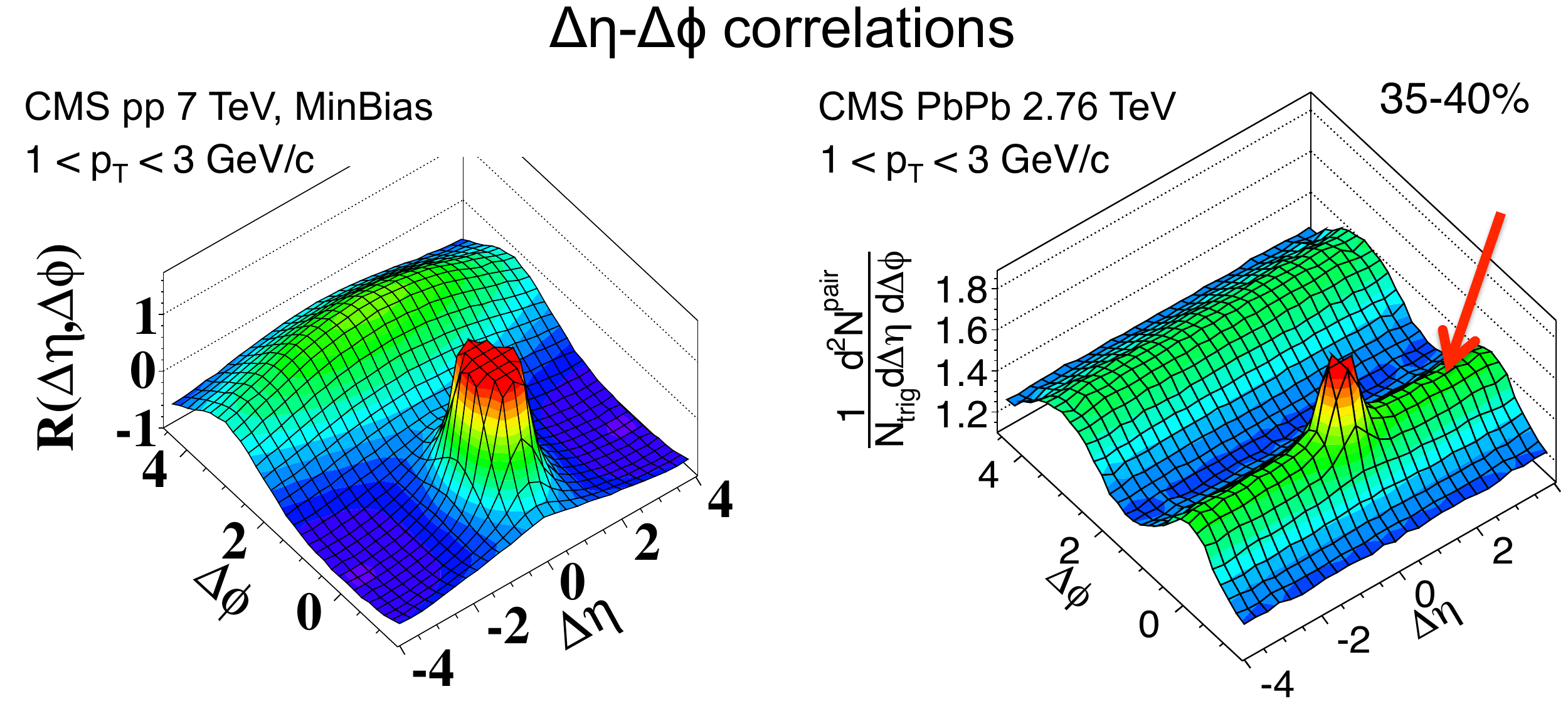}
\caption{ 
\label{fig:ridge_AA} 
Two-particle correlation function for charged hadrons with transverse momenta
$1<\pt<3$\GeVc\ in minimum bias \ppc collisions at \roots\ = 7\TeV\
\cite{Khachatryan:2010gv} (left) and in 35--40\% centrality \PbPb collisions at
\rootsNN\ = 2.76\TeV\ \cite{Chatrchyan:2012wg} (right), measured by the CMS
experiment.  
}
\end{figure}

Figure~\ref{fig:ridge_AA} shows one of the many persuasive pieces of evidence
for collective behavior in heavy-ion collisions.  The plot on the right shows
the $\Delta\eta$--$\Delta\phi$ correlation functions for pairs having
$1<\pt<3$\GeVc in semi-central \PbPb collisions at the LHC.  The right plot
shows the same observable for minimum bias \ppc collisions where a dijet-like
(or mini-jet) correlation structure can be seen; the narrow peak around
$(\Delta\eta, \Delta\phi) \sim (0, 0)$ comes from the showering and hadronization of the
leading parton, and the long-range away-side ($\Delta\phi \sim \pi$) structure
in $\Delta\eta$ represents correlations from the recoiling parton.  As the two
partons are produced back-to-back they necessarily have opposite rapidities and
will therefore populate the full acceptance in $\Delta\eta$.  In \PbPb
collisions in addition to the jet-like correlations, a pronounced near-side
($\Delta\phi \sim 0$) collimation extending over a long range in $\Delta\eta$
is observed and is now referred to as the ``ridge''.  This ridge-like
correlation in heavy-ion collisions such as \AuAu at RHIC and \PbPb at the LHC
is believed to be well understood:  The overlap area of a heavy-ion collision
at a finite impact parameter has an elliptic shape.  The larger pressure
gradients along the minor-axis of the ellipse lead to a larger flow in this
direction and therefore collimated production in both directions of this axis
creating a near- and away-side ridge.  Careful subtraction of the away-side jet
peak shows that such a double-ridge is present.  The absence of a ridge
structure in minimum bias \ppc collisions suggest the absence of collective
behavior in these systems even though event-by-event fluctuations may result in
highly eccentric initial states.  

\begin{figure}[t]
\center
\includegraphics[width=0.7\linewidth]{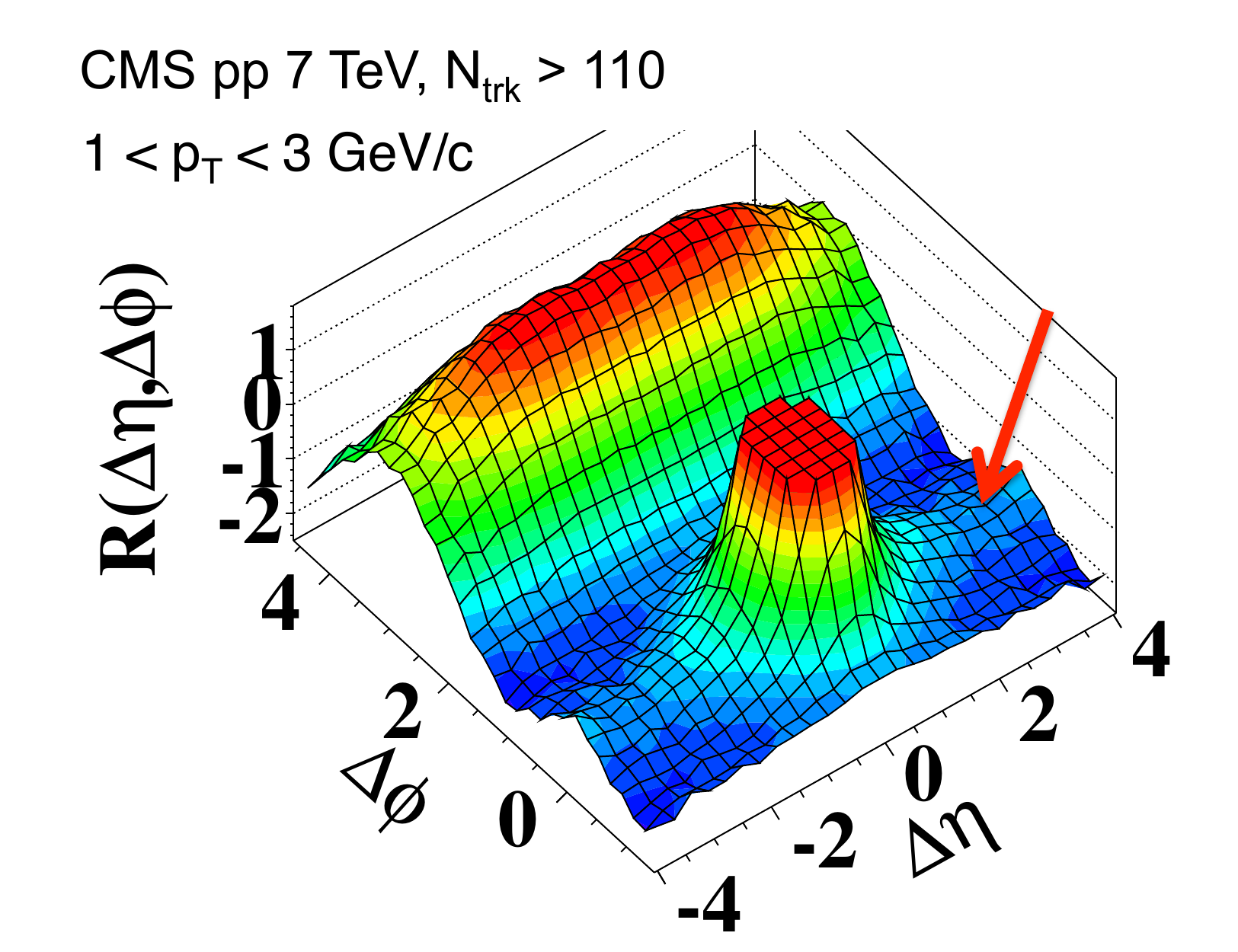}
\caption{ 
\label{fig:ridge_pp} 
Two-particle correlation function for particles having $1<\pt<3$\GeVc\ in
high-multiplicity \ppc collisions at \roots\ = 7\TeV, measured by the CMS
experiment~\cite{Khachatryan:2010gv}.
}
\end{figure}

In 2010 surprising indications for collectivity in \ppc collisions at the LHC
were observed when triggering on rare events with high multiplicities (large
number of final-state particles)~\cite{Khachatryan:2010gv}. The long-range
near-side ridge for \ppc events having charged multiplicity $N_{\rm trk}>110$
is shown in Fig.~\ref{fig:ridge_pp}.  This ridge-like structure, not present in
minimum bias \ppc collisions, is reminiscent of the two-particle correlation in
\AAc collisions.  Determining whether the \ppc ridge can be attributed to
collective flow effects will require a concerted effort by theorists and
experimentalists.  With a variety of theoretical proposals and limited
experimental data in high-multiplicity \ppc collisions (see
Ref.~\cite{Li:2012hc} for an early review) a conclusive explanation of the \ppc
ridge remains outstanding.  

With the discovery of the \ppc ridge it was natural to look for a similar
signal in \pA collisions.  First collisions of high-energy proton and ion beams
(\pA) were achieved at the LHC in 2011 at a center-of-mass energy of
5.02~TeV per nucleon pair.  The discovery of a ridge in high-multiplicity \pPb
collisions~\cite{Abelev:2012ola,Aad:2012gla,CMS:2012qk} did not come as a
complete surprise, but the fact that the strength of the correlation was almost
as large as the ridge in heavy-ion collisions was unexpected.  

The ridge correlations in \ppc and \pA collisions remain to be fully
understood, and the existence of similar structures in small colliding systems
such as d-Au~\cite{Adare:2014keg} and $^{3}$He-Au collisions
\cite{Adare:2014keg,Adare:2015ctn} at lower RHIC energies have stimulated both
experimental and theoretical communities to further investigate the properties
of the ridge.

This article provides a comprehensive review of the latest experimental results
and theoretical developments in our understanding of collective phenomena in
dense, high-multiplicity hadronic systems. Perspectives on future directions are
also discussed in light of future programs at major accelerator facilities such
as RHIC and the LHC.

%%% Local Variables: 
%%% mode: latex
%%% TeX-master: "../RidgeReview-ijmpe"
%%% End: 

\section{Collectivity and hydrodynamics in \AAc collisions}

Studies of multi-particle correlations have provided crucial insight into the
underlying mechanism of particle production in relativistic heavy-ion
collisions. The most prominent feature of multi-particle correlations
in AA collisions is due to  ``elliptic flow'', an azimuthal
anisotropy in momentum space induced by strong expansion of the initial
almond-shaped overlap area of two nuclei. \cite{Ollitrault:1992bk}  Elliptic
flow phenomena can be well described by relativistic hydrodynamic simulations
with a viscosity approaching the KSS bound $\eta/s=1/4\pi\approx 0.08$.  \cite{Gale:2013da} 
 In two-particle correlation measurements the elliptic flow
 generates an approximate $\cos(2\Delta\phi)$ component to the
two-particle correlation function that extends over a broad $|\Delta\eta|$
range~\cite{Alver:2008gk}.  The long-range in rapidity is a consequence of the
approximate boost invariance of the system.  Studies of elliptic flow have
been carried out over a wide range of energies and 
systems~\cite{Adcox:2004mh,Adams:2005dq,Back:2004je,Adler:2004cj,
Agakishiev:2011id,PHOBOS:PhysRevLett.98.242302,
Chatrchyan:2012ta,Aamodt:2010pa,ATLAS:2011ah} at both RHIC and the LHC. 

\begin{figure}[t]
\centering
\includegraphics[width=\textwidth]{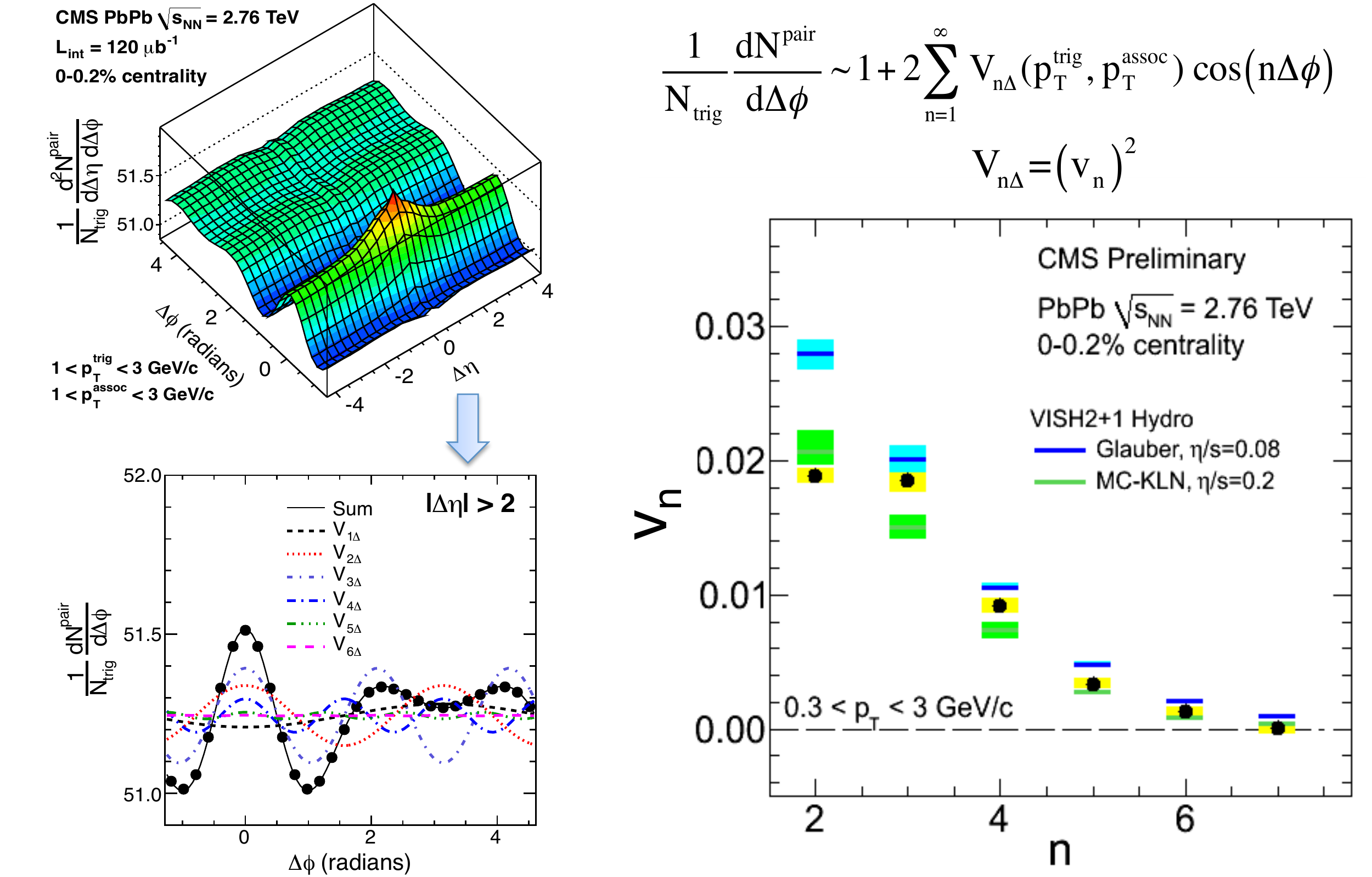}
\caption{Left: The 2-D and 1-D long-range $\Delta\phi$ two-particle correlation 
    functions in ultra-central (0--0.2\%)
    PbPb collisions at $\sqrt{s_{_{NN}}}=2.76$~TeV measured by the CMS collaboration. 
    Right: the extracted $v_n$ Fourier harmonics from long-range two-particle correlations 
    integrated over $0.3<\pt<3$\GeVc, compared with hydrodynamic calculations~\cite{CMS:2013bza}.}
\label{fig:2DCF}
\end{figure}

Event-by-event fluctuations in the collision geometry of the initial-state
introduce higher-order anisotropic flow components.  For example, ``triangular
flow'' results in a $\cos(3\Delta\phi)$ modulation of the two-particle
correlation.
\cite{Alver:2010gr,Alver:2010dn,Schenke:2010rr,Petersen:2010cw,Xu:2010du,Teaney:2010vd}
This contribution can dominate the structure of the entire $\Delta\phi$
correlation in very central \AAc collisions, as is shown in \fig{fig:2DCF} for
the top 0--0.2\% central \PbPb collisions at LHC. A prominent {\em double hump}
appears on the away-side.

By projecting the long-range component ($|\Delta\eta|>2$) of the 2-D
correlation function onto a 1-D $\Delta\phi$ axis the correlations can be
characterized by a Fourier series, $\sim 1+2 \sum_{n}
v_{n}^{2}\cos(n\Delta\phi)$, where $v_n$ denotes the single-particle anisotropy
harmonic of order $n$~\cite{Voloshin:1994mz}.  A dominant third-order Fourier
component is seen because in ultra-central \AAc collisions the impact parameter
is small enough such that the initial (average) ellipticity is negligible.  On
an event-by-event basis, the initial geometry is governed by fluctuations
resulting in comparable elliptic and triangular asymmetries.
 
\begin{figure}[t]
\centering
\includegraphics[width=0.47\textwidth]{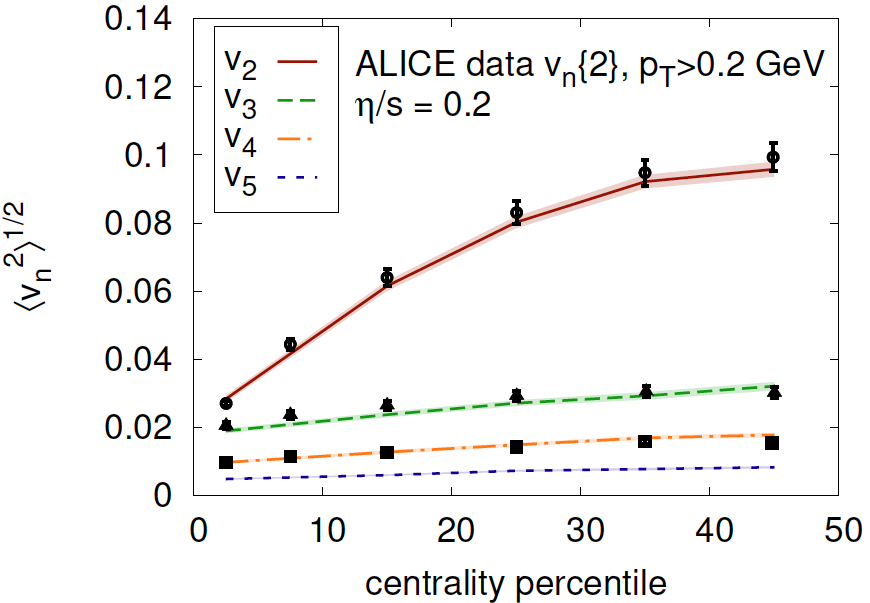}
\includegraphics[width=0.51\textwidth]{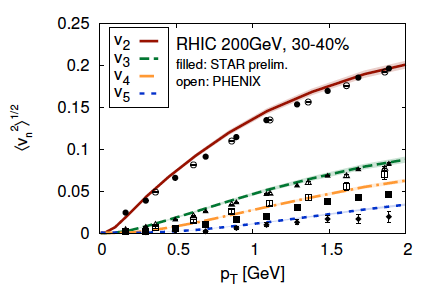}
\caption{Model calculations compared to measurements of the harmonic 
decomposition of azimuthal correlations produced in heavy ion collisions~\cite{Gale:2012rq}. 
The left panel shows model calculations and data for $v_n$ vs. collision 
centrality in Pb+Pb collisions at $\sqrt{s_{NN}}=2.76$ TeV. The right panel 
shows similar studies for the $p_T$ dependence of $v_n$ in 200 GeV Au+Au 
collisions. The comparison of the two energies provides insight on the temperature 
dependence of $\eta/s$. }
\label{fig:vn}
\end{figure}

Of course, any set of data can be decomposed into a Fourier series and the
existence of a large elliptic anisotropy cannot be taken as proof of
hydrodynamic behavior in and of itself.  Quantitative comparison with
theoretical calculations, in this case viscous relativistic hydrodynamic
simulations is necessary in order to draw any strong conclusions. On the right-hand
side of Fig.~\ref{fig:2DCF}, the extracted $v_2$--$v_6$ data in ultra-central
\PbPb collisions from CMS are compared to viscous hydrodynamic calculations
using two different initial condistions (the details of which will be discussed
later). These results indicate that the produced medium has a shear viscosity to entropy 
density ratio of $\eta/s\sim 0.08-0.2$. 

Higher-order flow components in \AAc collisions can provide
more stringent constraints on extracting both transport properties
(e.g., $\eta/s$) and initial-state models of heavy-ion collisions.
Indeed, the elliptic and higher-order flow phenomena have been extensively studied at
RHIC and the LHC over a wide range of collision centrality and particle \pt.
Fig.~\ref{fig:vn} shows the comparison of $\pt$-integrated $v_n$ as a function
of centrality in \PbPb collisions at the LHC (left) and $v_n$ as a function of
\pt\ for 30--40\% central \AuAu collisions at RHIC with hydrodynamic
calculations using the IP-Glasma initial-state model~\cite{Schenke:2012wb,Gale:2012rq}.  The LHC
data is well described by an $\eta/s=0.2$ and at RHIC by $\eta/s=0.12$ providing
indication of a temperature dependence. 
It has now been widely accepted that in relativistic heavy-ion collisions, a strongly-coupled medium is formed
exhibiting near-ideal fluid behavior.

%%% Local Variables: 
%%% mode: latex
%%% TeX-master: "../RidgeReview-ijmpe"
%%% End: 

\section{Experimental evidence of collectivity in small systems}
\label{sec:exppA}

This section provides a broad overview of experimental results in small
colliding systems, focusing on the recent results from \pPb collisions at the LHC.
While many of the experimental observables explored in this section will have
striking similarities with the results from \AAc collisions, one must be cautioned that
strong conclusions about the interpretation of these measurements in \AAc could
only be made after detailed models (in this case hydrodynamics) confronted the
data.  Clearly, the system size in \ppc or \pA collisions is significantly
smaller, and therefore the applicability of hydrodynamics must be scrutinized. 
This section collects the wealth of experimental data we have on small systems
and compares it to that in \AAc with interpretations postponed to section~\ref{theory}.

\begin{figure}[th]
  \begin{center}
    \includegraphics[width=0.65\linewidth]{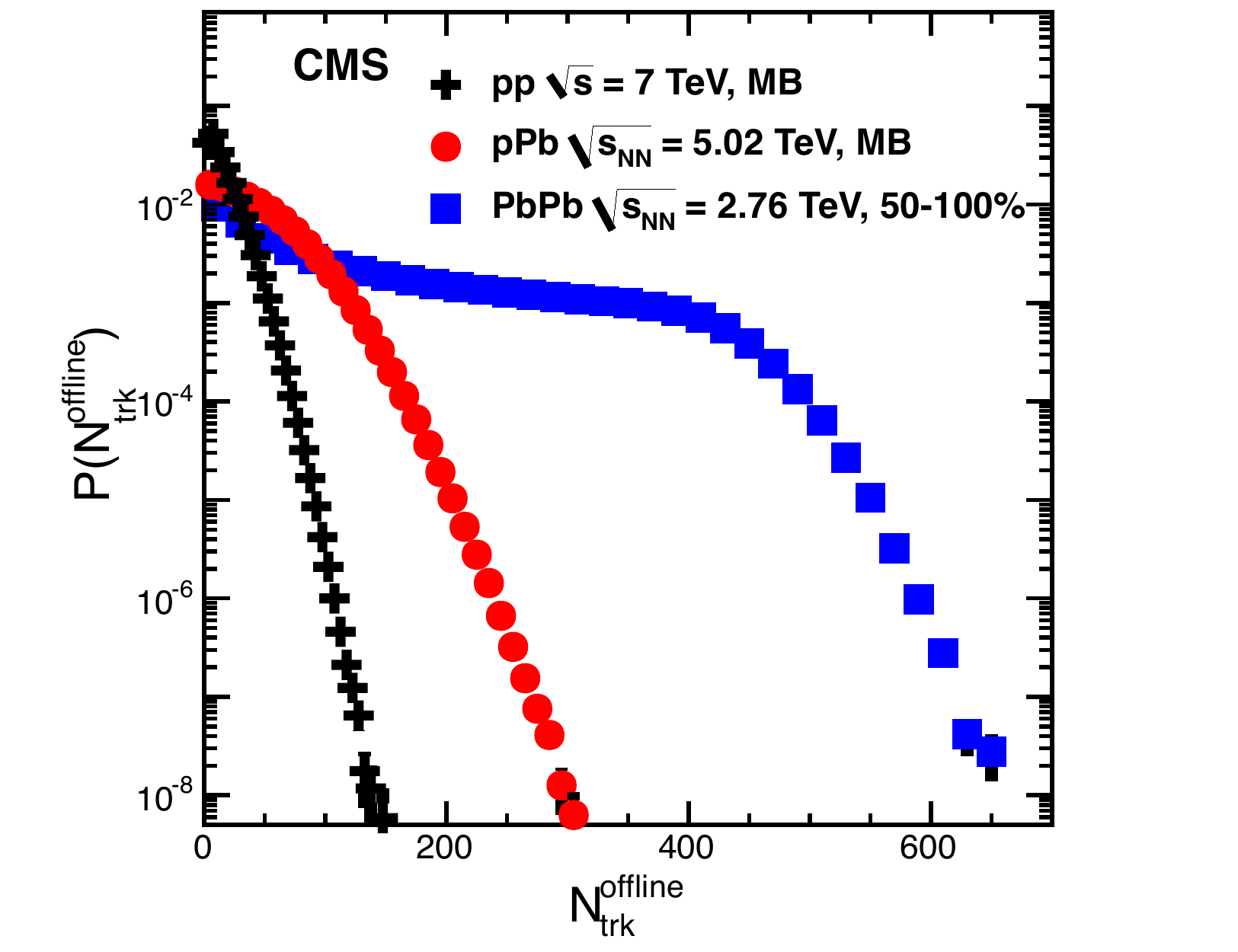}    
    \caption{ Multiplicity ($N_{\rm trk}$) distributions for MinBias \ppc, MinBias \pPb 
    and 50--100\% centrality \PbPb collisions~\cite{Khachatryan:2010gv,Chatrchyan:2013nka}.}
    \label{fig:mult}
  \end{center}
\end{figure}

An important point that should be kept in mind throughout this discussion is
that the nature of the very events we are studying in small systems are in a
class of their own.  They represent a fraction of the total cross-section, as can
be gleamed from figure~\ref{fig:mult} that shows the multiplicity distributions
for minimum bias \ppc, \pPb and peripheral 50--100\% central \PbPb collisions
from CMS~\cite{Khachatryan:2010gv,Chatrchyan:2013nka}. Here, multiplicity is
defined as the number of charged tracks within $|\eta|<2.4$ and $\pt >
0.4$\GeVc.  Events with 100--200 tracks (these are the high-multiplicity events
where a ridge signal is seen in \ppc and \pPb) are a common occurrence in
\PbPb.  In general, experimental signatures should not depend
only on the multiplicity; \ppc and \PbPb events producing 100 tracks have a
very different nature. 
 
\subsection{Two-particle correlations and azimuthal anisotropy}

In September 2012, first collisions of proton and lead have been delivered by
the LHC at \rootsNN\ = 5.02~TeV.  About two million minimum bias \pPb events
were collected in a few hours of beam time by each experiment. Experiments
with \pA were to serve as a control experiment for \AAc collisions providing
baselines for cold nuclear matter effects~\cite{Abreu:2007kv}. 
 
\begin{figure}[thb]
  \begin{center}
    \includegraphics[width=\linewidth]{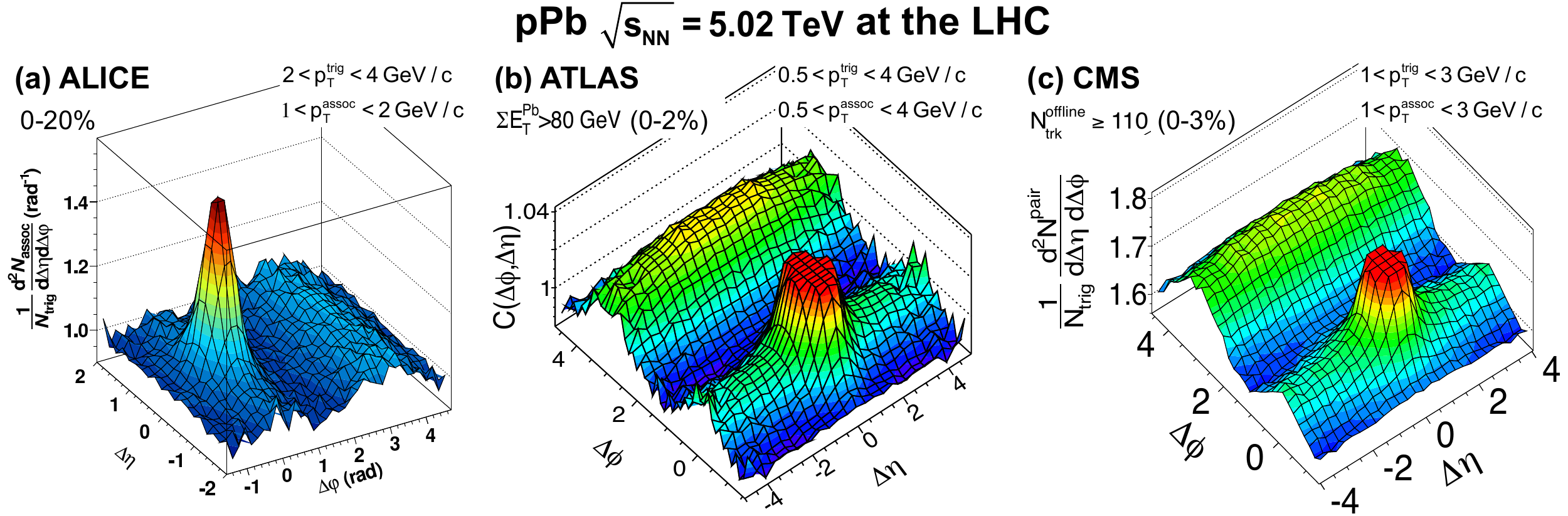}
    \caption{The 2D two-particle correlation functions in high-multiplicity 
    \pPb collisions at $\sqrt{s_{_{NN}}}=5.02$~TeV measured by the ALICE 
    (0--20\%, left)~\cite{Abelev:2012ola}, ATLAS (0--2\%, middle)~\cite{Aad:2012gla}
    and CMS (0--3\%, right)~\cite{CMS:2012qk} experiments.}
    \label{fig:pPb_2D}
  \end{center}
\end{figure}

However, with the recent discovery of the ridge in \ppc,
it was not clear what would be observed in high multiplicity \pA collisions.  The \pA pilot run
discovered a significant near-side ridge, as shown in Fig.\,\ref{fig:pPb_2D},
within the top few \% (CMS~\cite{CMS:2012qk},
ATLAS~\cite{Aad:2012gla}) or even 20\% (ALICE~\cite{Abelev:2012ola})
multiplicity class.  A 2013 high-intensity \pPb run provided access to higher
multiplicity classes and increased statistics.

\begin{figure}[thb]
  \begin{center}
    \includegraphics[width=0.38\linewidth]{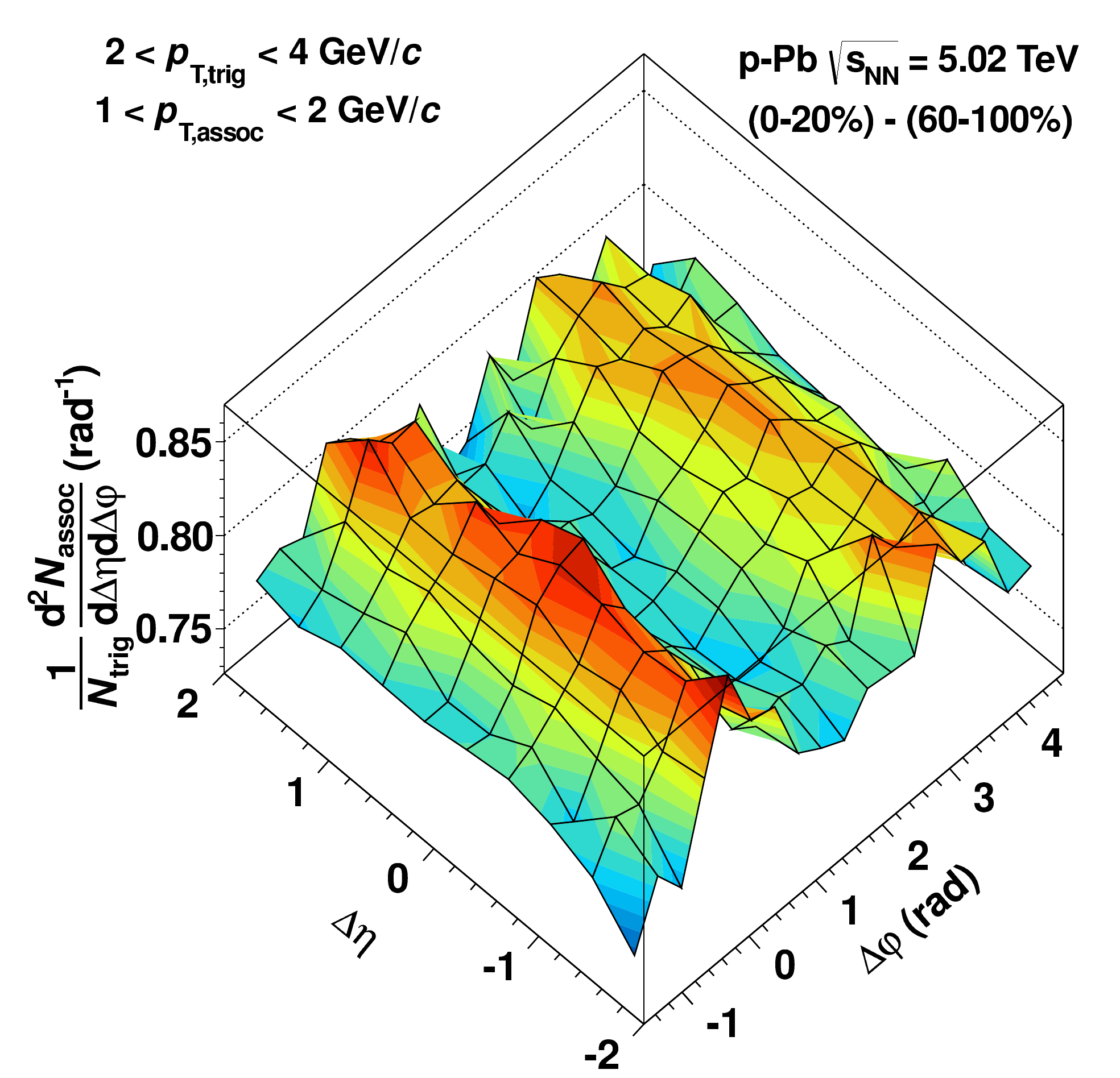}
    \hspace{0.3cm}
    \includegraphics[width=0.47\linewidth]{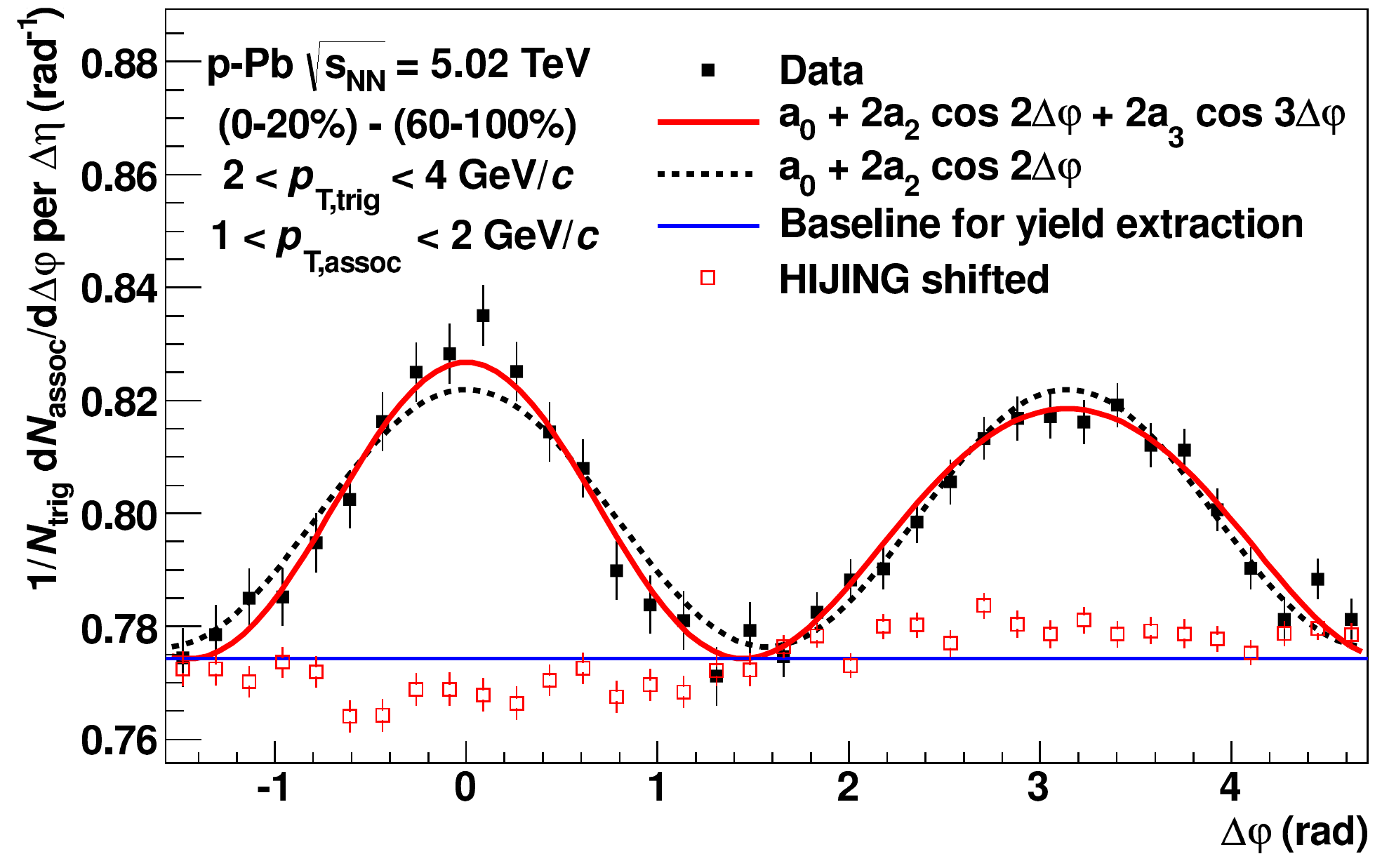}    
    \caption{The difference of two-particle correlation functions between
    high- and low-multiplicity \pPb\ collisions at $\sqrt{s_{_{NN}}}=5.02$~TeV 
    measured by the ALICE experiment~\cite{Abelev:2012ola}.}
    \label{fig:pPb_2D_doubleridge}
  \end{center}
\end{figure}

The magnitude of the (near side) ridge in \pPb collisions stands out as a prominent feature being about four times larger than the \ppc ridge for $\Ntrk > 110$ and becomes comparable to the away-side jet peak. In order to disentangle the ridge from the jet the ALICE and ATLAS collaborations subtracted the contribution of jets to the away-side correlation structure in high-multiplicity events, based on the estimates from low-multiplicity events.  Implicit in this procedure is the assumption that there is negligible modification of the jet due to the high multiplicities (while this is clearly not valid in \AAc collisions there has yet to be an observation of jet quenching in \ppc or \pA).
Following this subtraction procedure a ``double'' ridge structure having near- and away-side
correlations of similar strength has been observed, as shown in Fig.~\ref{fig:pPb_2D_doubleridge}
for the 2D correlation function and its projection to 1D in azimuthal angle.

\begin{figure}[thb]
\center
    \includegraphics[width=0.7\linewidth]{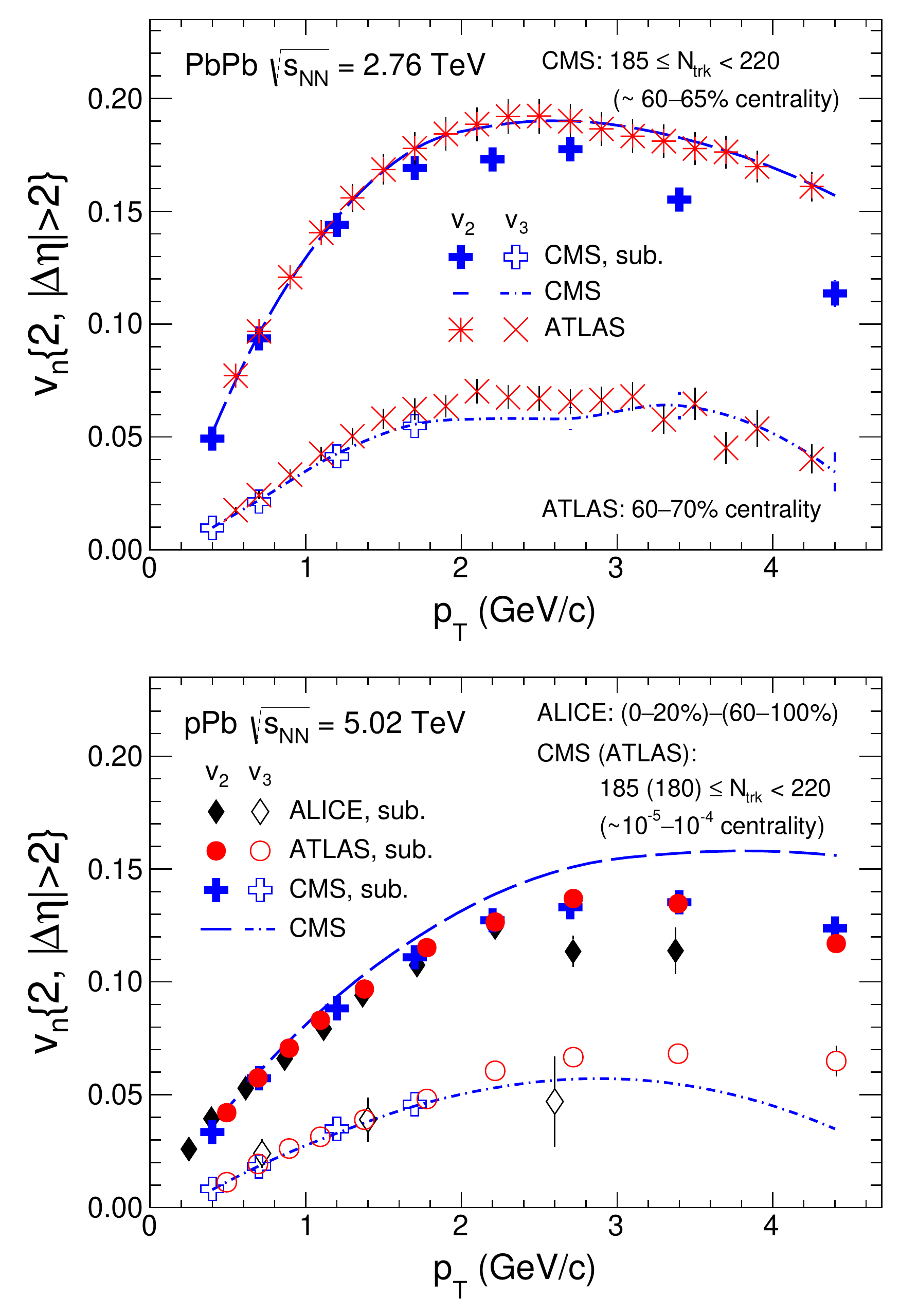}  
    \caption{ The second-order ($v_2$) and third-order ($v_3$) Fourier harmonics, 
    extracted from long-range two-particle correlations, as a function of $p_{T}$
    in \PbPb collisions at $\sqrt{s_{_{NN}}}=2.76$~TeV (top) and high-multiplicity \pPb collisions at 
    $\sqrt{s_{_{NN}}}=5.02$~TeV (bottom). The curves represent original $v_n$ data,
    while the markers denote the results after correcting for back-to-back jet correlations
    estimated from low-multiplicity events. 
    Data are obtained from Refs.~\cite{Aad:2014lta,Chatrchyan:2013nka,Abelev:2012ola,ATLAS:2012at}.
    } \label{fig:pPb_v2}
\end{figure}

Motivated by the last decade of study of flow harmonics in \AAc collisions the
``double'' ridge structure in \pPb collisions has been analyzed using the same
Fourier decomposition.  The second- and third-order anisotropy Fourier
harmonics, $v_2$ and $v_3$, are extracted from the long-range correlations as a
function of \pt\ in high-multiplicity \pPb collisions at \rootsNN\ = 5.02~TeV,
shown in Fig.~\ref{fig:pPb_v2} (bottom) from three LHC
experiments~\cite{Aad:2014lta,Chatrchyan:2013nka,Abelev:2012ola}.  The curves
shown from the CMS experiment are the results for $v_2$ and $v_3$ without
subtracting the away-side jet demonstrating that in high-multiplicity events
the contribution from jets to the Fourier harmonics are negligible at low
enough \pt .

Results of $v_2$ and $v_3$ in \PbPb collisions at \rootsNN\ = 2.76~TeV are also
shown in Fig.~\ref{fig:pPb_v2} (top), for a similar multiplicity range as for
\pPb collisions. ALICE, ATLAS and CMS results are in good agreement. Very
recently, a sizeable $v_2$ has also been extracted from \ppc collisions at
13\,{\rm TeV} by the ATLAS collaboration.\cite{Aad:2015gqa} 

\begin{figure}[thb]
\center
    \includegraphics[width=0.49\textwidth]{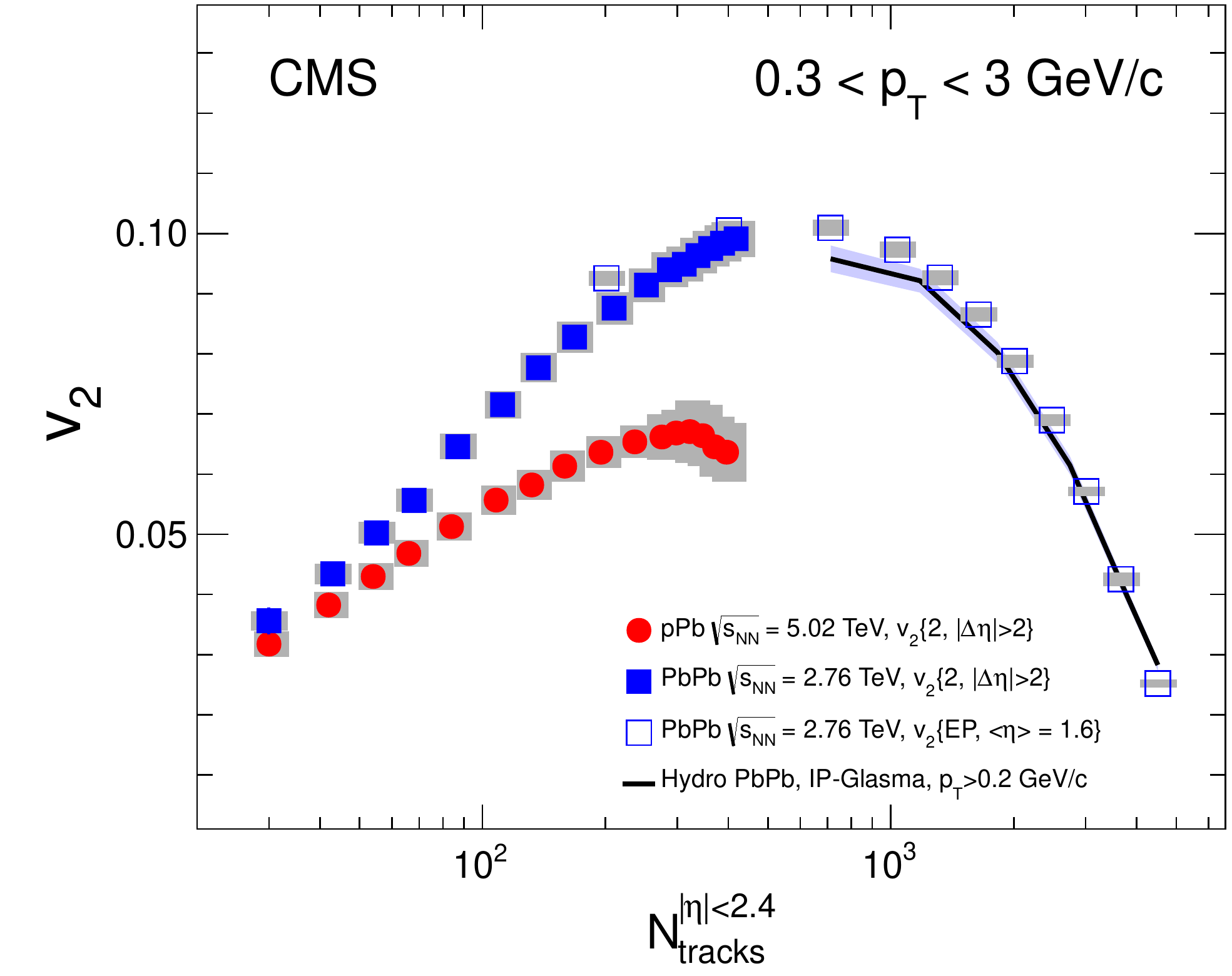}
    \includegraphics[width=0.49\textwidth]{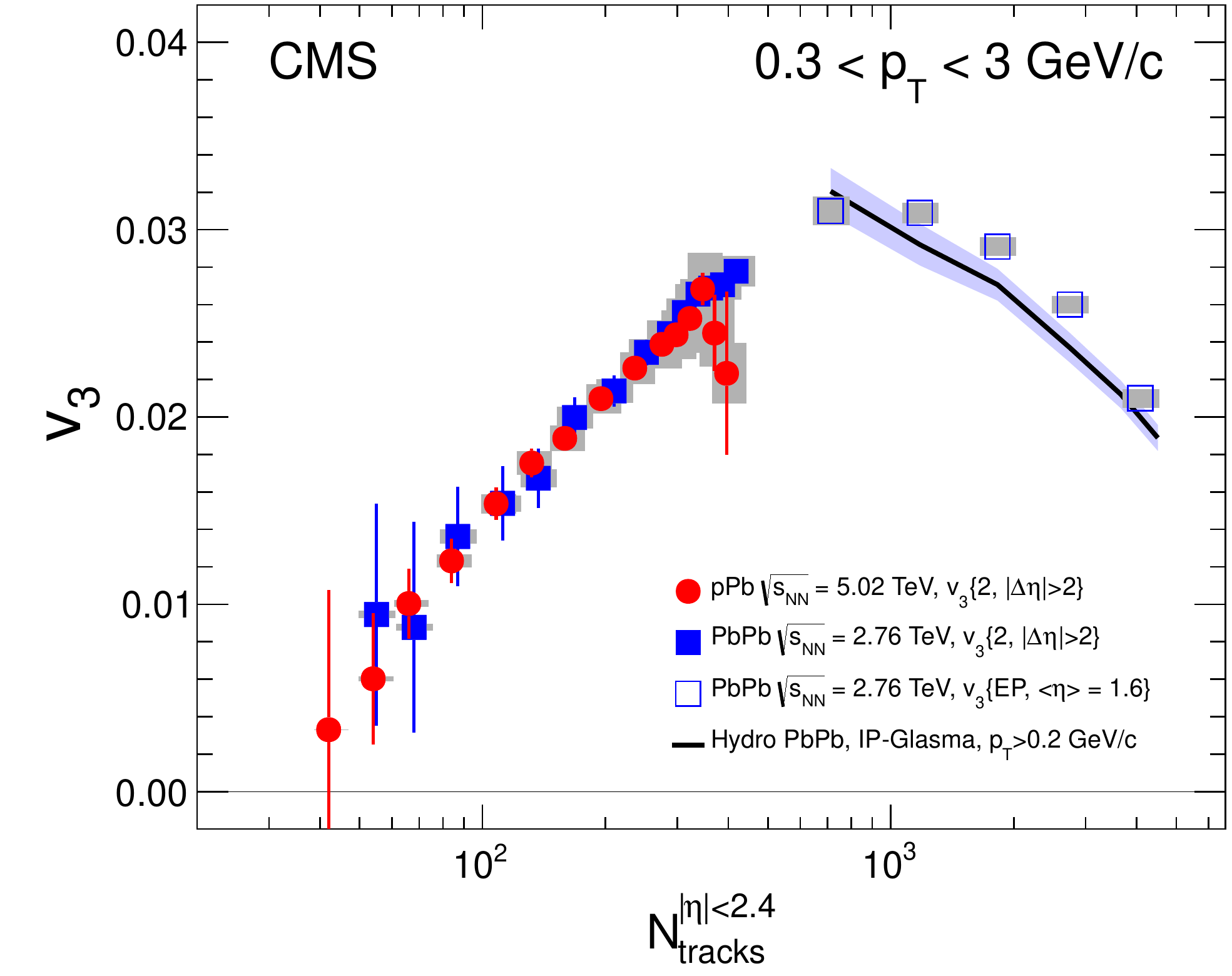}    
    \caption{ The $\pt$-averaged $v_2$ (left) and $v_3$ (right) for 0.3$<$\pt$<$3~GeV/c 
    as a function of multiplicity in pPb (from two-particle correlation method) and 
    PbPb (from event-plane method) collisions, measured by the CMS collaboration~\cite{Chatrchyan:2013nka}. 
    The solid line shows a hydrodynamic calculation
    using IP-Glasma initial conditions for \PbPb collisions~\cite{Schenke:2010rr}.
    } \label{fig:vNvsNtrk}
\end{figure}

As shown in Fig.~\ref{fig:pPb_v2}, the $v_2$ and $v_3$ values first 
rise with \pt\ up to around 3~GeV and then fall off toward
much higher \pt, a behavior that is very similar to \PbPb collisions. 
This may be an indication of
a common origin of the ridge phenomenon in all collision systems. In \PbPb collisions, 
the trend at low \pt\ is consistent with hydrodynamic predictions, while the decrease of higher \pt\
$v_n$ could be due to the lack of thermalization of more energetic probes. 
Surprisingly, the $v_3$ harmonics as a function of \pt\ for \pPb and \PbPb 
systems are almost identical. This striking 
similarity extends to the full multiplicity range, as one can see from 
Fig.~\ref{fig:vNvsNtrk} (right) for $v_3$ averaged over $0.3<\pt<3$~GeV/c in \pPb and 
\PbPb. Considering the drastically different initial-state geometry and system 
size of the two systems, this similarity and its implication remain to be fully
understood.  Arguments based on the universality of fluctuation-driven
initial-state
anisotropies\cite{Yan:2013laa,Basar:2013hea,Bzdak:2013rya,Bzdak:2013raa}
followed by the linear response of hydrodynamics may explain the coincidental
triangular flows seen in p+Pb and Pb-Pb collisions. 
The $v_2$ value at a given multiplicity (Fig.~\ref{fig:vNvsNtrk}, left) is larger 
in \PbPb than in \pPb; this is expected since we know in \PbPb there is a large 
initial eccentricity that after expansion generates a large $v_2$. 

\subsection{Multi-particle azimuthal correlations}

One of the key questions about the nature of the ridge and its collectivity is
whether it is only a two-particle correlation or if a correlation persists amongst
all produced particles.  The multi-particle cumulant technique was developed to
measure the strength of collective azimuthal anisotropy by correlating four or
more particles
simultaneously~\cite{Borghini:2001vi,Bhalerao:2003yq,Bilandzic:2010jr}.  It is
designed to extract cumulants of collective azimuthal correlations, while
suppressing non-collective short-range correlations, involving only a few
particles.  This approach has been widely used in studying \AAc collisions.
~\cite{Alt:2003ab,Adler:2002pu,Aamodt:2010pa,Abelev:2012di,Chatrchyan:2012ta,Chatrchyan:2013kba,Aad:2014vba} 

\begin{figure}[thb]
\center
  $\vcenter{\hbox{\includegraphics[width=0.5\linewidth]{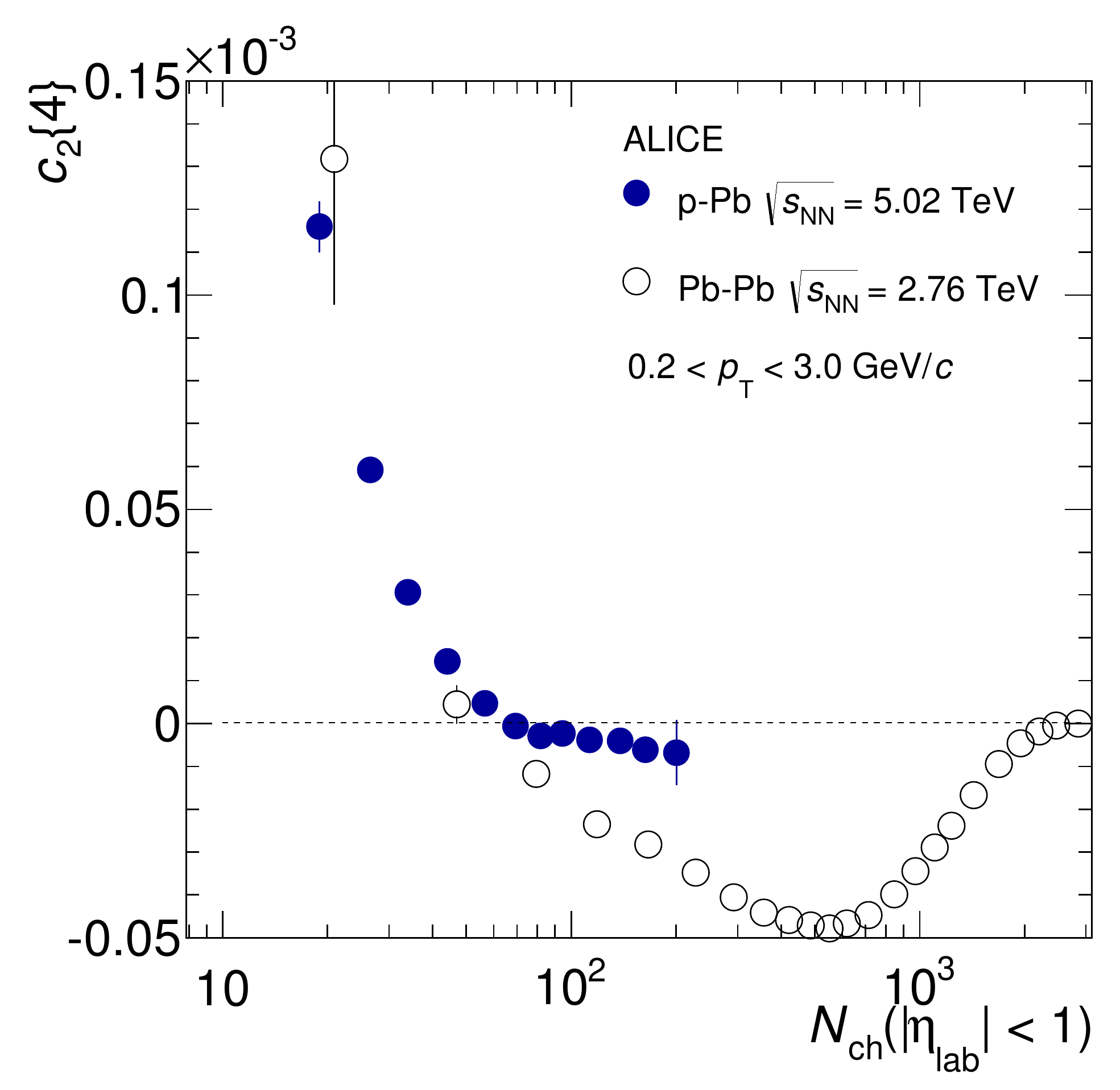}}}$
\hspace{0.5cm}
  $\vcenter{\hbox{\includegraphics[width=0.44\linewidth]{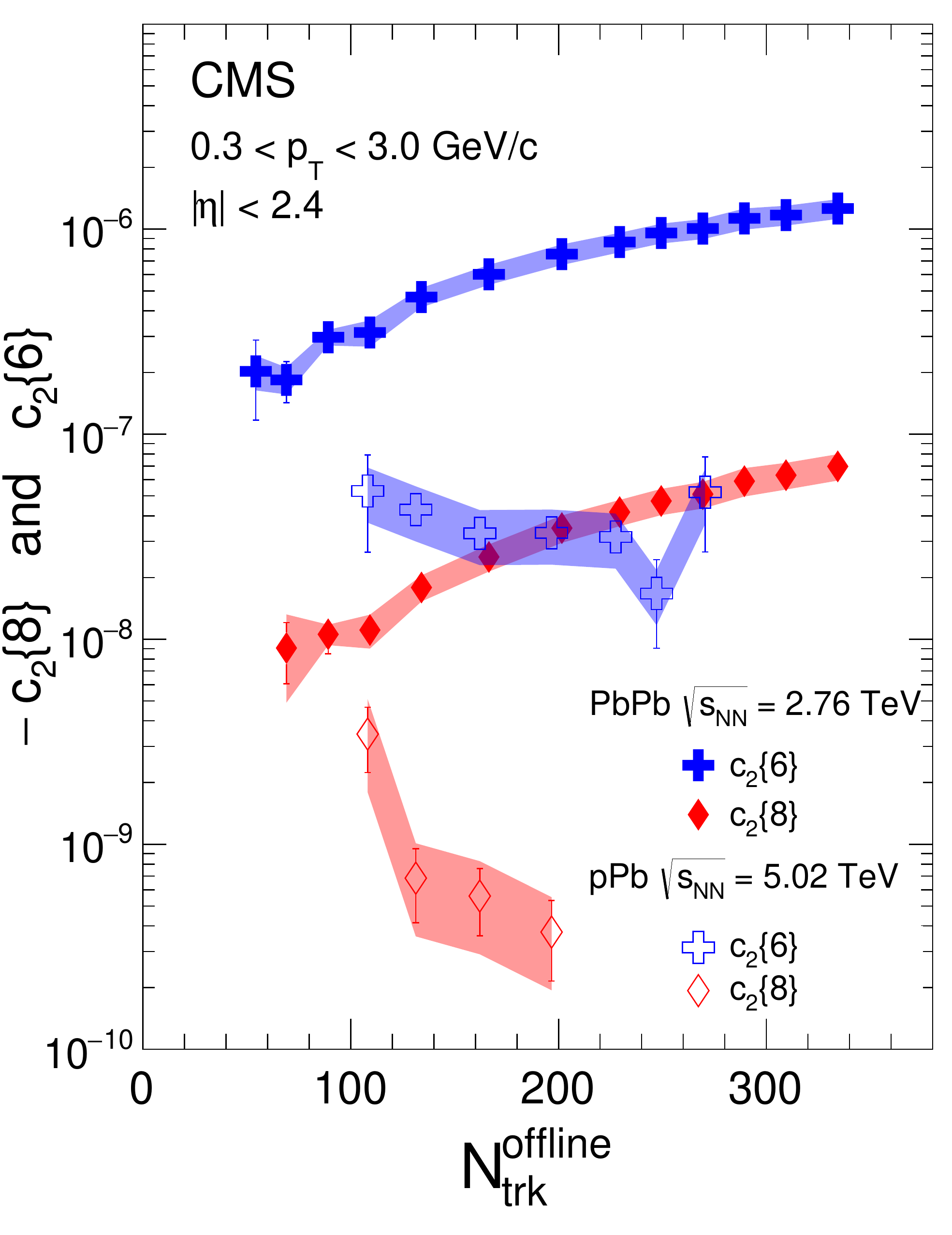}}}$
  \caption{ \label{fig:c2n} The four-($c_{4}$), six-($c_{6}$) 
  and eight-($c_{8}$) particle cumulants as a function of $N_{\rm trk}$, 
  averaged over $(0.2) 0.3<\pt<3.0$~GeV/c, in \PbPb collisions at \rootsNN\ = 2.76 TeV and \pPb at \rootsNN\ = 5.02 TeV.
  Data are obtained from Refs.~\cite{Abelev:2014mda,Khachatryan:2015waa}.
   }
\end{figure}

\begin{figure}[thb]
\center
\includegraphics[width=\linewidth]{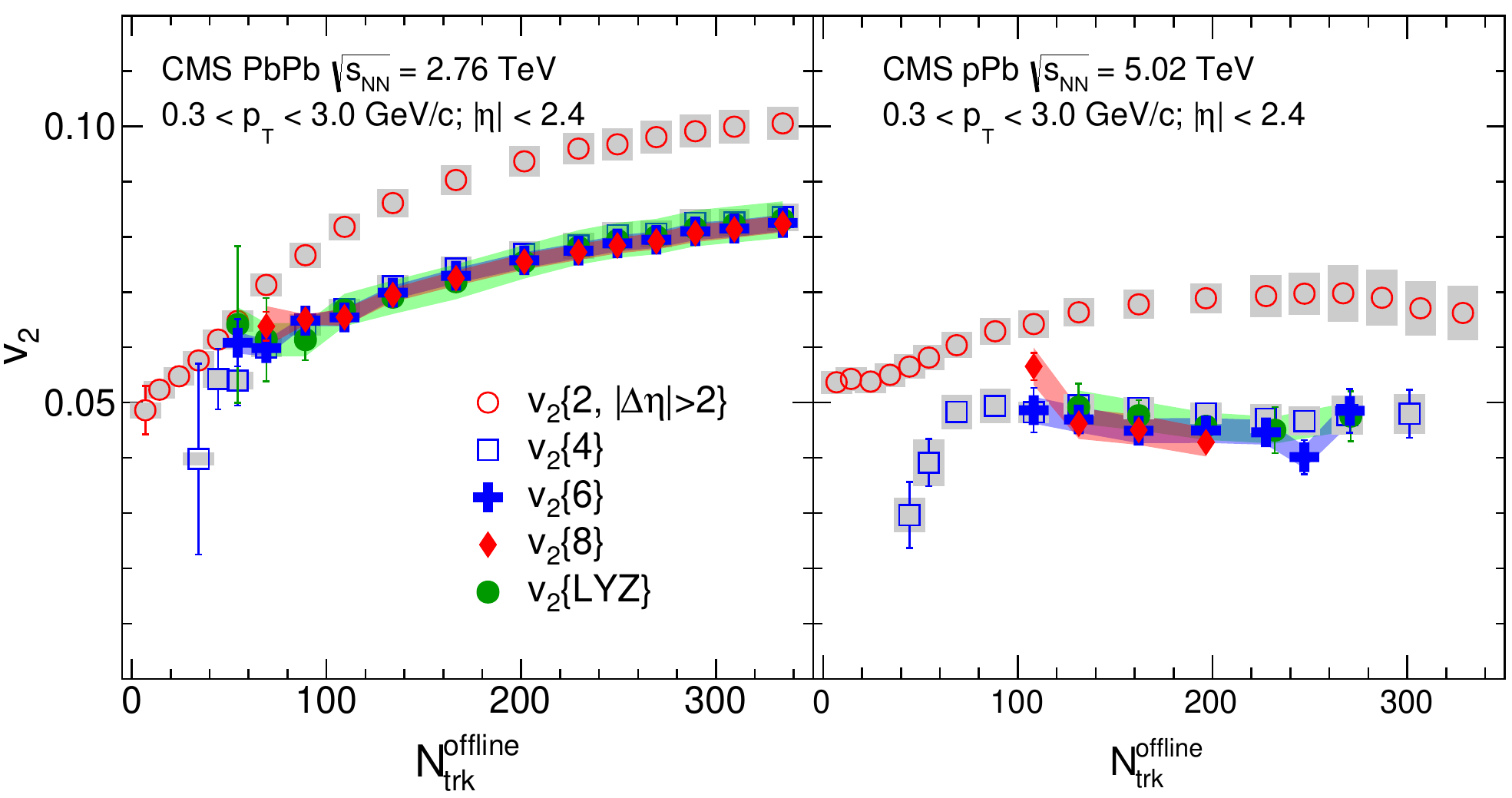}
  \caption{ \label{fig:v2n_3systems} The second-order Fourier harmonic, $v_2$, as a function of $N_{\rm trk}$
  obtained from two-, four-, six- and eight-particle cumulants, and the LYZ method, averaged over $0.3<\pt<3.0$~GeV/c, 
  in \PbPb at \rootsNN\ = 2.76 TeV (left) and \pPb at \rootsNN\ = 5.02 TeV (right)~\cite{Khachatryan:2015waa}.
   }
\end{figure}

Measurements of the four-particle cumulant have been performed 
by the ATLAS~\cite{Aad:2013fja}, CMS~\cite{Chatrchyan:2013nka} and ALICE~\cite{Abelev:2014mda} 
collaborations. It was later extended to six-, eight- and {\em all}-particle 
(so-called ``Lee-Yang Zeroes'' (LYZ)) cumulants~\cite{Khachatryan:2015waa}.
Fig.~\ref{fig:c2n} shows the measured multi-particle cumulants, $c_{4}$, $c_{6}$, and $c_{8}$,
as a function of multiplicity in \pPb and \PbPb collisions. 
These cumulants, $\cn{n}$, are calculated as follows:
\begin{equation}\begin{split}
\cn{2} =& \dmean{2},
\\
\cn{4} =& \dmean{4} - 2 \cdot \dmean{2}^2,
\\
\cn{6} =& \dmean{6} - 9 \cdot \dmean{4}\dmean{2} + 12 \cdot \dmean{2}^3,
\\
\cn{8} =& \dmean{8} - 16 \cdot \dmean{6}\dmean{2} -18 \cdot \dmean{4}^2 + 
%\\ &144 \cdot \dmean{4}\dmean{2}^2 - 144 \dmean{2}^4 , \label{eq:cn}
144 \cdot \dmean{4}\dmean{2}^2 - 144 \dmean{2}^4 , \label{eq:cn}
\end{split}\end{equation}

\noindent where two- and multi-particle azimuthal correlations are evaluated as:
\begin{equation}\begin{split}
\dmean{2} &\equiv
    \langle\langle e^{in(\phi_{1} - \phi_{2})} \rangle\rangle,
\\
\dmean{4} &\equiv
    \langle\langle e^{in(\phi_{1} + \phi_{2} - \phi_{3} - \phi_{4})} \rangle\rangle,
\\
\dmean{6} &\equiv
    \langle\langle e^{in(\phi_{1} + \phi_{2} + \phi_{3} - \phi_{4} - \phi_{5}
    - \phi_{6})} \rangle\rangle,
\\
\dmean{8} &\equiv
    \langle\langle e^{in(\phi_{1} + \phi_{2} + \phi_{3} + \phi_{4} - \phi_{5} - \phi_{6} - \phi_{7}
    - \phi_{8})} \rangle\rangle. \label{eq:corr}
\end{split}\end{equation}
\noindent Here $\langle\langle \cdots \rangle\rangle$ represents 
the average over all combinations of particles from all events.
The elliptic flow harmonics are related to the multi-particle cumulants by
\begin{eqnarray}
\vn{4} = \sqrt[4]{\cn{4}}\,,\,\,\,\,\,\,\,\,\, 
\vn{6} = \sqrt[6]{\frac{1}{4} \cn{6}}\,,\,\,\,\,\,\,\,\,\, 
\vn{8} = \sqrt[8]{-\frac{1}{33} \cn{8}}\,, \label{eq:vn}
\end{eqnarray}
which are shown in Fig.~\ref{fig:v2n_3systems}, for \pPb and \PbPb systems side-by-side as a function of multiplicity.

Within experimental uncertainties, $v_2$ values from various higher-order
cumulant methods are all consistent with each other in both \pPb and \PbPb as a
function of multiplicity lending support to the highly collective nature of
these systems.  The $v_2$ from two-particle correlations does not follow the
trends of the $v_2$ from higher cumulants.  Part of the deviation may be due to
the away-side jet which persists out to rapidity separations larger than the two unit
rapidity gap imposed in the measurement.  However, one would expect the
effect from mini-jets to be smaller at higher multiplicities.  The breaking of
$v_2\{2\}\neq v_2\{4\}$ is present in hydrodynamic models when event-by-event
dynamical fluctuations of $v_2$ contribute differently to two-particle and
higher-order cumulants.

It should be stressed that although the observed scaling in
Fig.~\ref{fig:v2n_3systems} is a necessary outcome of a hydrodynamic framework
it is not sufficient proof that hydrodynamics is the correct underlying theory.
The measurements are a milestone as they lend credence to the highly collective
nature of \pPb collisions and show that the ridge is a highly non-trivial
emergent phenomena.

\subsection{Identified particle spectra and correlations}

\begin{figure}[thb]
  \begin{center}
    \includegraphics[width=0.8\linewidth]{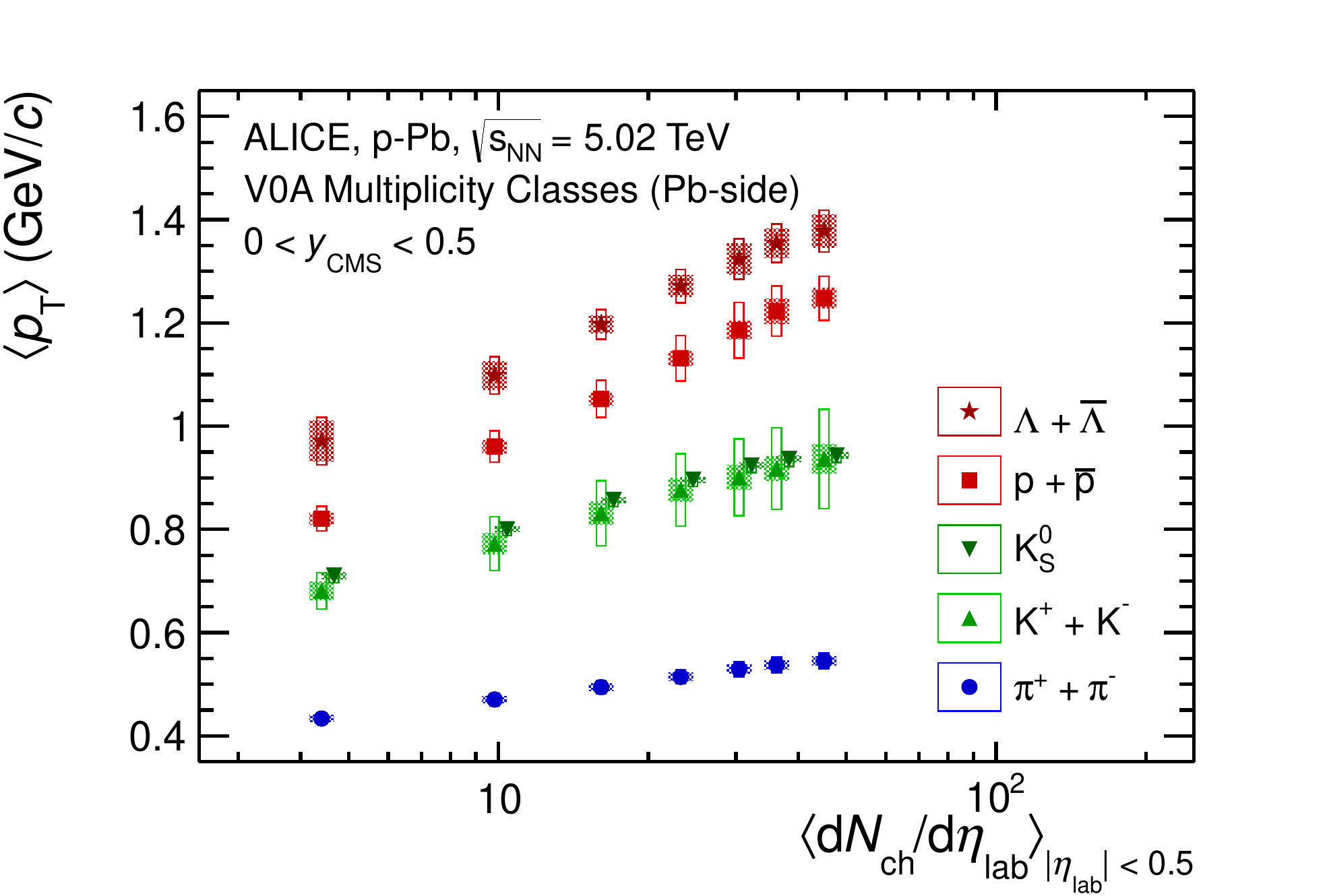}      
    \includegraphics[width=0.8\linewidth]{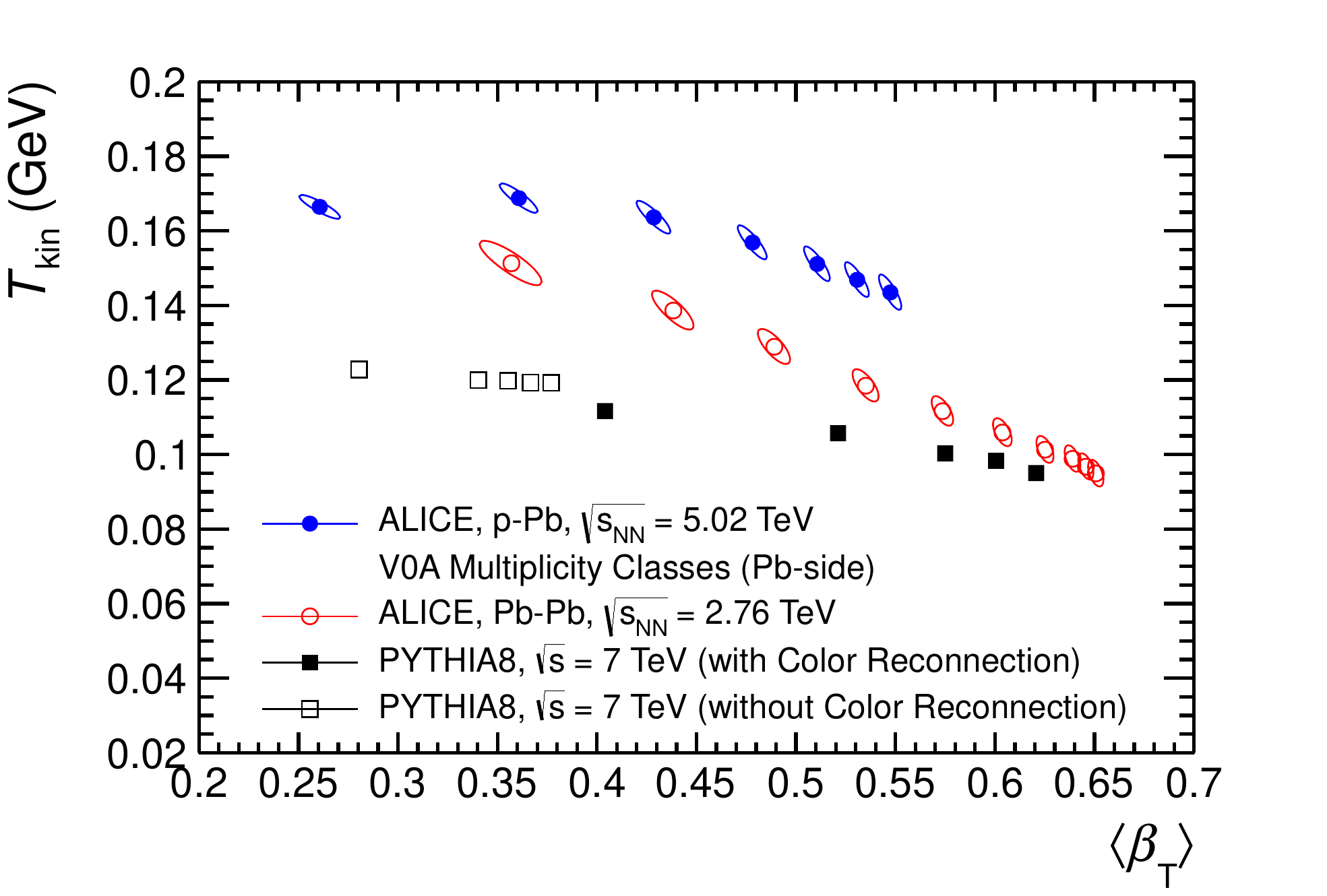}    
    \caption{ Top: Average transverse momentum of $\pi^{+}/\pi^{-}$, K$^{+}$/K$^{-}$,
    \Pp/\Pap, \PKzS\ and \PgL/\PagL\ particles as a function of multiplicity in pPb collisions at \rootsNN\ = 5.02 TeV.
    Bottom: parameters of blast-wave fits to identified particle spectra for pPb and PbPb collisions, and PYTHIA8 models~\cite{Abelev:2013haa}.
    } \label{fig:spectra}
  \end{center}
\end{figure}

Single-particle \pt\ spectra for various particle species as well as their multiplicity
dependence in each collision system provide rich information for constraining
the dynamics of particle production.
Multiplicity dependence of identified particle (PID) \pt\ spectra have been measured 
in \ppc, \pPb and \PbPb collisions at the LHC~\cite{Chatrchyan:2012qb,Chatrchyan:2013eya,Abelev:2013vea,Abelev:2013haa}.
In Fig.~\ref{fig:spectra} (top), the average \pt\ ($\left\langle \pt \right\rangle$) 
for $\pi^{+}/\pi^{-}$, K$^{+}$/K$^{-}$, \Pp/\Pap, \PKzS\ and \PgL/\PagL\ particles 
as measured by the ALICE collaboration is shown as a function of event multiplicity 
in \pPb collisions at \rootsNN\ = 5.02~TeV~\cite{Abelev:2013haa}.

As multiplicity increases, \pt\ spectra of all particle species become flatter
(or ``harder''). A similar trend is also found in MC models such as PYTHIA and
HIJING, HIJING, which may be due to the biasing towards higher $Q^2$ processes
when triggering on high-multiplicity events. However, the most distinct feature
of the data is not just the fact that $\left\langle \pt \right\rangle$
increases with multiplicity, but the increase is observed to be faster for
particles with a heavier mass. For example, the ratio of proton to pion \pt\
spectra is enhanced in the higher \pt\ region, going from low- to
high-multiplicity events.  Moreover, the CMS collaboration reported that the
amount of increase in $\left\langle \pt \right\rangle$ for each particle
species from low to high multiplicity events is approximately proportional to
the particle's mass~\cite{Chatrchyan:2013eya}.

\begin{figure}[thb]
\center
    \includegraphics[width=0.7\textwidth]{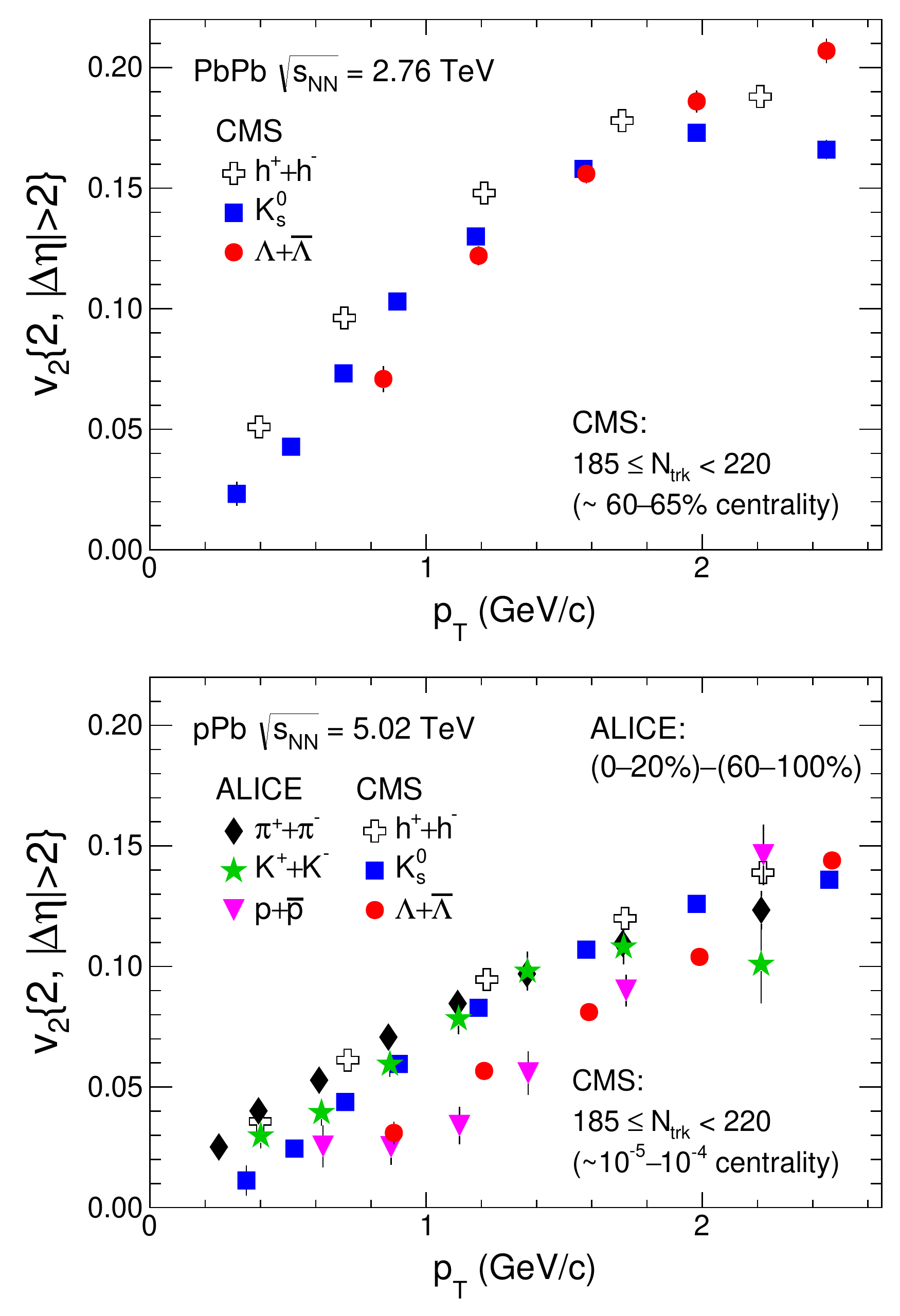}
    \caption{The second-order Fourier harmonic, $v_2$, for charged hadrons, $\pi^{+}/\pi^{-}$, K$^{+}$/K$^{-}$,
    and \Pp/\Pap\ from ALICE~\cite{ABELEV:2013wsa}, \PKzS\ and \PgL/\PagL\ from CMS~\cite{Khachatryan:2014jra}, 
    as a function of \pt\ for $185 \leq N_{\rm trk} < 220$ in \PbPb collisions at \rootsNN\ = 2.76~TeV and \pPb 
    collisions at \rootsNN\ = 5.02~TeV. 
    } \label{fig:PIDv2}
\end{figure}

Similar mass dependence of PID spectra in \AAc collisions has been extensively
studied.  In hydrodynamic models, it is attributed to the effect of a common
velocity field along the radial direction of the medium, which generates a
boost in particle momentum that is roughly proportional to the mass of the
particle.  A common framework for describing this characteristic mass
dependence of \pt\ spectra is the ``blast-wave''
model~\cite{Schnedermann_blastwave}, with parameters of a common kinetic
freeze-out temperature, $T_{\rm kin}$, and an average radial expansion
velocity, $\left\langle \beta_{\rm T} \right\rangle$ ($\beta_{\rm T}$ also
depends on the radius).

Fitting to the experimental spectra, $\left\langle \beta_{\rm T} \right\rangle$
and $T_{\rm kin}$ can be extracted, and are shown in Fig.~\ref{fig:spectra}
(bottom) for \pPb and \PbPb collisions for different multiplicity or centrality
ranges. In \PbPb, more central collisions tend to have a larger $\left\langle
\beta_{\rm T} \right\rangle$ (due to larger pressure gradients) and a smaller
$T_{\rm kin}$ (because of a larger system size resulting in a longer lifetime
of the hadronic rescattering stage). A similar trend is also observed in \pPb
collisions. Interestingly, if comparing \pPb and \PbPb systems at similar
multiplicities (about the same $T_{\rm kin}$), the radial flow velocity is
larger in the smaller \pPb system. If fixing the total energy or entropy, a
smaller QGP fluid possesses a stronger radial pressure
gradient~\cite{Shuryak:2013ke,Kalaydzhyan:2014zqa,Kalaydzhyan:2015xba}, which
is consistent with the data. To verify this picture, one important step will be
to carry out the same studies in high-multiplicity \ppc collisions, which have
an even smaller system size. 

While the hydrodynamic fluid picture provides a natural interpretation to the data, as shown in Fig.~\ref{fig:spectra} (bottom),
the PYTHIA model with color reconnections~\cite{Skands:2007zg,Schulz:2011qy} can also generate similar behavior as seen in the data. 
In PYTHIA an outgoing hard parton and the underlying event
are connected through color strings. As a result, strings
will be pulled out of the underlying event in the transverse direction by the fast-moving parton, and then 
fragment into final-state particles. The color reconnection method increases this effect by reconnecting partons close in phase space.
In this way, the ``radially boosted'' string provides an effective common velocity for the hadrons produced from it, just like a fluid cell in the hydrodynamic picture. Meanwhile, models based on gluon saturation can also qualitatively describe
the general trend of the data~\cite{McLerran:2013oju}. 

To further differentiate theoretical models in \pPb collisions, studies of
identified particles have been extended to two-particle correlations. A
long-range near-side two-particle correlation involving an identified particle
is also observed~\cite{ABELEV:2013wsa,Khachatryan:2014jra}.  Results for the
single-particle $v_2$ harmonic for $\pi^{+}/\pi^{-}$, K$^{+}$/K$^{-}$,
\Pp/\Pap, \PKzS\ and \PgL/\PagL\ particles as a function of \pt in \PbPb and
\pPb collisions at similar multiplicities are shown in Fig.~\ref{fig:PIDv2}.
For both systems, a particle species dependence of $v_2$ is observed. At a
given \pt, particles having a larger mass tend to exhibit a smaller $v_2$
anisotropy.  This mass ordering effect of $v_2$ was first seen in \AAc
collisions at RHIC and LHC
energies~\cite{STAR,PHENIX,Adler:2003kt,Adler:2001nb,Abelev:2014pua}, which can
again be understood as the effect of radial flow pushing heavier particles
toward higher-\pt\ ~\cite{Huovinen:2001cy,Kolb:2003dz,Shen:2011eg}.  Comparing
\pPb and \PbPb systems at similar multiplicities in Fig.~\ref{fig:PIDv2}, a
large mass splitting can be seen in the smaller \pPb system, which is again
consistent with the hydrodynamic picture~\cite{Werner:2013ipa,Bozek:2013ska}.
It remains to be seen whether other theoretical interpretations, such as string
fragmentation with color reconnections, can describe the features found in both
single-particle and two-particle correlation data, as well as their system size
dependence.

\begin{figure}[thb]
  \begin{center}
    \includegraphics[width=0.48\linewidth]{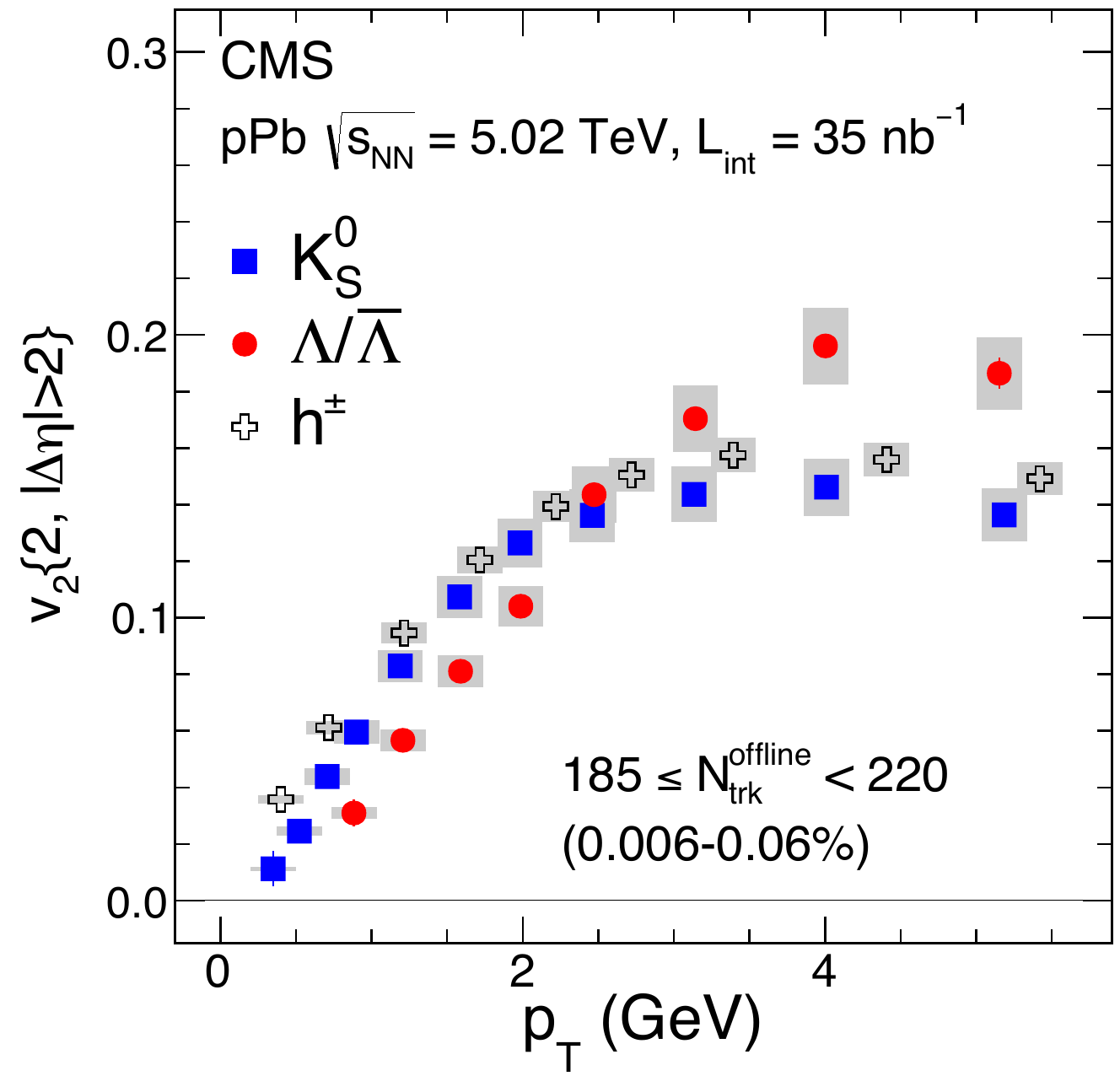}  
    \includegraphics[width=0.48\linewidth]{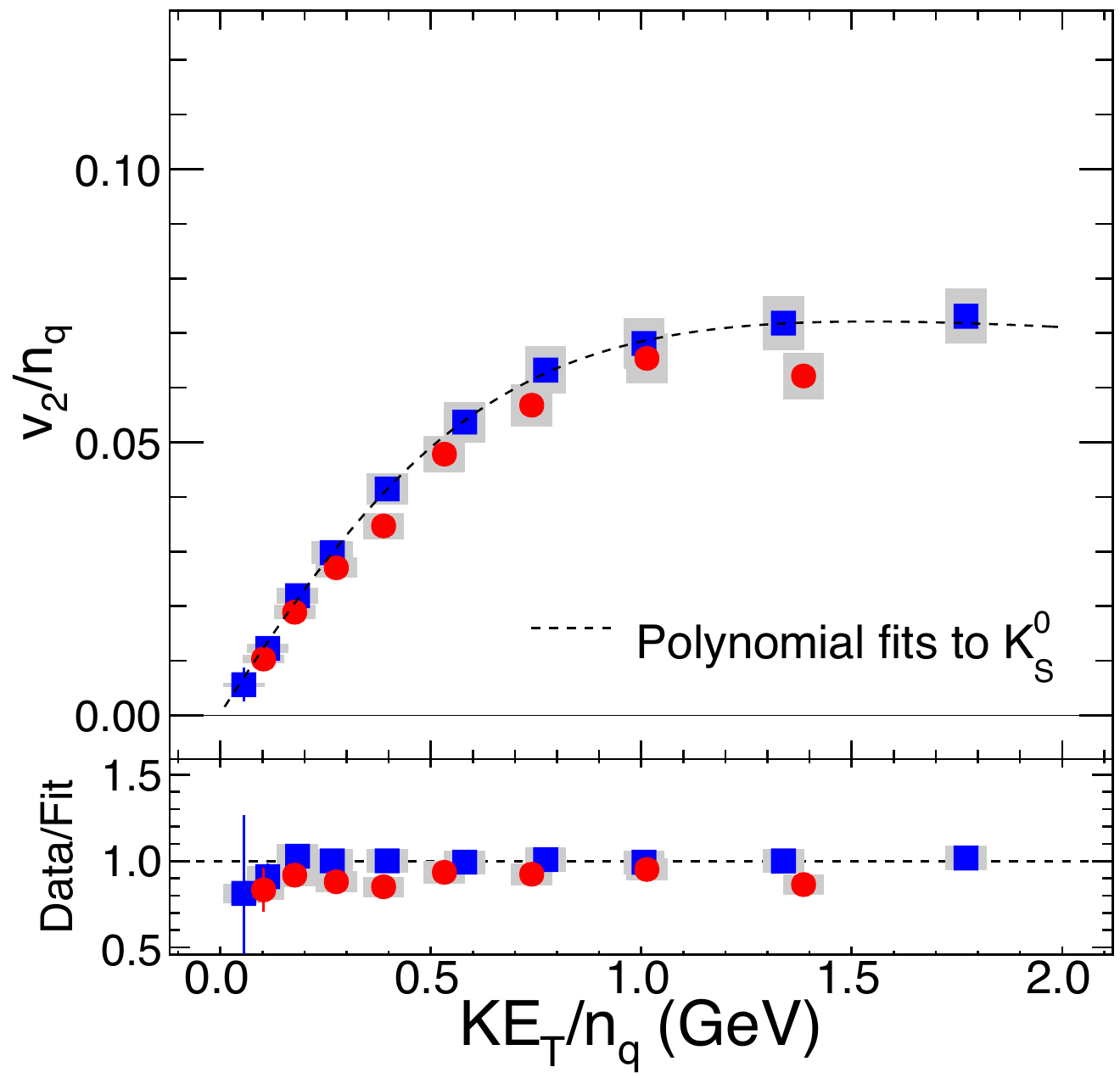}    
    \caption{ Top: The second-order Fourier harmonic, $v_2$, for \PKzS, \PgL/\PagL, and unidentified charged particles as a function of \pt\ 
    for $185 \leq N_{\rm trk} < 220$ obtained in \pPb collisions at \rootsNN\ = 5.02~TeV. Bottom: the $n_{q}$-scaled $v_2$ values 
    of \PKzS\ and \PgL/\PagL\ as a function of (\ket)/$n_{q}$. Ratios of $v_2$/$n_q$ to a smooth fit 
    function of $v_2$/$n_q$ for \PKzS\ as a function of (\ket)/$n_{q}$ are also shown~\cite{Khachatryan:2014jra}.
    } \label{fig:v2_ncq}
  \end{center}
\end{figure}

In heavy ion collisions, a scaling phenomenon of $v_2$ with the number of
constituent quarks ($n_{q}$) has been
discovered~\cite{Adams:2003am,Abelev:2007qg,Adare:2006ti}. The values of
$v_2$/$n_{q}$ are found to be very similar for all mesons ($n_{q}=2$) and
baryons ($n_{q}=3$) when compared at the same transverse kinetic energy per
constituent quark ($\ket$/$n_{q}$, where $\ket = \sqrt{\smash[b]{m^2 + \pt^2}}
- m$). This empirical scaling may indicate that final-state hadrons are formed
through recombination of quarks~\cite{Molnar:2003ff,Greco:2003xt,Fries:2003vb},
possibly providing evidence of deconfinement of quarks and gluons in these
systems.  This quark number scaling is found to be also valid within about 10\%
in high-multiplicity \pPb collisions as shown in Fig.~\ref{fig:v2_ncq} for
\PKzS\ and \PgL/\PagL\ particles. At similar multiplicities, the scaling holds
better in the smaller \pPb system than \PbPb system.  Although the idea of
quark coalescence is remarkably simple, the detailed dynamics of quark
recombination are far from fully understood.  The observed constituent quark
scaling in small systems may help elucidate this phenomena.

\subsection{Flow factorization}

When experimentally extracting the single-particle anisotropy $v_n$ from
multi-particle correlations the assumption is made that the multi-particle
momentum distribution can be factorized into a product of single-particle
distributions in each event, 
\begin{equation}
\label{eq:fact}
\frac{d^{3n}N}{d{\bf p_{1}}d{\bf p_{2}}\cdots d{\bf p_{n}}} = \frac{d^{3}N}{d{\bf p_{1}}}\frac{d^{3}N}{d{\bf p_{2}}}\cdots\frac{d^{3}N}{d{\bf p_{n}}}\,.
\end{equation}

The factorization assumption in equation 4 is broken by short-range
correlations, for example, from jets, resonances, and momentum conservation.
However, over a wide rapidity range (e.g., $|\Delta\eta|>2$) in \AAc
collisions, the predominant source of long-range correlations comes from an
expanding fluid-like system with a common preferred flow direction (i.e.,
short-axis of the elliptical overlap region, or ``event plane'').  Other
possible sources include back-to-back jet correlations on the away side but
their contributions are negligible, especially for central \AAc collisions, since
they are strongly suppressed by the large multiplicity ($1/N_{\rm trk}$) and
also by the effect of jet quenching. Therefore, the relation in
Eq.~\ref{eq:fact} holds and single-particle $v_n$ can be extracted from the
measurement of particle correlations.  The factorization assumption has always
been explicitly or implicitly applied in all methods of measuring anisotropic
flow. Any breakdown of factorization would be an indication of correlations not
originated from hydrodynamic flow.

\begin{figure}[thb]
\center
\includegraphics[width=\linewidth]{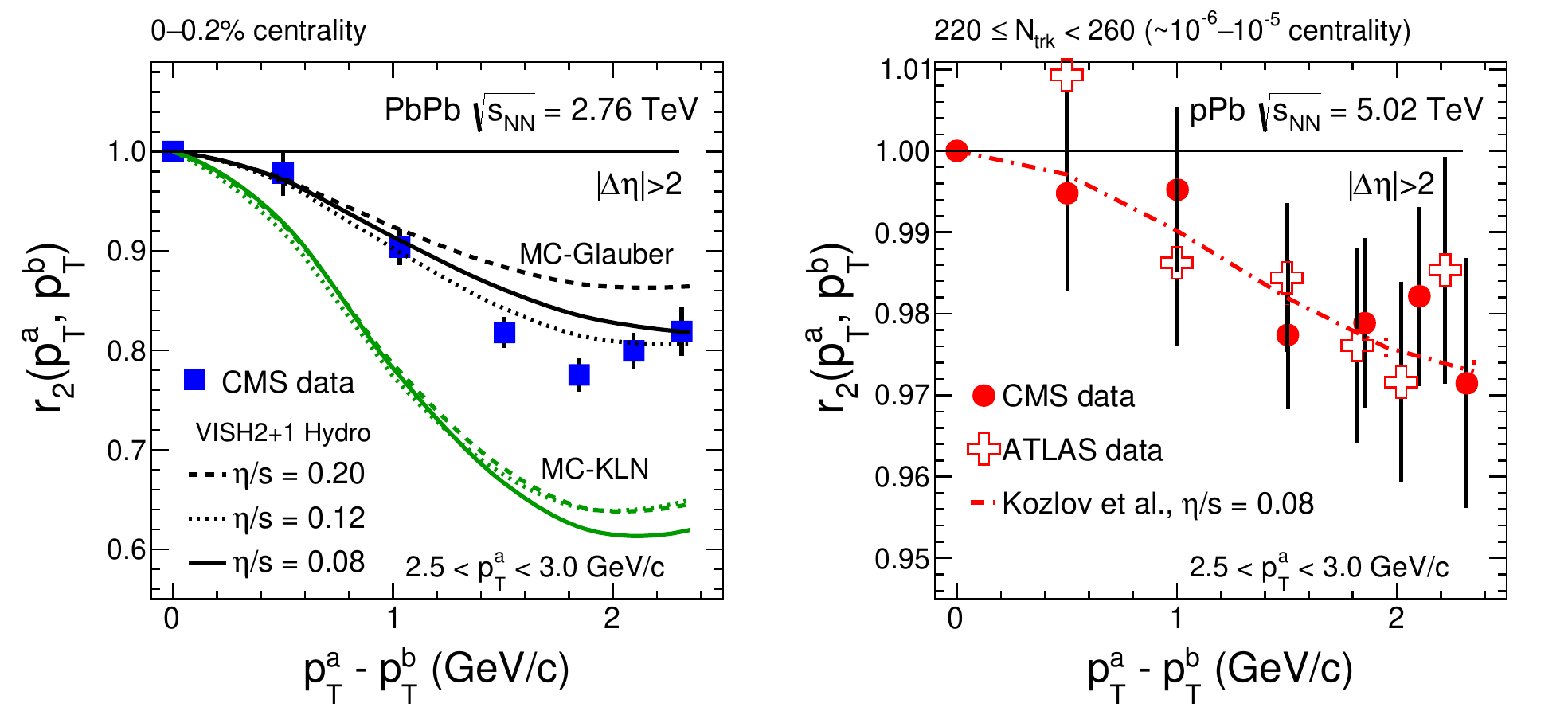}
  \caption{ \label{fig:flowfact} 
  Elliptic flow factorization ratio, $r_{2}(\pta,\ptb)$, as a function of \pta\ - \ptb\ , for
  $2.5<\pta<3.0$~GeV/c in ultra-central (0--0.2\% centrality) \PbPb collisions at \rootsNN\ = 2.76~TeV~\cite{Khachatryan:2015oea} (left)
  and high-multiplicity \pPb collisions at \rootsNN\ = 5.02~TeV~\cite{CMS:2013bza,Aad:2014lta,Khachatryan:2015oea} (right), compared to theoretical 
  calculations in hydrodynamic models \cite{Heinz:2013bua,Kozlov:2014fqa}.
   }
\end{figure}

However, it was realized recently that even if all particles were to  
originate from a hydrodynamically flowing medium, the lumpy
energy density distribution in the initial state would also break the 
factorization relation~\cite{Gardim:2012im,Heinz:2013bua}.
This is because particles produced at different \pt\ and $\eta$ do not in general share a common
flow direction, or event plane. The breakdown of the factorization relation 
as a function of particle's \pt\ and $\eta$ may provide information 
on the details of initial-state fluctuations, or more specifically, the
lumpiness of the initial-state geometry.

The \pt\ dependence of factorization breakdown has been quantified in terms 
of a factorization ratio,
\begin{equation}
    \label{r_n_def}
    r_{n}(\pta,\ptb) \equiv \frac{V_{n\Delta}(\pta,\ptb)}{\sqrt{V_{n\Delta}(\pta,\pta)V_{n\Delta}(\ptb,\ptb)}},
\end{equation}

\noindent where $V_{n\Delta}(\pta,\ptb)$ represents the Fourier coefficient 
of two-particle azimuthal correlations for a given (\pta,\ptb) range.
The data are shown in Fig.~\ref{fig:flowfact} for ultra-central (0--0.2\% centrality) 
\PbPb (left) and high-multiplicity \pPb (right) collisions.
If the factorization relation holds, $r_{2}(\pta,\ptb)$ will be constant at unity. Up to 20\% deviation from 
unity of $r_{2}$ has been observed in ultra-central \PbPb collisions, which 
is consistent with hydrodynamic calculations using a Glauber initial condition, 
while the MC-KLN initial condition predicted a much larger effect. \cite{Heinz:2013bua} This quantify
is found to be insensitive to the $\eta/s$ value, 
providing very powerful constraints to models of initial-state fluctuations.
Many follow-up theoretical 
studies showed that the magnitude of factorization breakdown is strongly correlated with
the granularity of initial-state
fluctuations~\cite{Kozlov:2014fqa,Floerchinger:2013rya,ColemanSmith:2012ka,Mazeliauskas:2015vea}. 
A lumpy initial state with larger radial excitations
would lead to a more significant breakdown of flow factorization, as seen in ultra-central 
\PbPb collisions. On the other hand, very little effect is found in high-multiplicity 
\pPb collisions (Fig.~\ref{fig:flowfact}, right). This may reflect
a relatively larger granularity of initial-state fluctuations in \pPb with respect to 
its smaller system size. Or in other words, a \pPb system is relatively smoother than
an ultra-central \PbPb system. Note that for both systems, there is almost no 
average geometry.
Any final-state anisotropy is entirely generated by fluctuations. 

\subsection{Femtoscopy from Hanbury-Brown-Twiss correlations}\label{sec:HBT}

\begin{figure}[thb]
\center
\includegraphics[width=\linewidth]{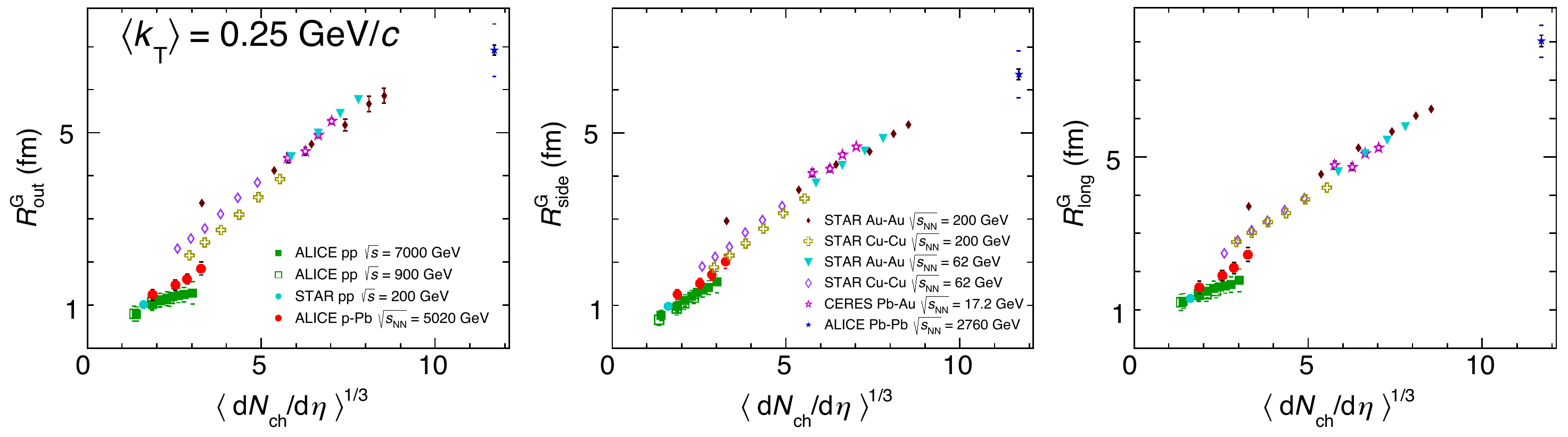}
  \caption{ \label{fig:hbt} Comparison of femtoscopic radii (extracted from a Gaussian fit 
  to the two-particle correlation function), as a function of charged-particle multiplicity, 
  measured for various collision systems and energies~\cite{Adam:2015pya}.
   }
\end{figure}

Hanbury-Brown-Twiss (HBT) correlations provide crucial space time information of the source at the freezeout surface. 
The extracted femtoscopic radii in various collisions systems from \ppc, \pA and \AAc at RHIC and the LHC 
are summarized in Fig.~\ref{fig:hbt} as a function of event multiplicity~\cite{Adam:2015pya}.
The femtoscopic radii follow a linear $N_{\rm trk}^{1/3}$ dependence for all systems, although
the slopes are somewhat different. At similar multiplicities, the femtoscopic radii 
(extracted from a Gaussian fit to the two-particle correlation function) in \pPb collisions
is about 15--20\% larger than that in \ppc collisions. Both are significantly smaller than the value
in \PbPb collisions, possibly due to a smaller initial system size. We will return to the discussion of HBT 
correlations when we present the details of hydrodynamic model calculations in later sections.

\subsection{Heavy flavor production}

Production of heavy flavor quarks in \ppc and \pPb collisions has also
been studied as a function of multiplicity. 
Studies of heavy quarkonia states (e.g., $J/\Psi$, $\Upsilon$) in heavy ion collisions 
can probe the possible onset of color screening from a deconfined
QGP medium~\cite{Matsui:1986dk}. An intriguing multiplicity dependence of the $\Upsilon$(2S) to $\Upsilon$(1S) yield ratio
from \ppc, \pPb to \PbPb collisions has been observed by the CMS collaboration~\cite{Chatrchyan:2013nza}.
The yield of the $\Upsilon$(2S) state is suppressed relative to the $\Upsilon$(1S) state as multiplicity
increases in \ppc and \pPb collisions. Although limited in multiplicity reach, the amount of 
relative suppression has a tendency of smoothly extrapolating to the values measured in \PbPb collisions
at higher multiplicities. This may be a hint that similar medium effects on the $\Upsilon$ states
become significant in \ppc and \pPb collisions as multiplicity increases. However, more studies are still needed to
exclude any possible bias introduced by the multiplicity selection.

\begin{figure}[thb]
\center
  $\vcenter{\hbox{\includegraphics[width=0.46\linewidth]{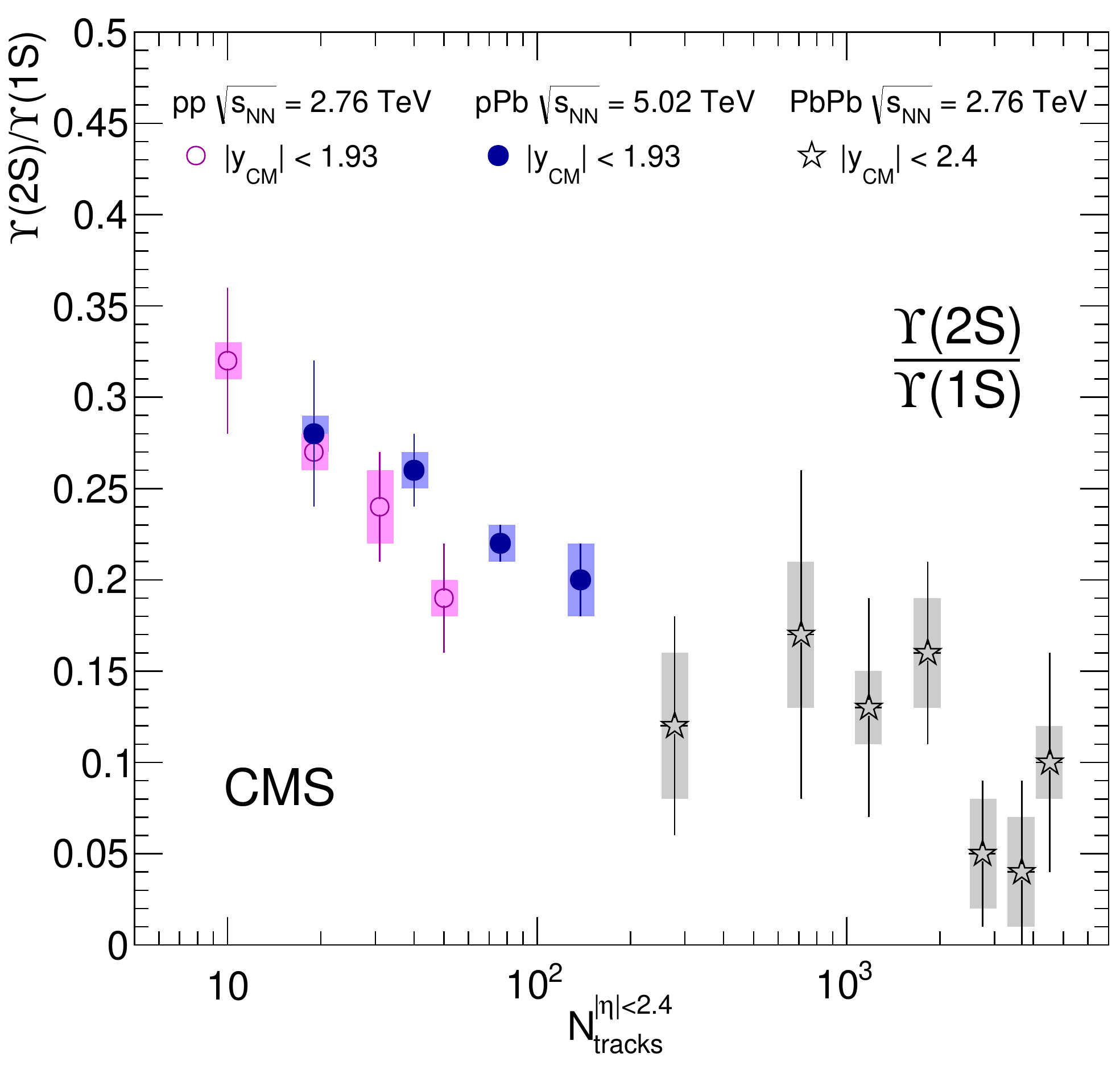}}}$
  \hspace{0.7cm}
  $\vcenter{\hbox{\includegraphics[width=0.46\linewidth]{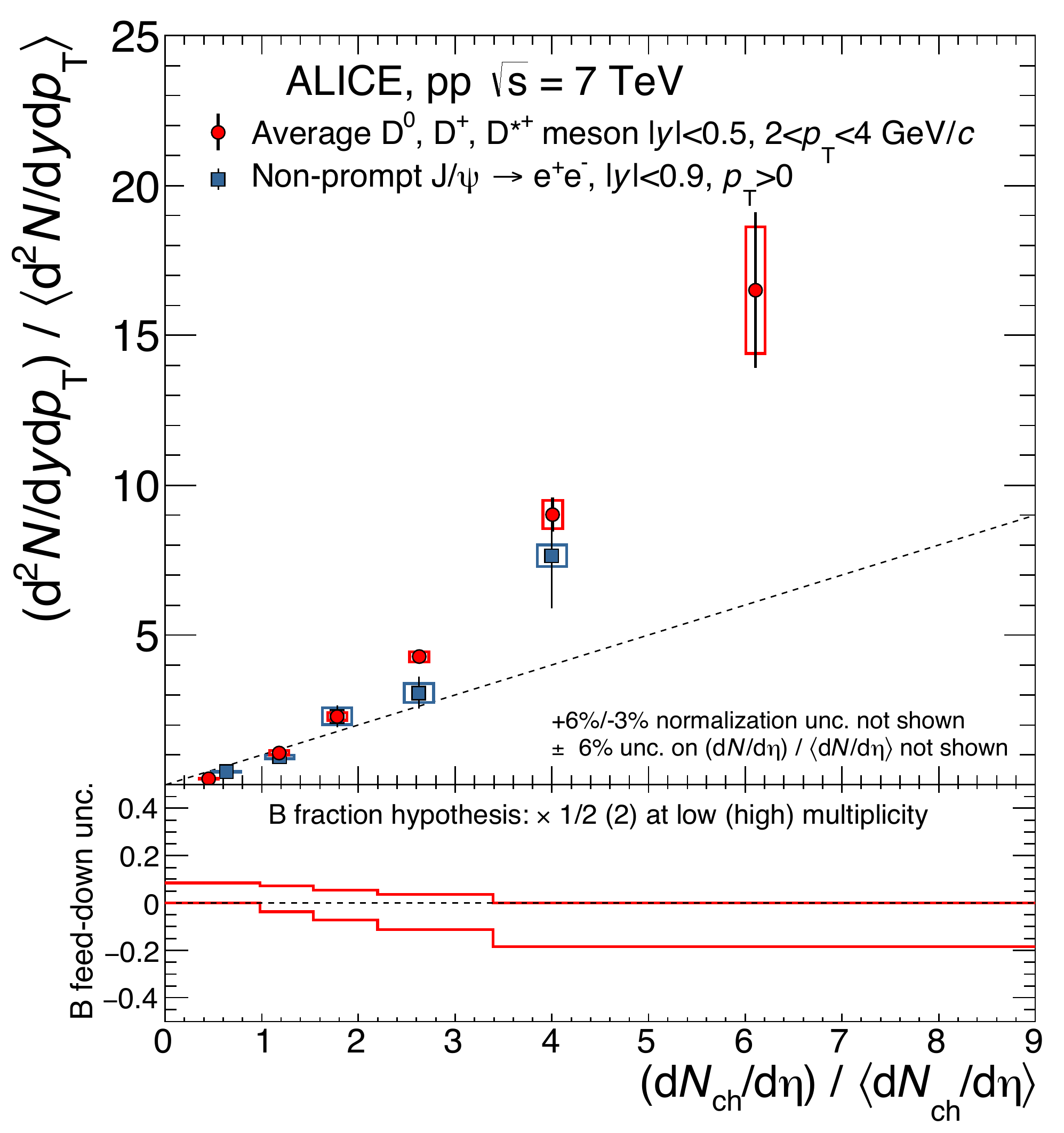}}}$
  \caption{ \label{fig:heavyflavor} 
  Left: ratios of $\Upsilon$(2S) to $\Upsilon$(1S) yields as a function of charged-particle 
  multiplicity in \ppc collisions at \roots\ = 7~TeV, \pPb at \rootsNN\ = 5.02~TeV and \PbPb at \rootsNN\ = 2.76~TeV~\cite{Chatrchyan:2013nza}.
  Right: average $D$ meson and non-prompt $J/\Psi$ relative (to minimum bias events) 
  yields as a function of the relative (to minimum bias events)
  charged-particle multiplicity at midrapidity in \ppc collisions at \roots\ = 7~TeV~\cite{Adam:2015ota}.
   }
\end{figure}

Open heavy flavor production has been measured in \ppc collisions as a function of multiplicity 
by the ALICE collaboration~\cite{Adam:2015ota}. The yields of open charm $D^{0}$, $D^{+}$
and $D^{*+}$ mesons as a function of multiplicity in \ppc collisions at \roots\ = 7\TeV\ are
shown in Fig.~\ref{fig:heavyflavor}. Both the yield (y-axis) and multiplicity (x-axis)
are normalized by the average values from minimum bias events. The open charm meson
yields first linearly increase with multiplicity, which is consistent with the expectation
of independent multiple parton interactions. At above 3--4 times the average minimum bias
multiplicity, the increase in yield becomes much faster than linear. Around a similar 
multiplicity region, the long-range ridge correlations also start becoming significant. 
These observations suggest the emergence of new dynamic processes in very high-multiplicity
\ppc collisions.

%\clearpage

%%% Local Variables: 
%%% mode: latex
%%% TeX-master: "../RidgeReview-ijmpe"
%%% End: 

\section{Theoretical interpretations}
\label{theory}
Since the 2010 discovery of the ridge in \ppc collisions by the CMS collaboration there
has been a flurry of theoretical work to uncover its origin.  Five years later,
as the dust has started to settle, the heavy-ion community has ruled out and
refined various models.  In this section we will provide an overview of some of
the theoretical ideas on the market, focusing on those that confront the data
on a quantitative level.

The dialogue in the heavy-ion community has been divided between initial-state
and final-state models depending on whether the momentum space anisotropy is established
at the moment of particle production or subsequently during strong final state interactions, 
or (while clearly not equivalent but sometimes used
interchangeably) non-hydrodynamic vs hydrodynamic descriptions.  
However, hydrodynamics is an effective description of the long wavelength, small
frequency limit of an underlying theory.  Concluding that a system behaves
hydrodynamically does not provide any information about the underlying physical
processes. Also, for some cases, it is debatable if the model is initial- or final-state.
With that being said, we will adopt the usual classification and identify models as
non-hydrodynamic (initial-state) or hydrodynamic (final-state) in order to 
classify different approaches conveniently. 

All theoretical models that we consider have a common theme that makes them
particularly interesting; correlations between pairs of rapidity-separated
particles must have formed at early times. \cite{Dumitru:2008wn}.  In
particular, if two correlated particles {\em freeze out} at a proper time
$\taufo$ then causality dictates that the correlation must have formed at an
earlier proper time $\tauzero$ constrained by 
\begin{equation}
\tauzero \leq \taufo \exp\left(-\frac{1}{2}\vert y_a - y_b\vert\right)\;,
\label{eq:corrst1}
\end{equation}
where $y_{a,b}$ is the momentum-space rapidity of particles $a,b$.  This
relation can be derived by considering two particles at space-time rapidities
$\eta_a$ and $\eta_b$.  Assume particle $a$ emits a signal at time
$t_a=\tauzero\cosh\eta_a$ which is subsequently received by particle $b$ at
time $t_b=\taufo\cosh\eta_b$.  The distance the signal must travel is $\vert
z_a-z_b\vert=\vert \taufo\sinh\eta_b-\tauzero\sinh\eta_a\vert$. Causality
restricts $\vert z_b-z_a\vert\leq c\left(t_b-t_a\right)$ and therefore
$\tauzero\leq \taufo\exp\left(-\vert\eta_b-\eta_a\vert\right)$.  

The factor of $1/2$ in \eq{eq:corrst1} is obtained by introducing a third
mutually correlated particle at rapidity $\eta_c$.  If this third particle were
to emit two signals at time $\tauzero$ particle $a$ and $b$ can be correlated
as long as $\tauzero\leq \taufo\exp\left(-\vert\eta_a-\eta_c\vert\right)$ {\em
and} $\tauzero\leq \taufo\exp\left(-\vert\eta_b-\eta_c\vert\right)$.  We arrive
at the most stringent condition for $\tauzero$ when the third particle,
particle $c$, is located at mid-rapidity, $(\eta_a+\eta_b)/2$, and find 
\begin{equation}
\tauzero\leq \taufo \exp\left(-\frac{1}{2}\vert \eta_a - \eta_b\vert\right)\;.
\label{eq:corrst2}
\end{equation}

\begin{figure*}[t]
\centering
\includegraphics[height=6.cm]{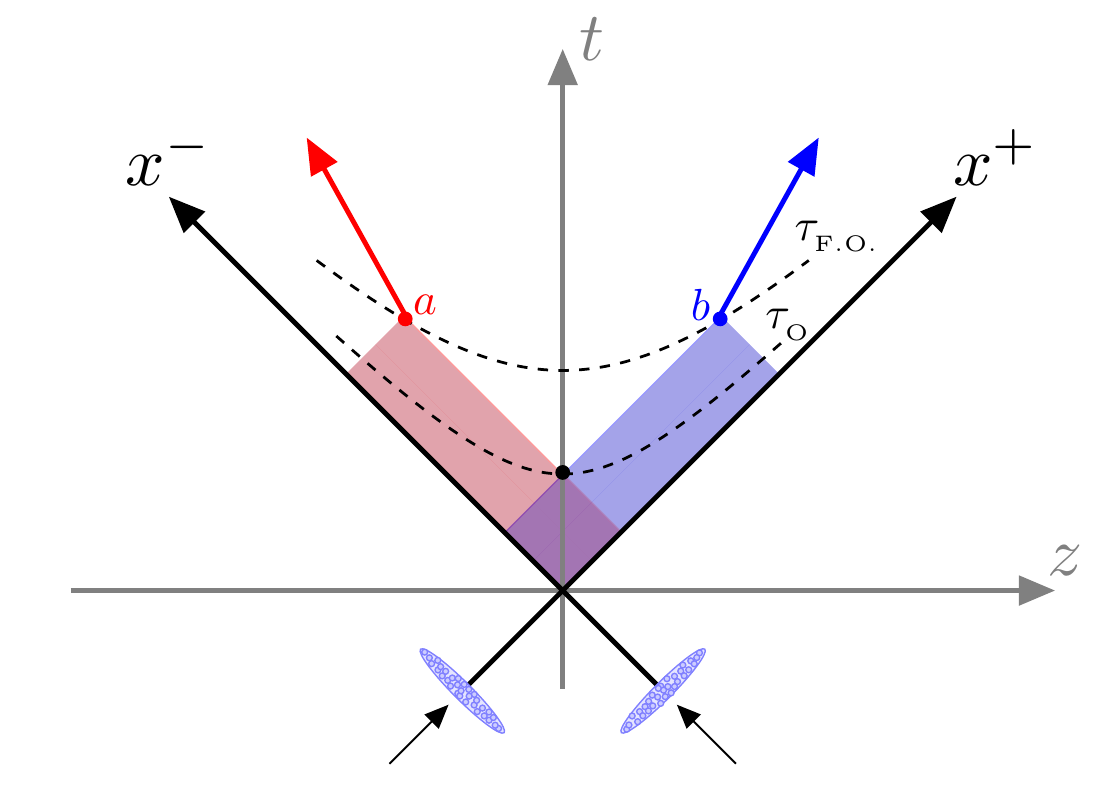}
\caption{Space-time diagram showing the causal relation between particles $a$
and $b$.  Particle $a$ and $b$ are causally connected to the red and blue
patches respectively.  The correlation between them must have formed in their
overlap at a proper-time before $\tauzero$.} 
\end{figure*}

The final step in arriving at the desired result of \eq{eq:corrst1}, is
equating the space-time rapidity with the momentum-space rapidity ({\em i.e.}
$\eta\sim y$ in the notation used throughout this section).  This is natural at
high energies; Lorentz contraction makes the nuclei infinitely thin in the
center of mass frame and there is no dimensionful scale in the longitudinal
direction.  Under a boost-invariant expansion the longitudinal flow of the
system has the scale invariant form $v_L=z/t$ and therefore $\eta=y$.

However, it is not a priori the case that this scaling is maintained by the
subsequent dynamics.  Once an additional scale is introduced the strict scaling
relation $v_L=z/t$ may no longer hold.  For example, in thermal
equilibrium the particle distribution has the form 
\begin{equation}
f\sim \exp\left(-\frac{m_\perp}{T}\cosh(y-\eta)\right)
\end{equation} 
A thermal smearing of about one unit in rapidity can be expected as long as
$m_\perp/T\gtrsim 1$.  On the other hand, if $p_T\ll T$ the correspondence
between $\eta$ and $y$ is completely lost.

\begin{figure*}[t]
\centering
\includegraphics[height=5.cm]{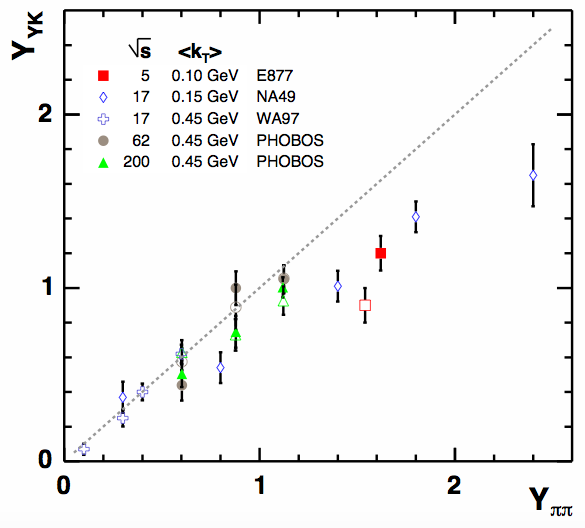}
\caption{Yano-Koonin source velocity, $Y_{YK}$ versus pair momentum
$Y_{\pi\pi}$ demonstrating approximate boost-invariant longitudinal
expansion.~\cite{Lisa:2005dd}} \label{fig:yykvsypp}
\end{figure*}

The best empirical evidence for a rigid correlation between the space-time
rapidity $\eta$ and momentum-space rapidity $y$ comes from studying the
Yano-Koonin-Podgoretski\u{\i} (YKP) parameterization~\cite{Yano:1978gk
,Podgoretsky:1982xu} of measured Hanbury Brown-Twiss (HBT) correlation
functions.  The relationship between the Yano-Koonin source velocity,
$Y_{YK}=\tanh^{-1}v_{YK}$, and pair momentum
$Y_{\pi\pi}=\frac{1}{2}\log\left(\frac{E_a+E_b+p_{za}+p_{zb}}{E_a+E_b-p_{za}-p_{zb}}\right)$
allows for testing the longitudinal expansion\cite{Wu:1996wk}.  For a
non-expanding source $Y_{YK}$ would be independent of $Y_{\pi\pi}$.  For a
source undergoing a boost invariant expansion $Y_{YK}=Y_{\pi\pi}$, and this
appears to be satisfied by the available data at RHIC as shown in
\fig{fig:yykvsypp}. 
 
There seems to be a compelling case that the bulk of particle production is
approximately boost invariant and the above causality arguments are well
founded. The proceeding discussion has far-reaching implications on our
understanding of long-range correlations.   

The PHOBOS collaboration has observed that the near-side ridge signal in
heavy-ion collisions persists up to six units in rapidity
separation.~\cite{Alver:2008aa,Alver:2009id} Using a very conservative estimate
of $\taufo\sim 10$~fm/c these correlations must have been established well
before $\tau\sim 0.5$~fm/c.  Long-range rapidity correlations thus probe early
space-time dynamics and are therefore particularly well suited for studying
novel QCD processes.  For example, it has been argued that
long range correlations may be a useful tool for probing high energy evolution
of multi-parton correlations in the nuclear wavefunction.~\cite{Dusling:2009ni}

%%% Local Variables: 
%%% mode: latex
%%% TeX-master: "../RidgeReview-ijmpe"
%%% End: 

\subsection{Final-state interactions: hydrodynamics in small systems}
\label{sec:hydropA}

\begin{figure}[ht]
\centering
\includegraphics[width=0.7\textwidth]{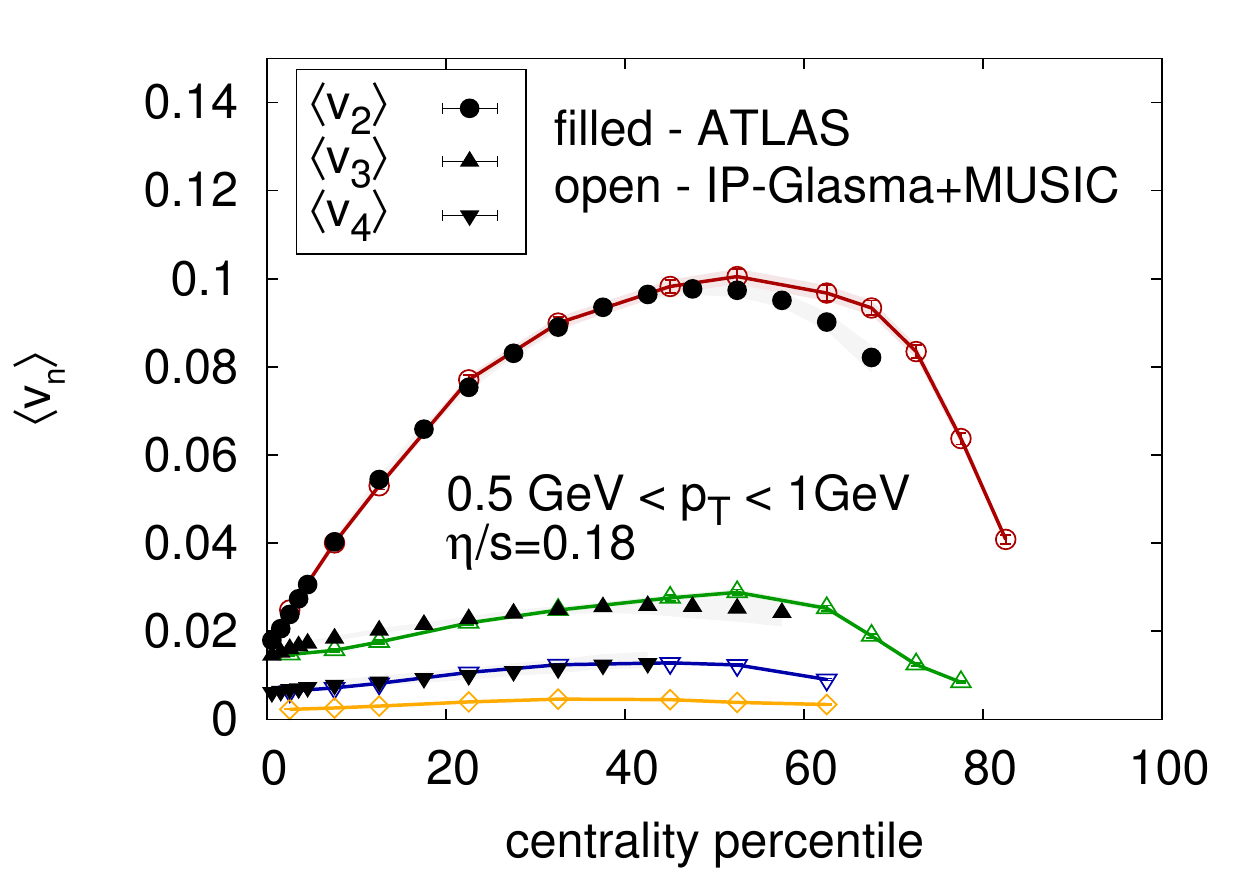}
\caption{Model calculations using a constant shear viscosity to entropy density ratio $\eta/s=0.18$ for $v_n$ vs. collision centrality in \PbPb collisions at $\sqrt{s_{NN}}=2.76$ TeV compared to experimental measurements\,\cite{Schenke:2014zha,Aad:2013xma}.}
\label{fig:hydro:vn}
\end{figure}

From the earliest days of RHIC, hydrodynamics has been very successful in describing the bulk properties of heavy-ion collisions. \cite{Teaney:2000cw,Hirano:2002ds,Kolb:2003dz,Huovinen:2003fa}  Ideal hydrodynamics, which neglects viscous corrections, was able to reproduce the trends seen in experimental data, in particular the centrality and transverse momentum dependence of particle spectra and elliptic flow.  Subsequent developments in the description of the initial state, in particular the inclusion of event-by-event fluctuations, and in the extension of the formalism to include off-equilibrium corrections like those introduced by a non-zero shear viscosity, have advanced the theoretical modeling significantly. \cite{Gale:2013da,deSouza:2015ena} Quantitative agreement of viscous relativistic fluid dynamics with a wide range of experimental data is excellent (see Fig.\,\ref{fig:hydro:vn}), and recent improvements such as the inclusion of bulk viscosity and microscopic transport for the later stages of the collision improve the agreement even more\cite{Ryu:2015vwa}. 

The similarity of experimental data for multi-particle correlations in small
systems with those in heavy ion collisions (see Sec.\,\ref{sec:exppA}) suggests
that the physical processes responsible for the observed correlations may be the
same in both systems. More explicitly, hydrodynamics could also be the
appropriate model to describe high multiplicity \ppc and \pA collisions. 
 
While the applicability of hydrodynamics in very small systems should
definitely be questioned (and we will discuss this issue in detail below), the
earliest discussion of (ideal) relativistic hydrodynamics to describe nuclear
collisions involved \ppc collisions
\cite{Landau:1953la,Belenkji:1956la}.  Following the early argument by Landau,
hydrodynamics should be applicable if at the moment of collision, a large
number of ``particles'' is created, the ``collision mean free path'' in the
created system is small compared with its dimensions, and a statistical
equilibrium is established.  We use quotation marks because the definition of
particle number and mean free path may be ambiguous in a strongly interacting
system. 
 
More recently hydrodynamic calculations aiming at the prediction and
description of experimental data for small collision systems at RHIC and LHC
have become available.
\cite{Werner:2010ny,Werner:2010ss,Bozek:2011if,Bozek:2012gr,Bozek:2013ska,Werner:2013tya,Werner:2013ipa,Schenke:2014zha} 
In particular, \pA and \dAu collisions at high energy were explored early on,
and predictions showed qualitative agreement with subsequently published
experimental data (albeit at different collision energies). 
 
In the following we will discuss the ingredients of these calculations,
emphasizing the differences between the calculations of various groups. We will
see in particular that results for anisotropic flow are highly sensitive to
assumptions about the initial state and its fluctuating structure. 

\subsubsection{Modeling of the initial state} \label{subsubsec:initial}

The most important ingredient in the hydrodynamic description of small
collision systems is the modeling of the initial geometry.  Fluctuations in the
initial energy density that deviate from sphericity propagate hydrodynamically
and generate non-zero flow harmonics.  In practice the initial state geometry
is generated by doing event-by-event calculations where fluctuations of density
distributions inside the incoming nuclei or hadrons are taken into account.  In
many calculations these fluctuations are dominated (or exclusively given) by
the random positions of nucleons in a nucleus. However, in collisions of a
single proton with a large nucleus the substructure of the proton may become
important. 

We present the details of the initial state model calculations that have been
employed for the description of \pA collisions below. These are the Monte Carlo
(MC) Glauber model \cite{Miller:2007ri}, the IP-Glasma model
\cite{Schenke:2012wb,Schenke:2012fw}, and EPOS \cite{Werner:2013tya}.

\paragraph{Monte Carlo Glauber model}
We follow the discussion in the literature to introduce the implementation of the Monte Carlo (MC) Glauber model for \pA collisions.\cite{Bozek:2013uha} 

In the MC Glauber model, which has been fairly successful in describing experimental data from heavy ion collisions \cite{Miller:2007ri}, the proton and the energy deposited per wounded (participating) nucleon are assumed to be spherical. Hence, in \pA collisions the fluctuating shape of the interaction region is solely due to the random positions of struck nucleons in the heavy nucleus.

The specific implementation we discuss here\cite{Bozek:2013uha} uses GLISSANDO
\cite{Broniowski200969}, where the entropy deposited per wounded nucleon (or
binary collision) fluctuates according to a gamma distribution in order that
the final multiplicity distribution reproduces the empirically observed
negative binomial.  The width of each Gaussian shaped (in transverse space)
contribution to the entropy density is chosen to be $0.4\,{\rm fm}$.

To determine whether a nucleon was wounded, a Gaussian wounding profile is used. This means that a nucleon undergoes a collision with probability
\begin{equation}
  P(b) = A\,\exp(-\pi A b^2/\sigma_{\rm NN})\,,
\end{equation}
where $b$ is the distance to the nucleon it is potentially colliding with, $\sigma_{\rm NN}$ is the nucleon nucleon cross section, and $A$ is a constant chosen to be $0.92$. 

\begin{figure}
\center
    \includegraphics[width=\linewidth]{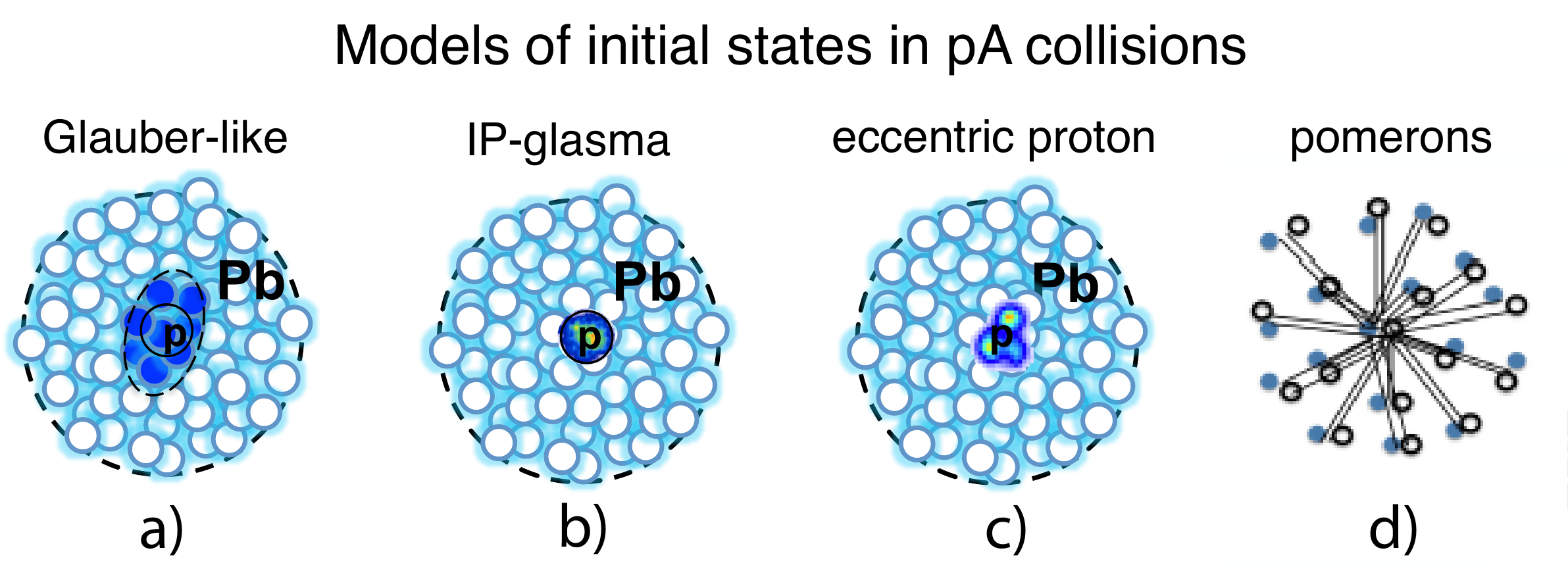}  
    \caption{ Typical geometric configurations generated by different initial state models for \pA collisions: a) MC-Glauber with wounded nucleons as sources b) IP-Glasma with round proton c) IP-Glasma with an eccentric proton d) Pomerons.
    } \label{fig:initialstate_pA}
\end{figure}

This method leads to a fairly large interaction region, much larger than the size of the incoming proton. This is illustrated in Fig.\,\ref{fig:initialstate_pA} a). If one chooses the alternative method of depositing energy around the interaction point of two nucleons, the average size is reduced by approximately 40\% \cite{Bozek:2013uha}.\footnote{Also using the black disk approximation, where a collision only occurs if $b<R=\sqrt{\sigma_{\rm NN}/\pi}$, leads to smaller interaction regions.}

In 3+1 dimensional simulations \cite{Bozek:2013uha,Kozlov:2014fqa}, the initial entropy density needs to be defined in the space-time rapidity direction as well. Typically one assumes the factorization of the transverse and longitudinal distributions. Assuming that the transverse distribution is independent of rapidity within some range around mid-rapidity is important to reproduce the experimentally observed ridge structure: 
Having the same initial transverse geometry at different space-time rapidities, leads to correlations between the flow pattern of particles emerging with different momentum rapidities.
The typical profile in space-time rapidity is given by a combination of a linearly dropping central region and two half Gaussians at large (positive and negative) rapidities:
\begin{equation}
  \rho(\eta_s) = \exp\left( -\frac{(|\eta_s|-\eta_0)^2}{2\sigma_\eta^2} \theta(|\eta_s|-\eta_0)\right) \left( 1 \pm \frac{\eta_s}{y_{\rm beam}}\right) \theta(y_{\rm beam} \pm \eta_s)\,.
\end{equation}
The linear drop away from the nucleon introduces a torque effect and some small fluctuation in the longitudinal direction.
The parameters typically chosen \cite{Bozek:2013uha,Kozlov:2014fqa} in calculations for \pA collisions at $\sqrt{s}=5.02\,{\rm TeV}$ are $\eta_0=2.5$, $\sigma_\eta=1.4$, and $y_{\rm beam} = 8.58$.  Another parameter is the initial time that the hydrodynamic calculation starts, typically chosen between $\tau_0=0.2\,{\rm fm}/c$ and $\tau_0=0.6\,{\rm fm}/c$.

Since in the hydrodynamic framework the final flow anisotropy is driven by the initial state eccentricity, we define it here and will refer to it later in the text when discussing how different initial state models generate different geometries.  The $n$'th order eccentricity is given by \cite{Alver:2010gr,Petersen:2010cw}

\begin{equation}\label{eq:eccen}
 \varepsilon_n = \frac{\sqrt{\langle r^n \cos(n\phi)\rangle^2+\langle r^n \sin(n\phi)\rangle^2}}{\langle r^n \rangle}\,,
\end{equation} 
where $\langle \cdot \rangle$ is the energy density (or entropy density) weighted average, and $r$ and $\phi$ are the polar coordinates in the plane transverse to the collision axis.  

\paragraph{IP-Glasma model}
The Impact Parameter (IP) dependent Glasma model \cite{Schenke:2012wb,Schenke:2012fw} is based on the color glass condensate framework \cite{Iancu:2003xm} and uses the IP-Sat model \cite{Kowalski:2003hm} to constrain the impact parameter and gluon momentum fraction dependence of the dipole cross section. This, together with fluctuating nucleon positions for heavy ions determines the initial distribution of color charges, which then enter the currents in the Yang-Mills equations that determine the initial gluon fields.  The assumption in this model being that at high energies the QCD coupling constant is small\footnote{The coupling is determined at the dynamically generated saturation scale $Q_s$, which grows with increasing energy.} and the occupation numbers of gluons are non-perturbatively large ($\sim 1/\alpha_s$).  In this limit the dynamics can be approximated classically by the Yang-Mills equations describing both the initial configuration of gluon fields and their dynamical evolution \cite{Kovner:1995ja,Kovchegov:1997ke,Krasnitz:1998ns,Krasnitz:1999wc,Krasnitz:2000gz,Lappi:2003bi}.
For heavy ion collisions at high energies these assumptions are likely to be valid. For smaller systems, such as \pA collisions, the approximations might be appropriate for high multiplicity events.
Details on the implementation in \AAc collisions \cite{Schenke:2012fw,Schenke:2013dpa} and \pA collision can be found in the literature \cite{Schenke:2013dpa,Schenke:2014zha}.
Here we list the main ingredients of the IP-Glasma model:

\begin{enumerate}
  \item IP-Sat parametrization of the saturation scale $Q_s(x,\mathbf{x}_T)$ and fixed parameters from DIS data. \cite{Bartels:2002cj,Kowalski:2003hm,Rezaeian:2012ji}
  \item Monte Carlo sampling of nucleon positions in the incoming nucleons and the impact parameter $\mathbf{b}$.
  \item Determination of the thickness function as a function of $\mathbf{x}_T$, $T(\mathbf{x}_T)$, in each nucleus using information from (2). Using (1), this results in the distribution of $Q_s(\mathbf{x}_T)$ at the collision energy of interest.
  \item Using a proportionality constant of order 1 that allows the conversion from $Q_s$ to the color charge density $g^2\mu$ leads to the color charge density distribution in each nucleus. This proportionality constant can fluctuate to account for the effect of gluon number fluctuations. See discussion below.
  \item Monte Carlo sampling of color charges $\rho^a$ (with color index $a$) for each nucleus from the color charge density distribution, assuming 
    \begin{equation}
      \langle \rho^a(\mathbf{x}_T)\rho^b(\mathbf{y}_T)\rangle = g^2\mu^2(x,\mathbf{x}_T) \delta^{ab} \delta^{(2)}(\mathbf{x}_T-\mathbf{y}_T) \, .
    \end{equation}
  \item Solution of the Yang Mills equation 
    \begin{equation}\label{eq:YM1}
      [D_{\mu},F^{\mu\nu}] = J^\nu\,,
    \end{equation}
    for each nucleus with the color currents given by 
    \begin{equation}\label{eq:current}
      J^\nu = \delta^{\nu \pm}\rho^a(x^\mp,\mathbf{x}_T)t^a\,,
    \end{equation}
    where $+$ and $-$ are for right moving and left moving nuclei, respectively. $t^a$ are the generators of SU(3).
  \item Numerical solution for the gluon fields after the collision at time $\tau=0^+$: \cite{Kovner:1995ja,Kovner:1995ts}
    \begin{align}
      A^i &= A^i_{(A)} + A^i_{(B)}\,,\label{eq:init1}\\
      A^\eta &= \frac{ig}{2}\left[A^i_{(A)},A^i_{(B)}\right]\,,\label{eq:init2}\\
      \partial_\tau A^i &= 0\,,\\
      \partial_\tau A^\eta &= 0\,,
    \end{align}
    where (A) and (B) denote fields of the two colliding nuclei.
  \item Solution of the source-free Yang-Mills equation to obtain the time evolution of the produced gluon fields.
  \item Landau matching of the gluon-field energy momentum tensor to obtain the relevant quantities for fluid dynamic calculations: energy density $\varepsilon(\mathbf{x}_T)$ and flow velocities $u^\mu(\mathbf{x}_T)$.
\end{enumerate}
Note that initial conditions obtained with this model are boost invariant, which is a good approximation for high energy collisions and rapidities not too far away from mid-rapidity.

We will now discuss specific features of the model relevant for small collision systems. Due to the fluctuating color charges, gluon multiplicity distributions in the IP-Glasma model already include negative binomial fluctuations~\cite{Schenke:2012wb,Schenke:2013dpa}. It was however discovered that for \ppc and \pA collisions additional fluctuations (as compared to the original model \cite{Schenke:2012wb}) of the color charge density were needed to describe the experimental multiplicity distributions (see also Fig.\,\ref{fig:nbd}). These fluctuations can be understood as fluctuations of the gluon number for a given value of the saturation scale $Q_s$. The comparison of gluon with hadron distributions is of course only qualitative and additional fluctuations from the hadronization mechanism are also expected to modify the result. Nevertheless, it is important to include the gluon number fluctuations to allow for rare fluctuations that generate the high multiplicity events of interest.

Turning to the initial geometry, the IP-Glasma model generates interaction regions in the transverse plane that are dominated by the smaller of the two colliding objects. This is because energy is only deposited where both nuclei's gluon fields are non-zero (otherwise the fields are pure gauge fields and indistinguishable from the vacuum). For a proton colliding with a large nucleus (not too close to its edge), the interaction region is thus strongly affected by the shape of the proton. This behavior leads to much smaller interaction regions than in the Monte Carlo Glauber model as illustrated in Fig.\,\ref{fig:initialstate_pA} b). Furthermore, if the proton is assumed to be round (up to fluctuations of the color charges $\rho^a$), initial eccentricities as defined in (\ref{eq:eccen}) are very small.

\paragraph{EPOS}
The EPOS model\cite{Drescher:2000ha,Werner:2012xh,Werner:2013ipa} has successfully been applied to describe experimental data from both \ppc and \pPb collisions.
EPOS includes both a fluctuating initial state and hydrodynamic evolution, as well as jets and the interaction between jets and the medium.
The initial state calculation is based on multiple scatterings, where initial state radiation is included. Thus, the model deals with parton ladders or ``Pomerons''. The Pomerons, whose number determines the multiplicity, are placed in between two colliding partons, which are distributed around the center of  nucleons according to an exponential distribution. In addition, the pomerons carry a transverse momentum (generated by the initial scattering) and thus 
move transversely even before the coupling to hydrodynamics occurs. EPOS does not assume boost invariance but produces a smoothly changing longitudinal energy density distribution. 
It should be noted that EPOS also needs to include saturation effects in order to not violate unitarity. It is done by introducing an energy dependent saturation scale for each Pomeron.\cite{Werner:2013ipa}.

\paragraph{Holographic pomerons} 

Following the AdS/CFT conjecture \cite{Maldacena:1997re} a reggeized and unitarized scattering
amplitude was derived for ${\cal N}=4$ SYM in the limit of large N, strong
coupling $g^2_{YM}N \gg 1$ and high energy \cite{Rho:1999jm}.  
In \cite{Basar:2012jb} it was recognized that a single closed string exchange contribution 
to the eikonalized dipole-dipole scattering amplitude yields a Regge behavior of the elastic amplitude,
with features consistent with experimental data.

This Regge behavior is driven by worldsheet instantons, which describe the Schwinger mechanism for string pair creation
by an electric field, where the longitudinal electric field at the origin of this non-perturbative mechanism is induced
by the relative rapidity of the scattering dipoles. 

This stringy Schwinger
mechanism can generate quantum fluctuations of virtual strings, which take the form of thermal
fluctuations with an effective temperature related to the inverse string
circumference and therefore impact parameter; at the center of the string
$T_{\rm eff}^{-1}\sim 2\pi|{\bf b}_\perp|$.  If the impact parameter is small enough
there may be a region in the middle of the string where the effective
temperature is above the deconfinement temperature suggesting that the string
description should be replaced by a plasma phase.\cite{Basar:2012jb}.

It has been argued that near this critical impact parameter the increased
quantum fluctuations reduce the string tension and self-interactions among the
strings cause an implosion and formation of a {\em string
ball}\cite{Shuryak:2013sra}.  A more quantitative study of string
self-interactions and string ball formation was explored
in\cite{Kalaydzhyan:2014tfa,Kalaydzhyan:2014zqa}.  At effective temperatures
larger than $T_c$ the string ball may form a strongly coupled high density
plasma that expands hydrodynamically.  One of the key features of the
holographic pomerons that may distinguish them from other pictures of plasma
formation is the sudden onset of hydrodynamic behavior at a critical impact
parameter.  However, no explicit hydrodynamic calculations have been performed
using this framework.

\subsubsection{Viscous hydrodynamics and its limits of application}
Hydrodynamics is an effective theory for the long wavelength modes of a system
able to describe the interactions of the low momentum part of the particle
spectrum if, from a kinetic theory point of view, the mean free path of the
particles is significantly smaller than the system size.  This lead many to
criticize the use of viscous fluid dynamics in small colliding systems and
indeed one should always check whether hydrodynamics is being applied within
its domain of validity.  In this section we will first introduce the basic
ingredients of viscous relativistic hydrodynamics and quantify the domain of
validity more explicitly.  

The viscous relativistic fluid dynamic equations for small systems are
identical to those employed to describe heavy ion collisions. Various recent
reviews cover the details of both the equations and their derivation, as well
as the implementation in numerical codes. \cite{Heinz:2013th, Gale:2013da,
deSouza:2015ena}

Here we summarize the most important aspects. Relativistic viscous hydrodynamics is formulated as a gradient expansion to take into account deviations from local thermal equilibrium or ideal hydrodynamics.  Due to the acausal behavior of the relativistic equations expanded to first order in gradients, numerical implementations require the inclusion of terms at least of second order in the gradients.
Commonly used are Israel-Stewart type equations \cite{Israel:1976tn,Israel:1979wp} that were derived using the 14 moment method and the second moment of the Boltzmann equation to determine the hydrodynamic equations for the dissipative currents. Improvements that provide a more systematic expansion were derived more recently. \cite{Denicol:2012cn}

Many simulations concentrate on shear viscosity, but some also include bulk viscosity, and possibly the coupling between shear and bulk terms. In fact, the effect of bulk viscosity has been found to be large \cite{Ryu:2015vwa} for certain observables, including spectra and anisotropic flow. However, a large uncertainty arises from off-equilibrium corrections to the particle spectra when converting hydrodynamic quantities to particles at freeze-out, in particular when including bulk viscous corrections.

Large corrections to the equilibrium distribution functions (typically appearing at transverse momenta $p_T\gtrsim 2\,{\rm GeV}$) are one indication that the fluid dynamic approximation we are using breaks down. Another, more systematic quantification of the validity of viscous fluid dynamics is provided by the study of Knudsen and Reynolds numbers. The Knudsen number is defined as the ratio of a microscopic to a macroscopic scale in the system.
Various choices for these scales are possible, and a variety of them has been studied in both heavy ion collisions and smaller systems. \cite{Niemi:2014wta} For example, the macroscopic scale can be defined as the inverse of the expansion rate $L_{\rm macro}^\theta = \theta^{-1} = (\partial_\mu u^\mu)^{-1}$, with $u^\mu$ the flow velocity, while the microscopic scale can be described by the shear relaxation time, $\l_{\rm micro} = \tau_\pi$, which in a dilute gas is proportional to the mean free path, $\tau_\pi \sim \lambda_{\rm mfp}$. So, $Kn_\theta = \tau_{\pi} \theta$, or alternatively we can define $Kn_{\varepsilon}$, with $L_{\rm macro}^{-1} = \sqrt{\nabla_\mu \varepsilon_0 \nabla^\mu \varepsilon_0}/\varepsilon_0$, where $\varepsilon_0$ is the energy density. 

\begin{figure}
\center
    \includegraphics[width=\linewidth]{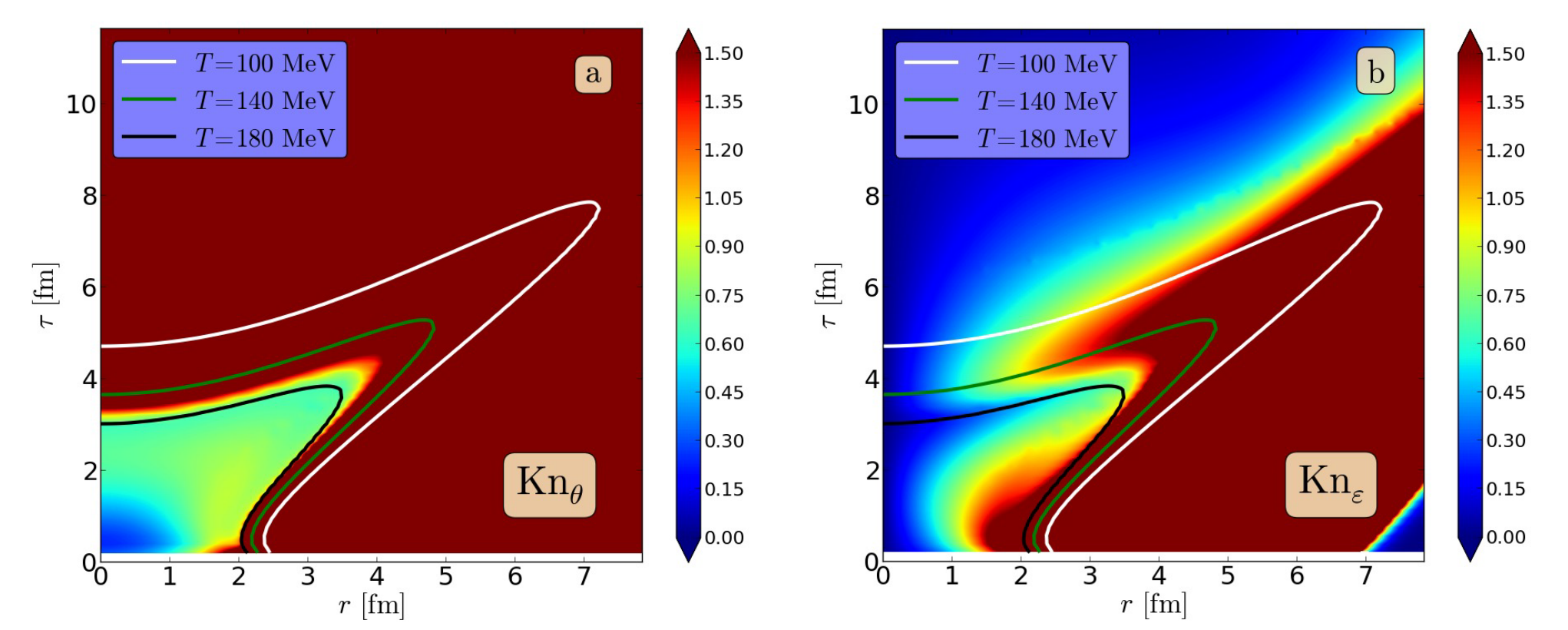}  
    \caption{ Space-time evolution of the Knudsen numbers in \pPb collision at the LHC, with $\eta/s=0.08$ in the QGP phase and a temperature dependent $\eta/s$ in the hadron gas phase, and $dN_{\rm ch}/d\eta = 270$. (a) $Kn_\theta$ and (b) $Kn_\varepsilon$. \cite{Niemi:2014wta}
    } \label{fig:Knudsen}
\end{figure}

For hydrodynamics to be valid, the Knudsen number should stay significantly below one at all points in space time. Whether this condition is fulfilled was tested for various choices for $L_{\rm macro}$ \cite{Niemi:2014wta} and we show examples for \pA collisions in Fig.\,\ref{fig:Knudsen}. 

One can see that around $T = 100\,{\rm MeV}$ the fluid dynamical description is
no longer applicable, with all $Kn_\theta$ values being above 1.5.  The results
shown are for a constant $\eta/s = 0.08$ in the QGP phase -- if a temperature
dependent $\eta/s$ was used in the QGP phase, the situation would worsen and
fluid dynamics would be out of its domain of applicability even in the early
stages of the evolution.  While large Knudsen numbers indicate the breakdown of
hydrodynamics it does not preclude large final state effects.  However, it does
mean that another framework outside of hydrodynamics might be more appropriate to model
the system when the Knudsen number is large.

\subsubsection{Results from hydrodynamics in small systems}
We now discuss the results obtained for the Fourier harmonics in small collision systems within various model calculations, beginning with the existing predictions for proton-proton collisions at the LHC and continuing with proton/deuteron on heavy ion collisions. We will then go into more detail with the discussion of more recent, very sophisticated calculations and their results for mean transverse momentum, Fourier harmonics of charged hadrons and identified particles, as well as HBT radii.

\paragraph{Proton-proton collisions}
Wide ranges of predictions for elliptic flow in proton-proton collisions were published early on - in particular in expectation of LHC results at 7 and 14 GeV. We list several calculations and predicted values for the elliptic flow in \ppc collisions in the following.

Ideal hydrodynamic calculations \cite{Luzum:2009sb} indicated that $v_2 \sim 0.035$ in \ppc collisions, and that any value above $0.02$ in minimum bias collisions would indicate an extremely small viscosity (below the KSS bound of $1/4\pi$) of the medium.

In a string percolation model that includes a directional dependence of the sources, a sizable $v_2$ was predicted for \ppc collisions \cite{Bautista:2009my}. Various scenarios including different shapes of the proton have been studied\cite{d'Enterria:2010hd} and estimates of $v_2$ for \ppc collisions depended strongly on the assumed shape, ranging from zero to almost $1.5\%$ for the more realistic models. Similarly strong dependencies on the proton profile were found in hydrodynamic simulations \cite{Prasad:2009bx}, where very large values for the minimum bias $v_2$, ranging from $6\%$ to $25\%$ are quoted.

High multiplicity events were predicted to have an integrated $v_2$ of around 10\%, but again with a wide spread depending on the initial state configuration, when using hydrodynamics after initializing with flux tubes obtained in a constituent quark model. \cite{Bozek:2009dt} A similar model using partonic interactions determines eccentricities compatible with $v_2$ values ranging from 10\% to 20\%, depending on model parameters, when assuming a given relation between eccentricities and elliptic flow values determined from hydrodynamic calculations. \cite{CasalderreySolana:2009uk}

Mimicking quantum fluctuations by introducing an eccentricity to the interaction region, and then performing hydrodynamic simulations, a $v_2$ value of 3\% was found if the induced eccentricity has average value of about 17\% in another study. \cite{Ortona:2009yc}.

Using DIPSY, a Monte Carlo event generator based on the QCD dipole model, to generate the initial conditions and a fixed relation between $v_2$ and eccentricity, a value of $6\%$ for $v_2$ in \ppc collisions at LHC was estimated, being fairly independent of the multiplicity. \cite{Avsar:2010rf}

In the EPOS framework that performs hydrodynamic expansion based on flux-tube initial conditions as described above, both the experimentally determined  mean transverse momentum and measures of the source size were reproduced. \cite{Werner:2010ny} Furthermore, the ridge structure was reproduced within this model. \cite{Werner:2010ss} This indicates that the elliptic flow component is correctly described by EPOS in \ppc collisions.

The conclusion to be drawn from this extremely wide range of predictions is the strong model dependence, in particular on the initial state for hydrodynamic calculations, that one faces when computing $v_2$ in proton-proton collisions.

\paragraph{Proton-heavy ion collisions}
Early predictions for flow in \dPb and \pPb collisions from hydrodynamic simulations were also presented. \cite{Bozek:2011if} In this work MC-Glauber initial conditions, as described in Section \ref{subsubsec:initial}, were used together with 3+1 dimensional viscous hydrodynamic simulations to obtain results for particle spectra and flow harmonics as functions of transverse momentum and pseudo rapidity.

\begin{figure}
\center
    \includegraphics[width=0.7\linewidth]{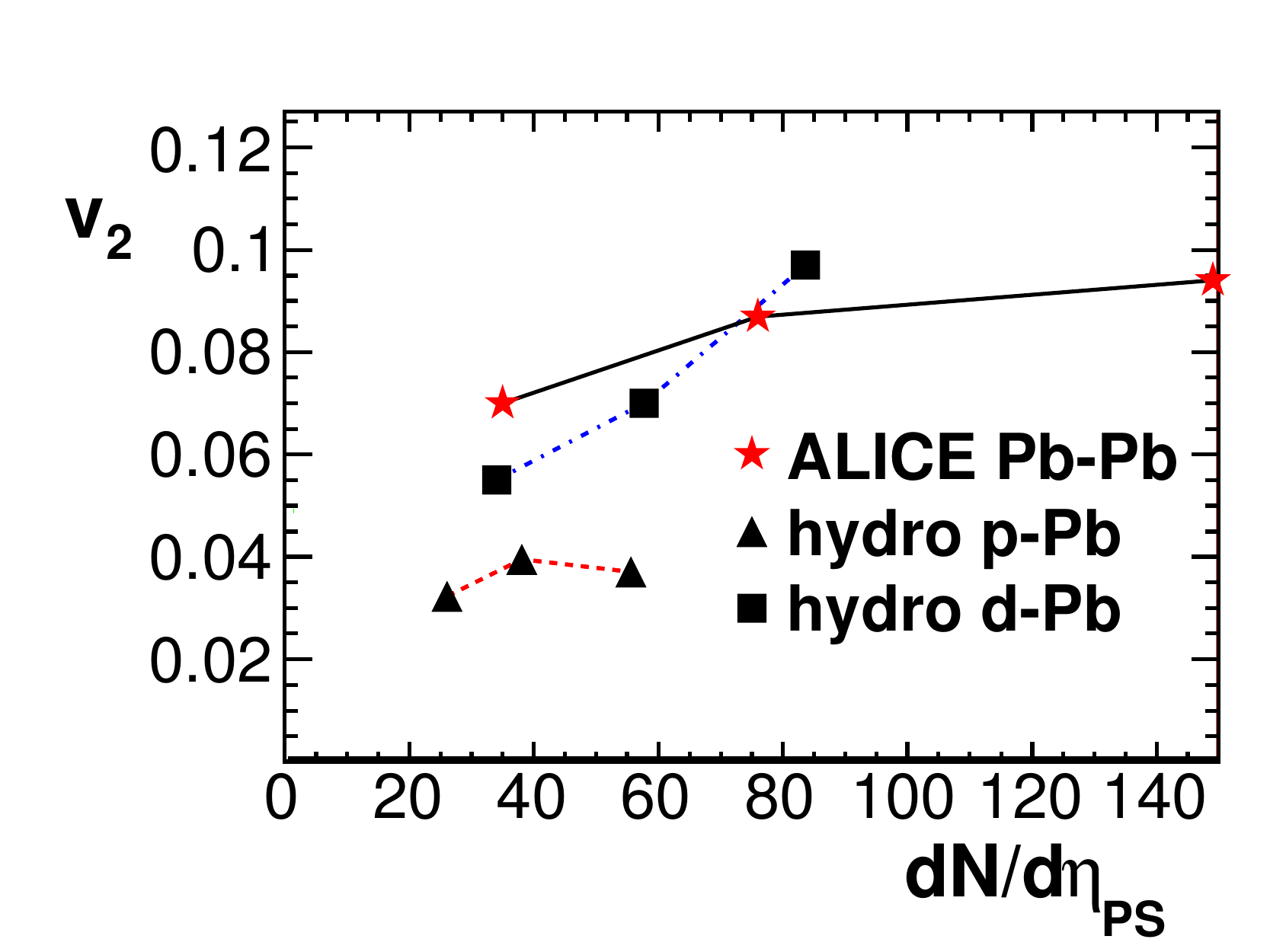}  
    \caption{ Predictions for the elliptic flow coefficient $v_2$ from a 3+1D viscous 
    relativistic hydrodynamic simulation for \pPb (at $\sqrt{s}=4.4\,{\rm TeV}$) 
    and d+Pb (at $\sqrt{s}=3.11\,{\rm TeV}$) collisions, compared to ALICE results for Pb+Pb collisions.~\cite{Bozek:2011if,ALICE:2011ab}. \label{fig:v2Bozek1} } 
\end{figure}

The elliptic flow is found to be 3\%-4\% in \pPb collisions, with little centrality dependence (Fig.\,\ref{fig:v2Bozek1}), significantly smaller than for \PbPb collisions with the same multiplicity. 
For the d-Pb system, the elliptic flow is significantly larger, increasing for central collisions, and reaching almost 10\%.

The ridge structure in two-particle correlations was observed in similar hydrodynamic calculations for \pPb collisions at $\sqrt{s}=5.02\,{\rm TeV}$. \cite{Bozek:2012gr}
For the largest rapidity gaps considered, $v_2$ integrated over $0.1\,{\rm GeV} < p_T < 2\,{\rm GeV}$ was found to be $4\%$, while the corresponding $v_3$ was about half as big.

As discussed before, in these calculations the initial interaction region turns out to be rather large when using scaling of the initial energy or entropy density with the participating nucleons. The size and lifetime of the collective source formed in central \pPb collisions is found to be $3-4\,{\rm fm}$.

More quantitative results on the size of the emission source using HBT techniques were subsequently presented. \cite{Bozek:2013df} Here, an alternative model for the initial distribution was also considered.  This ``compact'' variant locates the sources in the center of mass of nucleon-nucleon collisions. The average rms radii for the two sets of initial conditions considered were $1.5$ and $0.9\,{\rm fm}$, respectively.

Results obtained for HBT radii as a function of multiplicity were presented in comparison to heavy ion and \ppc data \cite{Bozek:2013df}.
It was found that the \pPb results for $R_{\rm side}$, $R_{\rm out}$, and $R_{\rm long}$ were much closer to the heavy ion results at a given multiplicity 
than to the \ppc results. $R_{\rm out}$ is reduced by 25\% when using the compact initialization, while the other radii are little affected.
We note that the difference in the initial size of the system has a large effect on the final measured HBT radii. In particular, it was shown \cite{Schenke:2014zha} that the IP-Glasma model (see Section \ref{subsubsec:initial}) produces significantly smaller initial energy density distributions in the transverse plane than either of the MC Glauber models discussed above. In particular, initial sizes in \pPb collisions were demonstrated to be much closer to those in \ppc collisions than \PbPb collisions. \cite{Schenke:2014zha} One would expect that this leads to similar differences for HBT radii obtained in hydro calculations using the different initial state models.

\begin{figure}
\center
    \includegraphics[width=0.8\linewidth]{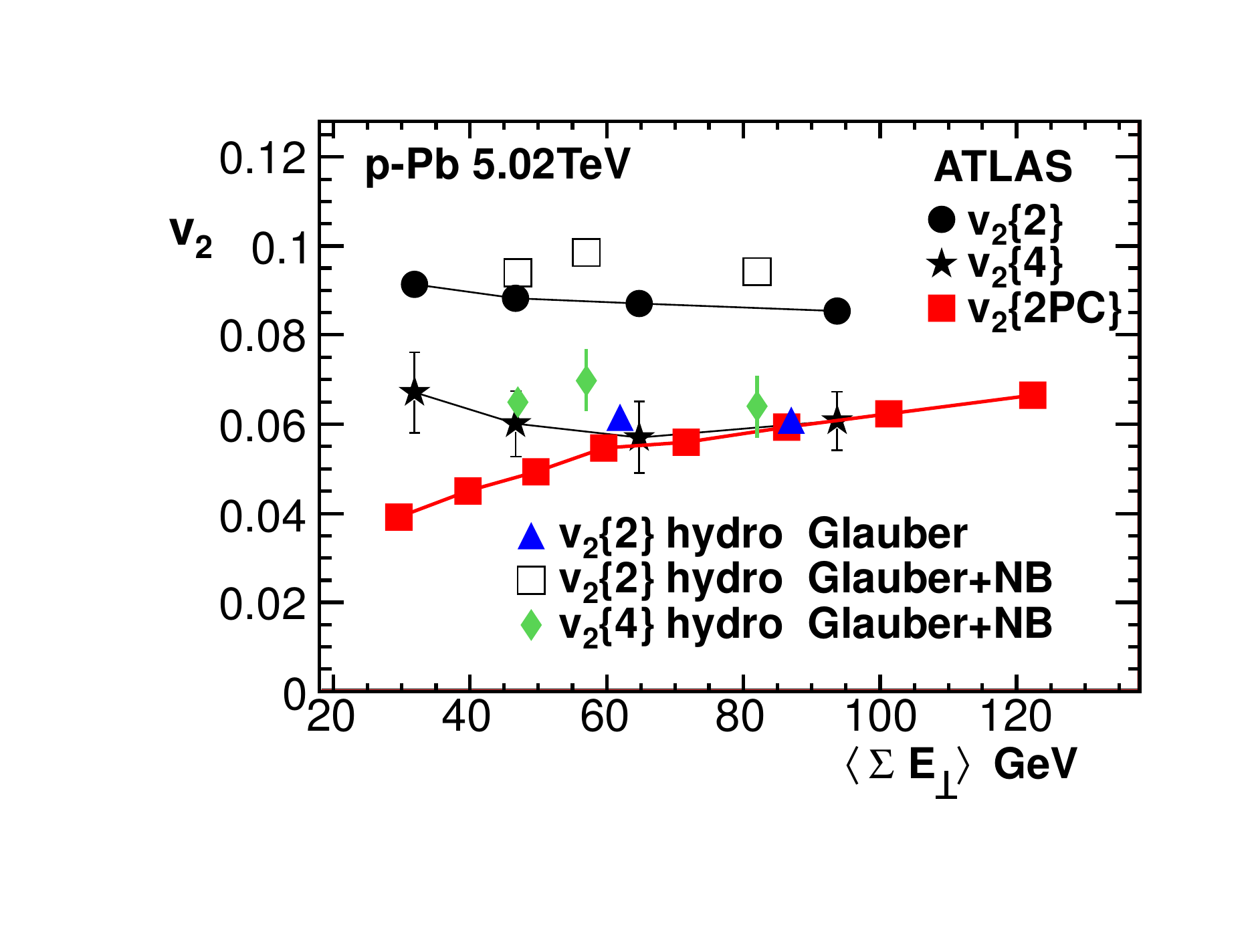}  
    \caption{ The elliptic flow coefficient of charged particles for $|\eta| < 2.5$, $0.3\,{\rm GeV} < p_T < 5.0\,{\rm GeV}$ from the cumulant method $v_2\{2\}$ and $v_2\{4\}$, and from the di-hadron correlation function measured by the ATLAS Collaboration \cite{Aad:2013fja}, compared to hydrodynamic calculations. \cite{Bozek:2013uha}. Glauber+NB includes negative binomial fluctuations in the initial entropy density distribution.
    } \label{fig:v2Bozek2}
\end{figure}

Very detailed analyses of flow in \pPb collisions using the MC Glauber model including negative binomial fluctuations, which are necessary to reproduce the multiplicity distribution, and in particular its tail at high multiplicity, have been carried out following the first exploratory works. \cite{Bozek:2013uha} Here, the authors employ hydrodynamic simulations followed by statistical freeze-out. The negative binomial fluctuations in the initial state, which cause the deposited entropy per participant nucleon to fluctuate, turn out to be necessary in order to describe the experimental data by increasing the initial eccentricities and thus the resulting $v_2\{2\}$ and $v_2\{4\}$. The successful description of ATLAS data \cite{Aad:2013fja} in this case is presented in Fig.\,\ref{fig:v2Bozek2}.

Comparing to predictions discussed above and shown in Fig.\,\ref{fig:v2Bozek1}, we conclude that (differences from the employed method to determine $v_2$ aside), large differences result from using different initial state models.

More detailed analyses have demonstrated that various features of the data are naturally reproduced by hydrodynamic calculations, even on a quantitative level. For example, the mass splitting of the mean transverse momentum \cite{Bozek:2013ska,Kozlov:2014fqa} and the $v_2$ of identified particles \cite{Bozek:2013ska,Werner:2013ipa} in \pPb collisions was presented using different initial state descriptions. 

We note that hadronic cascade models, such as UrQMD, also reproduce the characteristic mass splitting. UrQMD could however not describe the centrality dependence of the elliptic flow or get a real valued $v_2\{4\}$. \cite{Zhou:2015iba}
The latter is naturally achieved in hydrodynamic calculations, where all higher order cumulants (from 4, 6, 8 particles, etc.) are equal since all particles are correlated with one another.

The IP-Glasma model produces an initial transverse energy distribution having the shape of the assumed proton shape, and early calculations using a spherical proton therefore yielded a very small $v_2$ in \pA collisions.  \cite{Schenke:2014zha} 
However, the process of energy deposition is less ambiguous as in the MC Glauber model, where very different results can be achieved by using different prescriptions for the initial energy/entropy density distributions.

Within the IP-Glasma + hydrodynamics description it is thus clear that the results for azimuthal anisotropies in \pPb collisions are sensitive to the shape of the proton and its fluctuations event by event. \cite{Schenke:2014zha} 

In conclusion, all aspects of the experimental data are at least qualitatively and for the most part also quantitatively described by hydrodynamic calculations. Unfortunately, in addition to the unknown transport parameters, results are highly sensitive to the assumptions made in modeling the initial state.  Furthermore, one must scrutinize the applicability of hydrodynamics in these very small systems due to large Knudsen numbers throughout the evolution.

Some support for the applicability of hydrodynamics comes from calculations at large coupling, where it was shown that viscous hydrodynamics describes early on the evolution obtained using AdS/CFT correspondence in a system of colliding shock waves. \cite{Chesler:2015bba} 

\begin{figure}
  \center
    \includegraphics[width=\linewidth]{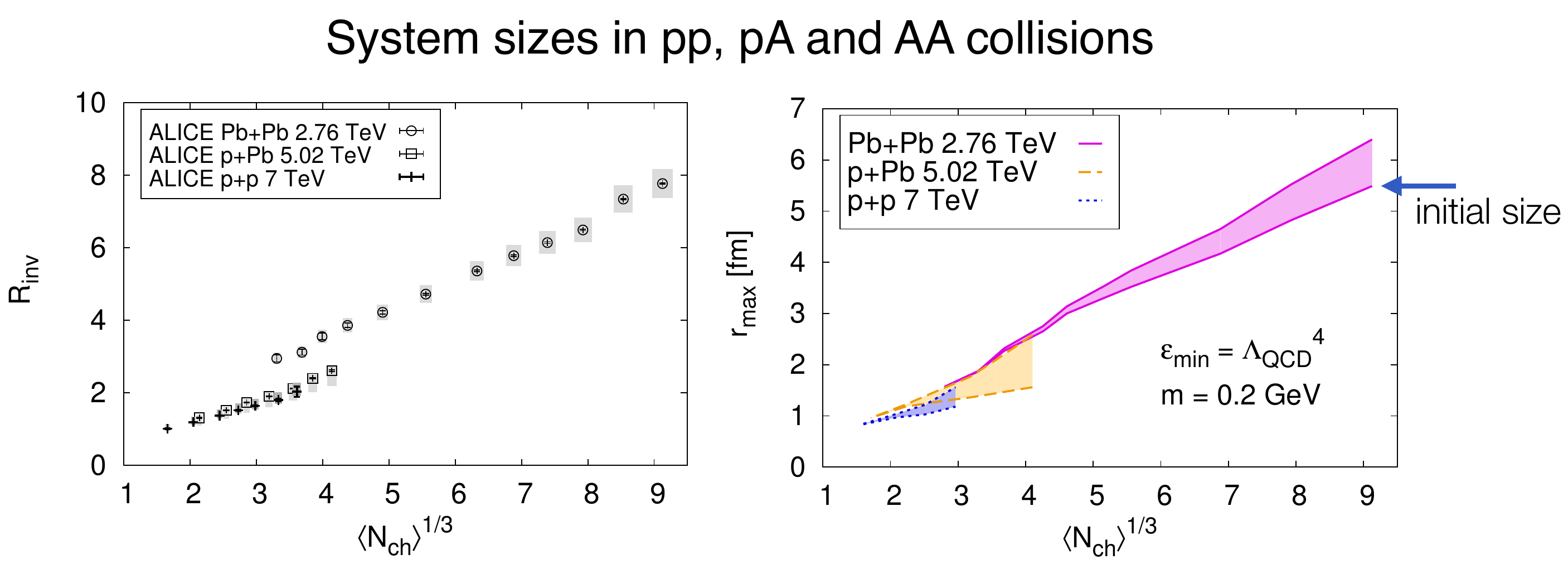}  
    \caption{ The experimental radius $R_{\rm inv}$ in \ppc, \pPb, and \PbPb collisions from the ALICE collaboration \cite{Abelev:2014pja} (left) compared to the radius $r_{\rm max}$ in the IP-Glasma (+\textsc{Music}) model \cite{Schenke:2014zha}. The lower end of the band indicates the initial size, the upper end the maximal size reached during hydrodynamic expansion.
Initial sizes are similar in \ppc and \pA collisions, while A+A collision systems at the same multiplicity are generally larger. This is a trend reflected also in the experimental HBT radii shown on the left.
    } \label{fig:initialstate_size}
\end{figure}

\paragraph{HBT radii from hydrodynamics}
As for the results on the azimuthal anisotropy, results for the system size, mainly characterized by HBT radii (discussed in Section \ref{sec:HBT}), vary largely depending on the initial state model. While the IP-Glasma model produces initial sizes that are similar in \ppc and \pA collisions, but large in \AAc for the same multiplicity, the MC-Glauber model produces larger \pA systems, closer to those in \AAc collisions \cite{Bozek:2013df}, as illustrated in Fig.\,\ref{fig:initialstate_pA}. 
While HBT calculations within the IP-Glasma model have yet to be performed,
the general trends of the initial size of the system and the measured HBT radii agree very well. In Fig.\,\ref{fig:initialstate_size} we compare the experimental data on the left to results for $r_{\rm max}$ on the right. We define $r_{\rm max}$ as the (angle averaged) radius where the system reaches a minimal threshold energy density $\varepsilon_{\rm min} = \Lambda_{\rm QCD}^4$. This radius by definition depends on the choice of $\varepsilon_{\rm min}$. This choice however only affects the overall normalization of $r_{\rm max}$; it does not affect the dependence of $r_{\rm max}$ on the number of charged particles $N_{\rm ch}$. There is also some dependence on the infrared cutoff $m$, which has been chosen to be on the order of $\Lambda_{\rm QCD}$. This dependence can mostly be absorbed into an overall normalization constant \cite{Schenke:2014zha}.

In summary, the system sizes obtained from experimental HBT measurements are generally compatible with the fluid dynamic picture, while the details, once again, depend strongly on the initial state and its size.

\paragraph{Deuteron/$^3$He-heavy ion collisions}
Small systems with drastically different geometries from \pA collisions could help distinguish whether the origin of the azimuthal anisotropies measured in \pA collisions could indeed stem from the hydrodynamic response to the initial geometry.

\begin{figure}
\center
    \includegraphics[width=0.8\linewidth]{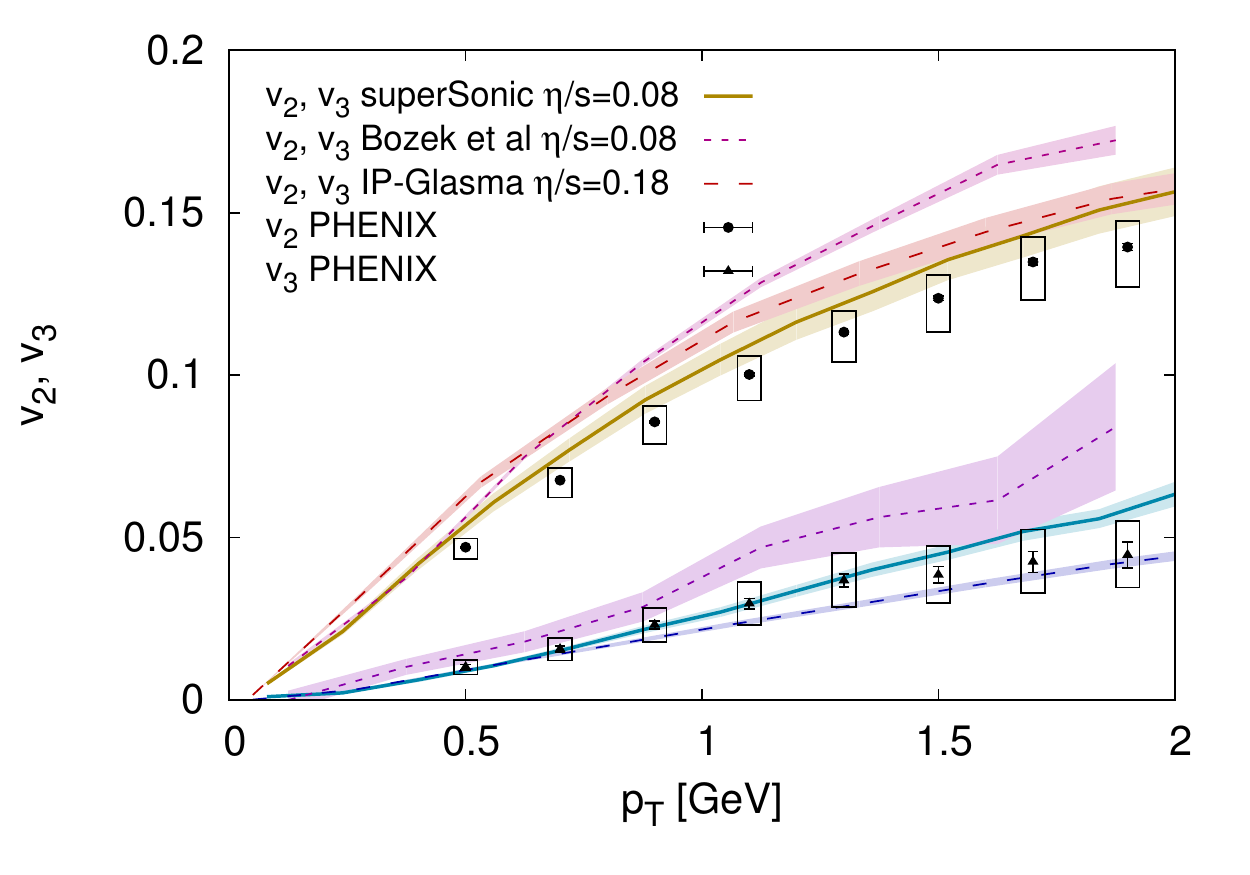}  
    \caption{ Experimental results for $v_2$ and $v_3$ in $^3$He+Au from PHENIX \cite{Adare:2015ctn} compared to three different hydrodynamic calculations:
      A prediction using MC Glauber initial conditions \cite{Bozek:2015qpa}, results from superSonic including pre-flow \cite{Romatschke:2015gxa},
      and IP-Glasma+\textsc{Music} calculation \cite{Schenke:2014gaa}, with increased viscosity of $\eta/s=0.18$ compared to the original prediction.\label{fig:vn3HeAu}} 
\end{figure}

Therefore, the PHENIX experiment has measured $v_2$ in d-Au and $v_2$ and $v_3$ in $^3$He-Au collisions. Hydrodynamic calculations \cite{Nagle:2013lja,Schenke:2014gaa} predicted a somewhat enhanced $v_3$ for the $^3$He-Au system because of the enhancement of significantly triangular events. 
Indeed, hydrodynamics can describe the $v_n$'s also in these two systems \cite{Bozek:2014cya, Bozek:2015qpa,Romatschke:2015gxa}. While some predictions \cite{Nagle:2013lja} had to be amended with a ``pre-flow'' component, others \cite{Schenke:2014gaa} need an adjustment of the shear viscosity. 
Both calculations then lead to good agreement with the experimental data \cite{Adare:2013piz,Adare:2015ctn}.
In Fig.\,\ref{fig:vn3HeAu} we show a compilation of results for $v_2$ and $v_3$ in $^3$He-Au collisions. While some of the calculations, which are predictions, can be improved by varying the shear viscosity, general agreement with the experimental data is very good, for the very different initial conditions.
superSonic and IP-Glasma+\textsc{Music} seem to have a somewhat too large ratio of $v_2/v_3$.
It would be desirable to also experimentally measure $v_3$ in d-A and \pA collisions to see whether it is indeed enhanced in $^3$He-A collisions as expected from the initial geometry and confirmed in hydrodynamic calculations \cite{Romatschke:2015gxa}.

\subsubsection{Parton transport models}
Parton transport models describe final state interactions microscopically, including various kinds of processes (elastic, inelastic) and different implementations of cross sections (isotropic, pQCD, ...). If these cross sections are large enough, the microscopic description should approach the hydrodynamic limit, and we expect similar results for flow observables as in (viscous) hydrodynamic simulations.

\paragraph{AMPT} The AMPT model \cite{Lin:2004en} is based on HIJING \cite{Wang:1991hta} to generate the initial state and the parton cascade ZPC \cite{Zhang:1997ej} that simulates the final state interactions.
One important ingredient that allows AMPT to produce enough final state collectivity is the so called string melting mechanism, which converts all initial minijets and soft strings into quarks and anti-quarks. These undergo elastic scatterings with a partonic cross-section which is controlled by the strong coupling constant and the Debye screening mass. After hadrons are formed using a coalescence model, they also interact further.

\begin{figure}
\center
    \includegraphics[width=\linewidth]{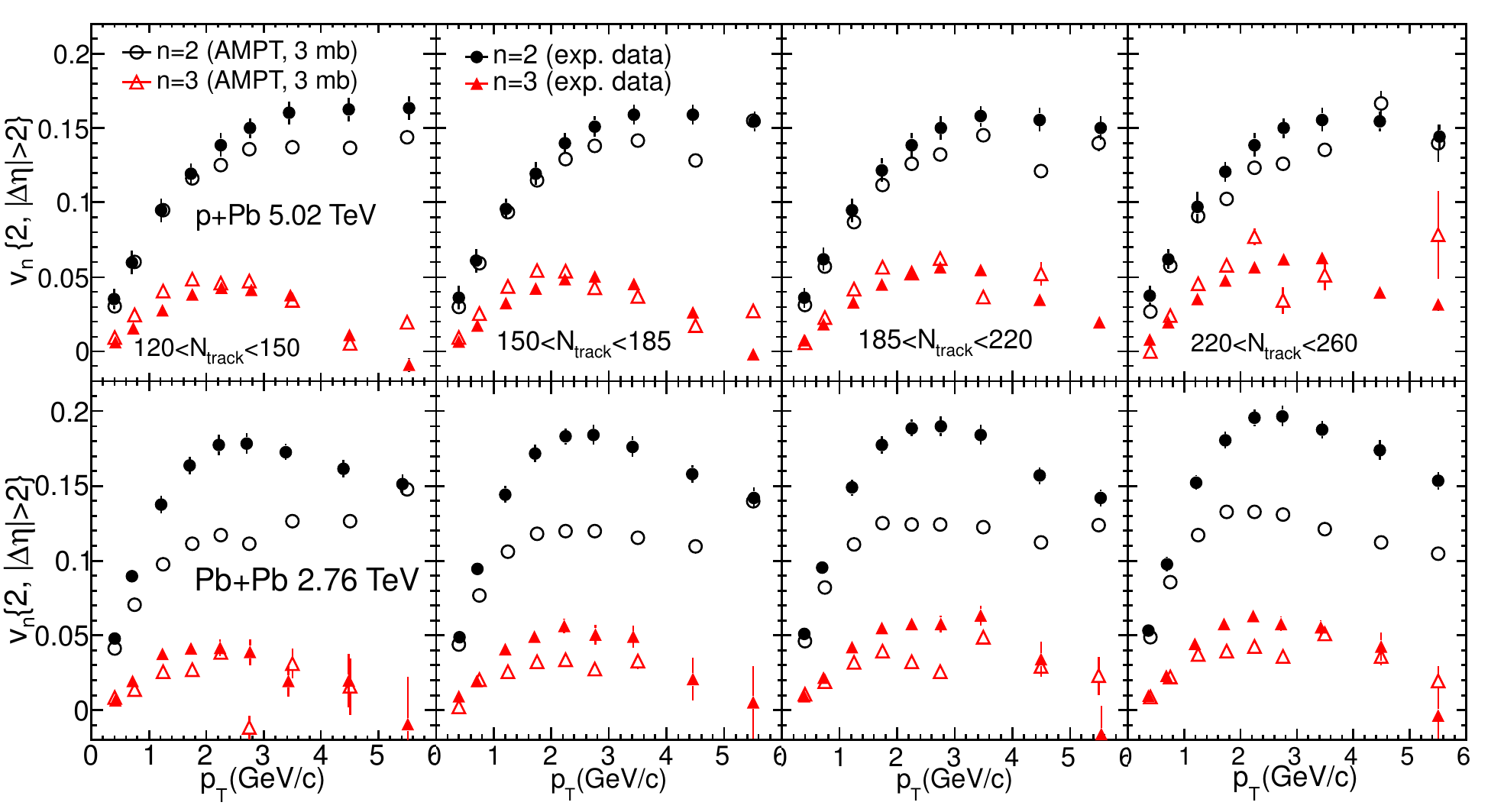}  
    \caption{The transverse momentum dependence of the elliptic, $v_2$, and triangular, $v_3$, anisotropy coefficients in \pPb (upper panel) and \PbPb collisions (lower panel) from the AMPT model (open symbols) compared to experimental data from CMS (filled symbols). Figure adapted from \cite{Bzdak:2014dia}.
    } \label{fig:bzdakAMPT}
\end{figure}

\pPb (and \ppc) collisions have been studied within the AMPT model in \cite{Ma:2014pva,Bzdak:2014dia}. It was found that with a modest elastic cross section of $\sigma = 1.5 -3\,{\rm mb}$ long range two-particle correlations were well described and the computed $v_2$ and $v_3$ agreed well with the experimental data from \pPb collisions (while $v_2$ in \PbPb collisions was underestimated by the model) - see Fig.\,\ref{fig:bzdakAMPT}.

A particularly interesting observation of this study \cite{Bzdak:2014dia} was that on average each parton undergoes only two collisions (and the hadronic cascade had a negligible effect on the results). It is surprising that such a small amount of interactions could lead to the build-up of collective flow.

A possible explanation of the physical process in the AMPT model was given in a recent work \cite{He:2015hfa}, where the details of what partons carry the anisotropy and how it depends on the number of their scatterings were analyzed. It was found that an escape mechanism seems to be responsible for the produced anisotropy in AMPT. The main ingredient of this is the initial spatial anisotropy of the medium and the resulting greater probability for partons to escape along the direction in which the medium is shorter.

This effect is drastically different from the build-up of hydrodynamic flow, and if it is the main effect, it is somewhat surprising that the transverse momentum dependence of the $v_n$ coefficients is so well reproduced by AMPT.

\subsection{Conclusions on final state interactions in small systems}
In conclusion we can state that it is possible to describe all characteristic features measured in \ppc and p/d/$^3$He-A collisions with models based on the collective response to an initial state geometry. In particular hydrodynamic models can reproduce the azimuthal anisotropies of charged hadrons $v_n$, the mass splitting of the mean transverse momentum and $v_2$ for identified particles, and the HBT radii. However, for all observables there is a significant uncertainty related to the little knowledge we have about the initial state in small systems. Depending on the assumptions made about how the initial shape of the system is generated, final results can vary dramatically. Apart from this caveat, the Knudsen number can take on rather large values in viscous hydrodynamic simulations of small systems. Thus, the quantitative results could be plagued with errors due to running hydrodynamics at least partly outside its domain of applicability.

%\clearpage

%%% Local Variables: 
%%% mode: latex
%%% TeX-master: "../RidgeReview-ijmpe"
%%% End: 

%\interfootnotelinepenalty=10000

\subsection{Multi-particle production in the Color Glass Condensate}

\paragraph{Introduction to the Color Glass Condensate}
\label{sec:introcgc}

A first attempt at understanding the nature of multi-particle correlations
should come from studying the production processes dictated by QCD.
Traditionally this is done in the framework of collinear factorization 
where the strong interaction processes are factorized into a product of cross
sections convoluted with parton distribution functions.  This approach is valid
in the Bjorken (or short distance) limit, 
\begin{equation}
\Lambda_{\rm QCD}\ll Q \sim x\sqrt{s}\sim \sqrt{s}
\end{equation} 
where $Q$ is the kinematic energy scale of the scattering process, and $x$ the
Bjorken $x$, {\em i.e.}, the longitudinal momentum fraction of the parton in
the infinite momentum frame.  Requiring the independence of an energy scale
$\mu$ separating of the long distance physics encoded in the parton
distribution functions from the short distance QCD matrix elements leads to the
DGLAP
(Dokshitzer-Gribov-Lipatov-Altarelli-Parisi)\cite{Gribov:1972ri,Altarelli:1977zs,Dokshitzer:1977sg}
renormalization group equation.  By construction the DGLAP equation resums
collinear logarithms of the form $\as\ln\left(Q^2/\Lambda_{\rm QCD}^2\right)$.

\begin{figure*}[t]
\centering
\includegraphics[height=7.cm]{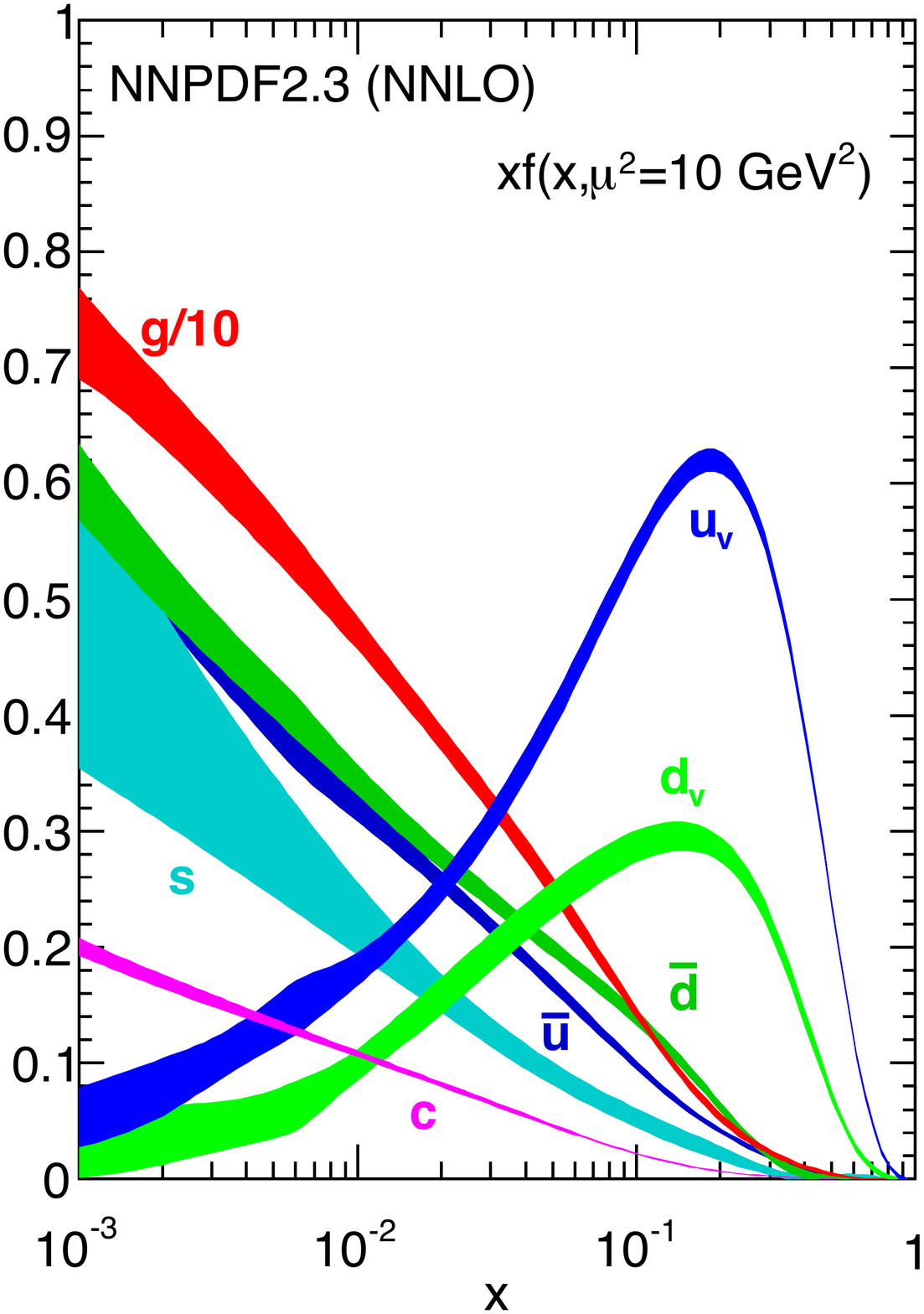}
\raisebox{0.75cm}{\includegraphics[height=6.cm]{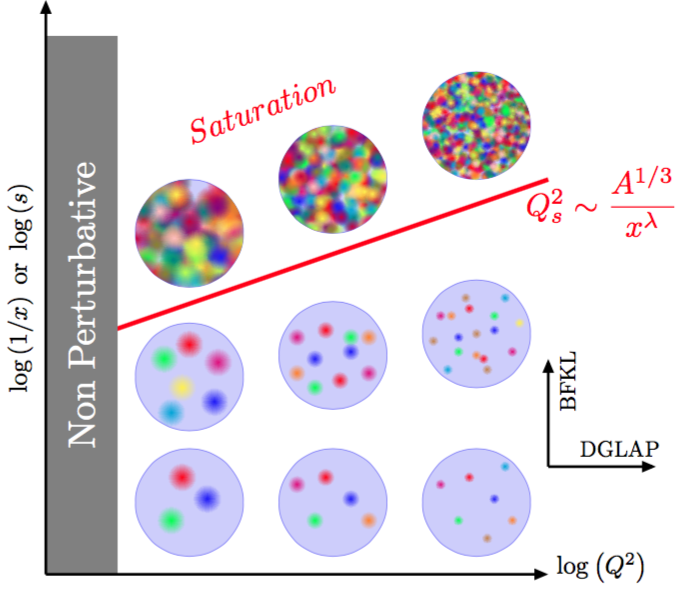}}
\caption{Left: Longitudinal momentum fraction, $x$, times the unpolarized parton distributions obtained in NNLO NNPDF2.3 global analysis \cite{Ball:2012cx} at scale $\mu^2=10$ GeV$^2$.  Figure adopted from \cite{Agashe:2014kda}.  Right: Evolution of the saturation scale $Q_S$ and the schematic structure of a nucleus with $Q^2$ along the horizontal axis and Bjorken $x$ along the vertical axis. The increase of the gluon number at a fixed transverse size scale along the vertical axis eventually leads to the phenomenon of saturation.}
\label{fig:landscape}
\end{figure*}

However, most of the particle production and the bulk of QCD phenomena occurs
at low $Q$ and therefore we consider QCD in the Regge-Gribov (or high energy)
limit,
\begin{equation}
\Lambda_{\rm QCD}\ll k_\perp \sim x\sqrt{s}\ll \sqrt{s}
\end{equation}
where another class of large logarithms, $\as\ln(1/x)$, must be resummed, in
this case by the Balitski-Fadin-Kuraev-Lipatov (BFKL)
equation. \cite{Kuraev:1977fs,Balitsky:1978ic}  The rapid rise in the gluon
distribution $xG(x,Q^2)$ at small $x$ for a fixed $Q^2$ (see
\fig{fig:landscape}), growing approximately as a power of $x^{-0.3}$ 
can be understood within the BFKL framework.  The rapid growth is moderated by the fact that gluons
having transverse momenta less than a semi-hard saturation scale
$\Qs(x)$ have a maximum occupation of $1/\as$.  A parametric estimate for the onset of saturation can be
made by considering the maximum number of partons that can be packed in
a proton of size $\Sp$.~\cite{Gribov:1984tu}  The condition that the transverse area occupied by a
parton ($\sim \Sp/xG$) is on the order of the gluon fusion cross-section
$\sigma\sim \as/Q^2$ defines a saturation scale 
$\Qs^2\simeq \as\;xG(x,\Qs^2)/\Sp$.
The argument can be extended to nuclei; A nucleus with atomic number $A$ has transverse size $\Sp\sim A^{2/3}$ and its parton distribution scales approximately as $xG_{\textrm{nucleus}}\sim A\cdot xG$ resulting in 
\begin{equation}
\Qs^2(x)\simeq \frac{\as A}{\Sp}xG(x,\Qs^2)\sim \frac{A^{1/3}}{x^\lambda}
\end{equation}
thereby enhancing saturation effects by a factor as large as $A^{1/3}\approx 6$ in nuclear targets such as lead ($A=208$).  The evolution of the saturation scale as a function of $Q^2$ and $x$ is schematically shown in the {\em phase diagram} of high energy QCD in \fig{fig:landscape}. 

The Color Glass Condensate is an effective field theory description of the saturated gluons in the Regge-Gribov limit.  More details can be found in the recent review article\cite{Gelis:2010nm}.  The degrees of freedom consist of strong color sources $\rho^a\sim 1/g$ of the nuclei and the resulting classical gauge fields.  
The strong color sources vary event-by-event and are governed by a gauge invariant probability distribution, $W$ with expectation values of operators schematically written as
\begin{equation}
\left<\mathcal{O}\right>\equiv \int \left[D\rhop\right] \left[D\rhot\right] W[\rhop]W[\rhot]\mathcal{O}\left[\rhop,\rhot\right]\,.
\end{equation}
The requirement that physical observables be independent of the scale separating the sources from fields results in the JIMWLK renormalization group equation.  There is no
known analytic solution to the JIMWLK evolution equation and while numerical
solutions are available \cite{Rummukainen:2003ns,Dumitru:2011vk}, their use in phenomenology remains limited. \cite{Lappi:2015vta}
There exists a kinematic window in $x$ where the nuclear wave-function can be derived in closed form.  
In the McLerran-Venugopalan
model\cite{McLerran:1993ni,McLerran:1993ka,McLerran:1994vd} $x$~must be large
enough, $\alpha_s\log(1/x)\ll 1$, such that quantum corrections entering at
leading logarithmic order can be neglected {\em and} $x$~must be small enough, $x\ll A^{-1/3}$,
such that the small-$x$ partons couple coherently to the large-$x$ sources.  In
this limit, the weight functional has the following Gaussian form
\begin{equation}
W[x,\rho]=\exp\left(-\int d^2\xp\left[\frac{\rho^a(\xp)\rho^a(\xp)}{2\mu_A^2}\right]\right)\,,
\label{eq:mvmodel}
\end{equation}
where $\mu^2_A=g^2A/(2\Sp)$ is the average color charge squared per unit
area per unit color degree of freedom.  While the MV model is a useful
phenomenological tool at moderate values of $x\sim 10^{-2}$, at smaller $x$, as
probed by the kinematics at the LHC, quantum corrections at leading logarithmic order in
$1/x$ must be incorporated.  The difficulties with the JIMWLK equation arise
from its hierarchical structure ({\em e.g.} the evolution of the two-point
function depends on the four-point function).  This can be circumvented by
working in the large $\Nc$ approximation (in addition to the already imposed
large nucleus approximation).  In this limit the expectation value of a
four-point function can be recast as the product of correlators of two point
functions and, at leading order in $\alpha_s$, the JIMWLK equation is
simplified to the LO Balitsky-Kovchegov \cite{Balitsky:1995ub,Kovchegov:1999ua}
(BK) equation 
\begin{eqnarray}
\frac{\partial T(\rb{},Y)}{\partial Y} 
=\int_{\rb{1}} {\mathcal K}(\rb{},\rb{1})
\left[ T(\rb{1},Y) + T(\rb{2},Y) - T(\rb{},Y) - T(\rb{1},Y)\,T(\rb{2},Y)\right],
\label{eq:BK-LO}
\end{eqnarray}
In the above expression $\rb{2}\equiv \rb{}-\rb{1}$ and $T(\rb{},Y)$ is the quark--anti-quark
dipole scattering amplitude from a nucleus.
In the context of the BK equation, the kernel ${\mathcal K}$ amounts to resumming all terms
of the form $\left(\alpha_s\ln(x_0/x)\right)^n$ arising at any order in
perturbation theory.  It is well known that running coupling corrections
qualitatively modify the small~$x$ evolution and analytic results determining
the scale of the coupling that enters the evolution kernel are known
\cite{Balitsky:2006wa,Kovchegov:2006vj}.  The form of these running coupling
corrections amount to two pieces; a piece resembling a running coupling
correction to the LO BK Kernel\cite{Lipatov:1996ts}, and a ``subtraction term''
that introduces new structures.  In what follows we use the ``Balitsky
prescription''~(\eq{eq:NLO-BFKL-kernel}) for the evolution kernel,
\begin{equation}
{\mathcal K}_{\rm Bal.}(\rb,\rb{1},\rb{2}) = \frac{\alpha_s(\rb{}) N_c}{\pi}\left[ \frac{\rb{}^2}{\rb{1}^2 \rb{2}^2} +\frac{1}{\rb{1}^2}\left(\frac{\alpha_s(\rb{1}^2)}{\alpha_s(\rb{2}^2)}-1\right)+\frac{1}{\rb{2}^2}\left(\frac{\alpha_s(\rb{2}^2)}{\alpha_s(\rb{1}^2)}-1\right)\right] \, ,
\label{eq:NLO-BFKL-kernel}
\end{equation}
as it was shown that the ``subtraction term'' is numerically less
important~\cite{Albacete:2007yr} and will be ignored in this work.
Numerical results\cite{Lappi:2015fma} for the full NLO BK
equation\cite{Balitsky:2008zza} have recently become available demonstrating a
sensitive dependence on initial condition with results not necessarily positive
definite calling for a better understanding of the NLO BK equation before it
can be used for phenomenology. A recent analysis that resums double logarithms seems to 
improve the situation at NLO.\cite{Iancu:2015vea}

\begin{figure*}[t]
\centering
\includegraphics[height=4.cm]{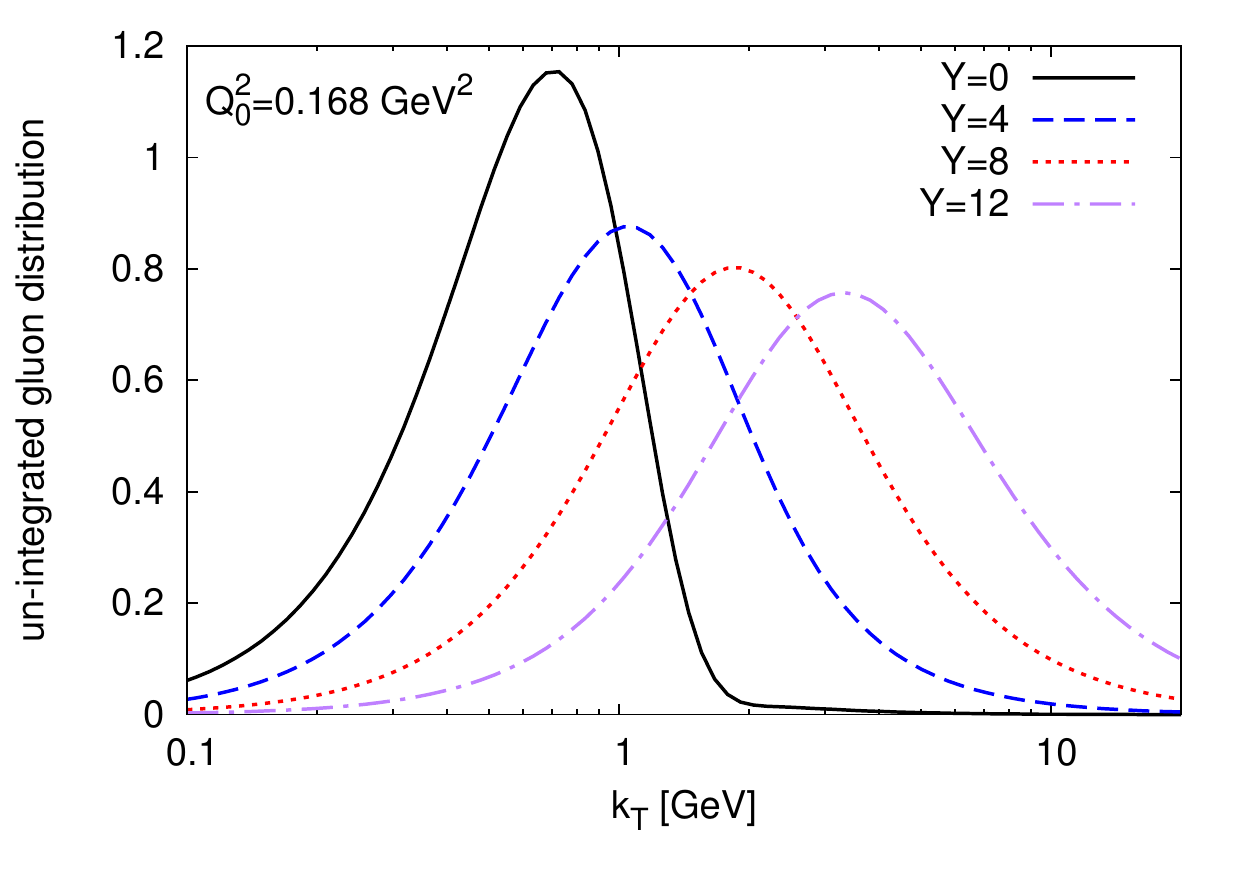}
\raisebox{0.25cm}{\includegraphics[height=4.cm]{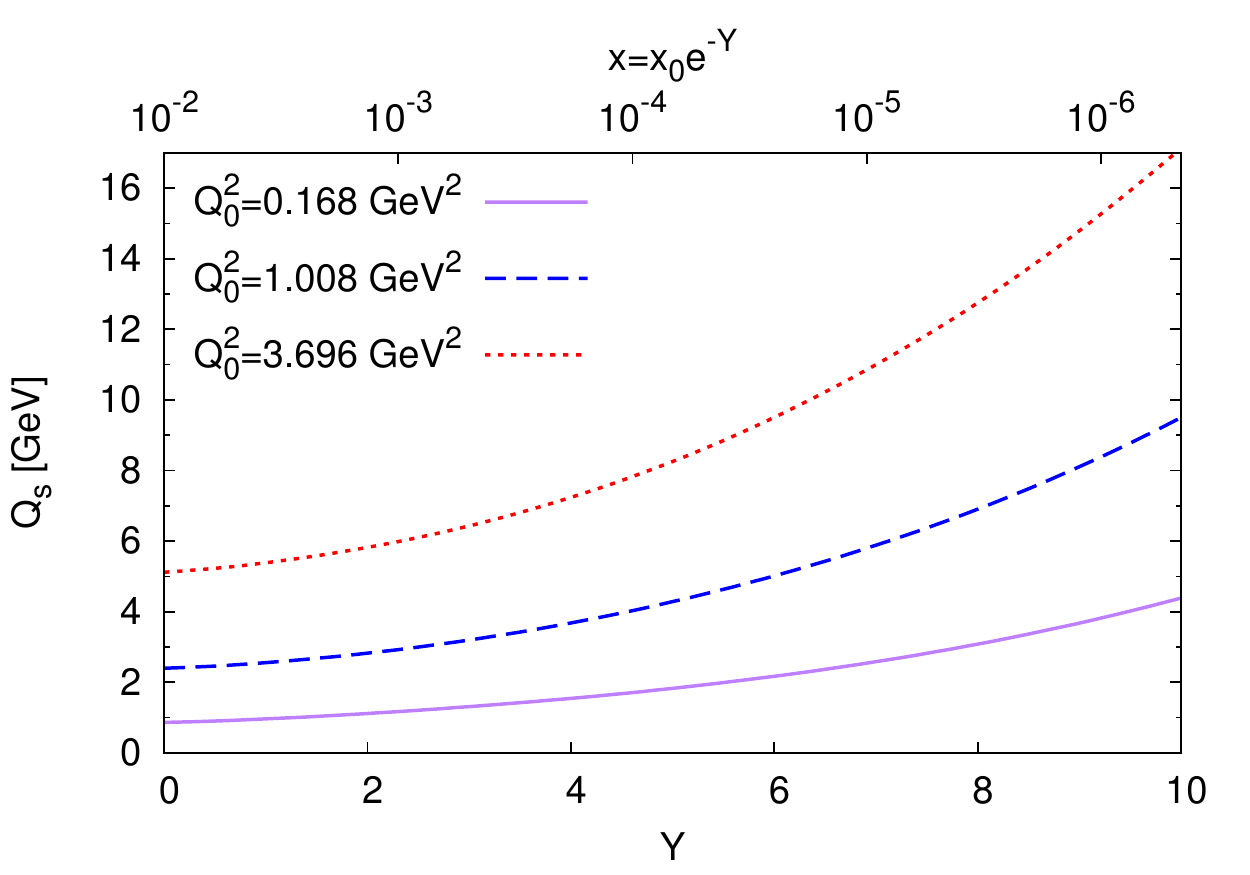}}
\caption{Left: Un-integrated gluon distribution in the adjoint representation, plotted as $\as\phi/(\Sp\Nc)$, with initial condition ($Y=0$) corresponding to a min. bias proton along with subsequent evolution by rcBK to ($Y=4,8,12$).  Right:  The saturation scale defined as the location of the maximum of $k_\perp\phi(Y,k_\perp)$ as a function of $Y=\ln(x_0/x)$ for initial conditions $Q_0^2=0.168, 1.008, 3.696$~GeV$^2$. }
\label{fig:rcBK}
\end{figure*}

Solution of the running coupling BK equations (rcBK) requires a suitable initial condition at large-$x$.  This review will focus on the results using an MV~model-like initial condition with anomalous dimension $\gamma$,
\begin{equation}
T(\rb{},Y=0) = 1-\exp\left[-\left(\frac{r^2Q_{s0}^2}{4}\right)^\gamma\ln\left(\frac{1}{r\Lambda_{QCD}}+e\right)\right].
\label{eq:mvic}
\end{equation}
The value of $\gamma$ and initial saturation scale (in the fundamental representation), $Q_{s0}^2$ have been constrained by global fits to available small-$x$ ($x\leq 0.01$) lepton-proton scattering data\cite{Albacete:2009fh}.  This work will use the parameter set h' from \cite{Albacete:2010sy} having parameters $\Lambda_{QCD}=0.241$~MeV, $\gamma=1.119$, $Q_{s0}^2=0.168$~GeV$^2$.  

At large $\Nc$ the dipole scattering ampliutde can be identified with the
unintegrated gluon distribution in the adjoint representation\cite{Fujii:2006ab} 
\begin{equation}
\Phi(Y,\kp)=\frac{\Nc\kp^2}{4\as}\int d^2\rb{}\; e^{i\kp\cdot\rb{}}\left[1-T(\rb{},Y)\right]^2\,.
\end{equation} 
For completeness we also point out that at large $\kp$
the Wilson lines can be expanded in powers
of the sources yielding a relation between the average squared color charge density and unintegrated gluon distribution
\begin{equation}
\Phi(Y,\kp)=g^2\pi\left(\Nc^2-1\right)\frac{\mu^2(Y,\kp)}{\kp^2}\,.
\label{eq:phitomu}
\end{equation}
The numerical solution of the rcBK equation is shown in the left plot of \fig{fig:rcBK}.  The solid black curve is the initial condition defined in ~\eq{eq:mvic} and its 
evolution to $Y=4,8,12$ is shown as the dashed, dotted, dash-dotted curves respectively.  As we will argue at length below,
the solitonic structure is crucial to our understanding of the ridge
phenomenon.  The non-monotonic behavior of the gluon distribution kinematically
constrains gluons to be produced in similar directions.  The transverse
momentum dependence of the ridge will be controlled to a large extent by the
maximum gluon occupation at the relevant values of $x$ probed in the processes.

The right plot of \fig{fig:rcBK} shows the location of the maximum of the
quantity $k_\perp\Phi(k_\perp)$ as a function of $Y$ or equivalently $x=x_0
e^{-Y}$.  The quantity $k_\perp\Phi(k_\perp)$ represents the number of gluons
at a given $k_\perp$ and its maximum value can be identified with a saturation
momentum though it should not be confused with the saturation scale that enters
into the initial condition in \eq{eq:mvic}.  The latter is the {\em
initial} saturation scale at $x=x_0=10^{-2}$ in the fundamental representation.  

The solid purple curve in the right plot of \fig{fig:rcBK} corresponds to the
wavefunction shown in the plot on the left ({\em i.e.} having initial condition
$Q_{s0}^2=0.168$~GeV$^2$ which we will refer to as ``minimum bias'').  The rcBK
equation has no impact parameter dependence and the fits to lepton-proton
scattering data (from which the value of $Q_{s0}^2=0.168$~GeV$^2$ is obtained)
assumes that the impact parameter dependence factorizes from the dipole
amplitude. Since the median impact parameter dominates inclusive scattering we
will use this initial condition for proton configurations prototypical of
minimum bias.

In order to mimic rare fock space configurations generated in high-multiplicity
events the initial saturation scale will be adjusted to account for the
increased multiplicity.  The right plot of \fig{fig:rcBK} also shows the
evolution of the saturation momentum for configurations involved in high
multiplicity events.  The initial condition having $Q_{s0}^2=1.008$~GeV$^2$
corresponding to $6\times$ the value of a min. bias proton will be argued to be
the relevant scale need to describe high-multiplicity events in p-p collisions
at the LHC.  The initial condition with $Q_{s0}^2=3.696$~GeV$^2$ ($22\times$
the min. bias proton) is used to describe a {\em hot-spot} configuration in the
nucleus probed in central \pPb events at the LHC.  This large value is expected
as it takes into account the additional $A^{1/3}$ enhancement from nuclear
coherence. 

\subsubsection{Two-gluon production for $\pp\gtrsim \Qs$: Glasma Graphs}

In the presence of strong classical sources the naive power counting of Feynman
diagrams can change drastically.  We consider two mechanisms for the production
of a pair of mid-rapidity hadrons.  The first proceeds via the production of
two gluons from a single $t$-channel gluon exchange as shown in the upper left amplitude in
table~\ref{tab:pcounting}.  We refer to this diagram as the Jet Graph, since
the gluon pairs that are produced in this channel are predominately
back-to-back in azimuthal angle.  We should stress that we are interested in
mini-jets ({\em i.e.} soft dihadron correlations) and not jets in the usual
sense studied in collinear factorization.

The second class of diagrams shown in the lower left of Fig.\,\ref{tab:pcounting},
has two gluons produced from two $t$-channel gluon exchanges.  As drawn here the
diagram appears disconnected and one might conclude does not contribute to an
intrinsic two-particle correlation.  This is not the case because the averaging
over the color sources introduces non-trivial connections within the
seemingly disconnected diagram.

\begin{table}
\begin{tabular}{m{1.3cm} | m{3.7cm}  m{0.8cm} | m{3.7cm}  m{0.8cm} }
& \multicolumn{2}{c}{Low color charge density} & \multicolumn{2}{|c}{High color
charge density} \\
& \multicolumn{2}{c}{(min. bias)} & \multicolumn{2}{|c}{(high multiplicity)} \\
\hline\hline 
Jet Graph &
\vspace{0.4cm}
\includegraphics[height=3.5cm]{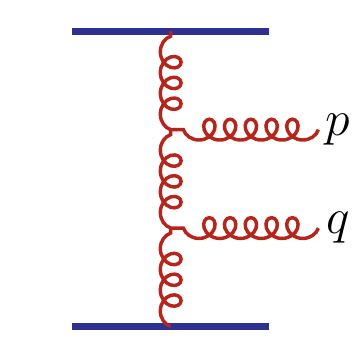}
& \hspace{+0.5cm}\scalebox{1.5}{$\alpha_s^4$} & 
\vspace{0.4cm}
\includegraphics[height=3.5cm]{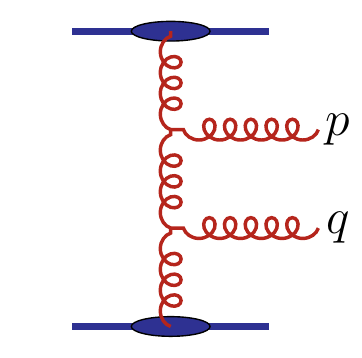}
& \hspace{+0.5cm}\scalebox{1.5}{$\alpha_s^0$} \\
Glasma Graph & 
\vspace{0.4cm}
\includegraphics[height=3.5cm]{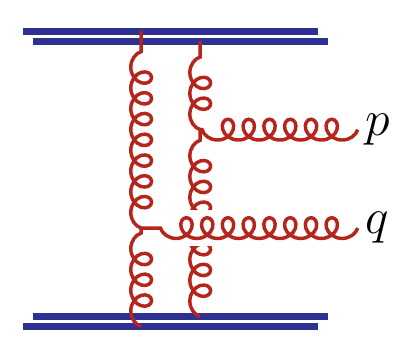}
& \hspace{+0.5cm}\scalebox{1.5}{$\alpha_s^6$} &
\vspace{0.4cm}
\includegraphics[height=3.5cm]{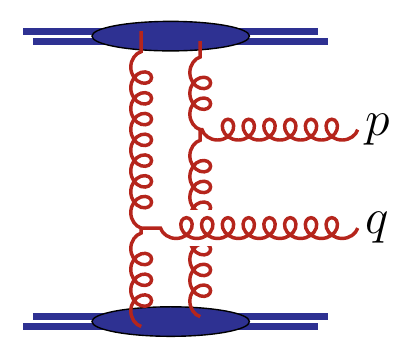}
& \hspace{+0.5cm}\scalebox{1.5}{$\alpha_s^{-2}$} \\
\end{tabular}
\caption{Power counting for multiparticle production in QCD in the dilute (left
column) and dense (right column) limit of color sources.  The top row shows the
diagram responsible for mini-jet production (hadrons produced primarily
back-to-back in azimuthal angle).  The bottom row shows the ``glasma" diagram responsible
for the near-side collimation that is long-range in rapidity.  Each three-gluon 
vertex or coupling to a bare parton is proportional to $g$.  Each gluon coupling 
to a {\em blob} contributes $1/g$.}
\label{tab:pcounting}
\end{table} 

In the dilute limit each vertex (both the three-gluon and the coupling to the
valence parton) come with a single power of the strong coupling constant $g$.
After squaring the jet amplitude the cross section scales as $\as^4$.  The
glasma diagram on the other hand has six such vertices and the cross-section
is proportional to $\as^6$ and is therefore suppressed by $\as^2$ in the dilute limit
of both projectile and target.

In the high energy limit the power counting changes dramatically.  By high
energy limit we mean that $k_\perp\ll \sqrt{s}$ such that there is
sufficient small-$x$ evolution that the gluon densities are non-perturbatively
large.  The high-energy limit is also reached for large nuclei or rare
collisions of hot-spot configurations of the projectile and/or target.  The two
tagged gluons now couple to the high occupancy $\sim 1/\as$ of larger-$x$
sources.  The effective coupling of a gluon to the valence region is now $1/g$
instead of $g$.  This is depicted in the right diagrams of table \ref{tab:pcounting}
where each connection to a {\em blob} couples with $1/g$.  In the dense-dense
limit the Jet Graph goes as $\as^0$ while the glasma graph scales as
$\as^{-2}$.  For high multiplicity collisions one therefore expects the glasma graph
as the main contributor to two-particle production.

These two classes of diagrams can also be treated in collinear factorization.
The Glasma graph would refer to double parton scattering and is 
suppressed by $\Lambda_{\rm QCD}^2/Q^2$ relative to a single parton scattering
diagram\cite{Manohar:2012pe,Diehl:2011tt}  (such as our Jet Graph).  The high-energy limit
of QCD therefore allows us to study novel QCD processes which would otherwise
be largely inaccessible by experiment.  

The motivation for the first study of two-particle correlations in the color
glass condensate framework \cite{Dumitru:2008wn} was geared towards explaining 
the ridge in heavy-ion collisions.  The basic idea is that the longitudinal
chromo-electric and -magnetic fields created in the collision of two large
nuclei radiate correlated particles approximately isotropically in the azimuthal
separation of the pairs.  The boost invariance of the classical field results
in boost invariant particle production naturally explaining the long-range
rapidity correlations.  Collimation in azimuthal separation, $\Delta\phi$ is
achieved by final-state effects such as strong radial flow due to hydrodynamic
expansion.  

The strength of the correlated two-particle distribution can be expressed as \cite{Dumitru:2008wn}  
\begin{equation}\label{eq:twopcorrIso}
\left<\frac{d^2N}{dy_p d^2\pp dy_q d^2\qp}\right>_{\rm conn.}=\kappa\frac{1}{\Qs^2\Sp}\left<\frac{dN}{dy_p d^2\pp}\right>\left<\frac{dN}{dy_q d^2\qp}\right>\,.
\end{equation} 
We should stress that the above is an intrinsic correlation expressed in terms of the single particle distribution\footnote{In general, the double inclusive distribution can be written as
\begin{equation}
\left<\frac{d^2N}{dy_p d^2\pp dy_q d^2\qp}\right>=\left<\frac{dN}{dy_p d^2\pp}\right>\left<\frac{dN}{ dy_q d^2\qp}\right>+\left<\frac{d^2N}{dy_p d^2\pp dy_q d^2\qp}\right>_{\rm conn.}
\end{equation}
}
and is essentially a consequence of dimensional analysis -- the correlations are
entirely classical in nature and there is only one scale, $\Qs$ characterizing
the strength of the fields.  The factor $1/(\Sp\Qs^2)$ has the physical
interpretation of the transverse area of a flux-tube from where the particles
are emitted divided by the transverse area of the system.  The coefficient
$\kappa$ can be calculated but it is a constant of $\mathcal{O}(1)$ and the
strength of the resulting correlation fares well with the experiment in a
blast-wave model\cite{Gavin:2008ev}.  The subsequent discovery of the
double-ridge (a symmetric collimation on {\em both} the near and away-side)
which was found after careful subtraction of the recoiling jet leaves a gap in
this picture.

It was later realized that the glasma graphs contain an intrinsic dependence on
the azimuthal separation irrespective of the presence of an additional
collimation mechanism \cite{Dumitru:2010iy} which is not apparent in
\eq{eq:twopcorrIso}. 
After the observation of a ridge in high-multiplicity
proton collisions a blast-wave analysis showed that the near-side ridge in p-p
could {\em not} be described by combination of isotropic production from glasma
flux tubes with subsequent transverse flow\cite{Dusling:2012iga}.  Instead, the
systematics of the p-p ridge were consistent with the intrinsic azimuthal
dependence from glasma graphs leaving little room for any additional transverse
expansion.

\begin{figure*}[t]
\centering
\includegraphics[width=12cm]{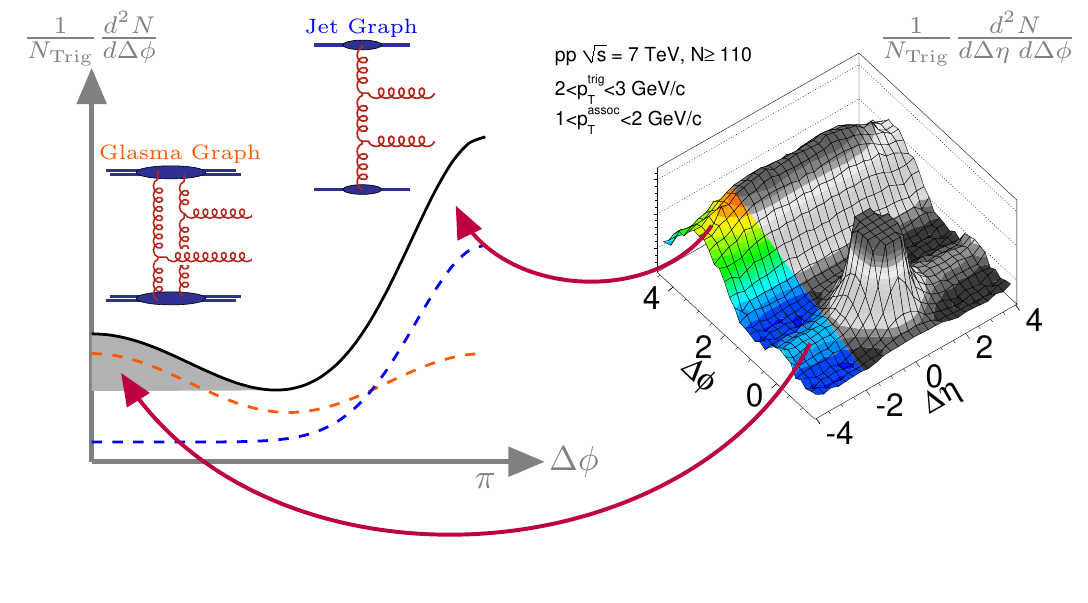}
\caption{Anatomy of a proton-proton collision.  The away-side peak, associated
with mini-jet production can be understood from the jet-graph  (two  gluons  produced
from  a  single  ladder)  as  shown  in  the  right  diagram  along  with  its
schematic contribution to the per-trigger-yield plotted in blue.  The Glasma
graph contribution (left diagram) is shown schematically by the orange curve.
The shaded gray region (extracted experimentally by the ZYAM procedure) is
referred to as the associated yield} 
\label{fig:ppsum}
\end{figure*}

We therefore have the following picture of the dihadron correlations in
proton-proton collisions (summarized graphically in Fig.\,\ref{fig:ppsum}).  The Jet
Graph generates particles predominately back-to-back ({\em i.e.} at relative
azimuthal angle of $\pi$) and when expressed as a per-trigger-yield
\footnote{The per-trigger-yield is the double-inclusive distribution
$\left<d^2N\right>$ divided by one power of the single inclusive distribution
$\left<dN\right>$.  Since the jet and single-gluon graphs proceed via a single
$t$-channel exchange they have the same number of connections to the larger-$x$
sources and therefore much of the centrality dependence cancels in the ratio.}
is approximately independent of multiplicity, an observation consistent with the
experimental data.

Due to a novel interference phenomenon generated from the intrinsic correlations
in the nucleus (the precise form of which will be discussed in the subsequent
sections) the glasma graph has enhanced production on angular separations of
$\Delta\phi=0$ and $\Delta\phi=\pi$.  The collimation at $\Delta\phi=0$ is
responsible for the near-side ridge while that at $\Delta\phi=\pi$ explains the
double-ridge (after appropriate jet subtraction).

The glasma graph has a much stronger centrality dependence than the jet graph.
In the dilute-dilute limit the per-trigger
yield
\footnote{The single inclusive spectra scale as $\as^3$, $\as$ and
$\as^{-1}$ in the dilute-dilute, dilute-dense, and dense-dense limits
respectively.} 
of the glasma graph scales with coupling as $\as^3$ while in the dense-dense
limit it goes as $\as^{-1}$.  This large enhancement is expected for high
multiplicity events at moderate $\pp\sim \Qs$.  

With this basic picture in mind we now go on to discuss two-gluon production in
the color glass-condensate framework.  The next two sections focus on the
perturbative $\pp\gg \Qs$ limit of di-jet production and glasma graphs
respectively and explore the mechanism behind the intrinsic azimuthal
correlations.  We will then give an overview of the relevant phenomenology of
this framework and discuss some open issues.   

\paragraph{Mini-jet production} In this section we explore in more detail our understanding of di-jet or
mini-jet production.   The away-side (or recoiling jet) is pervasive in all
dihadron analyses and serves as an important baseline.

Computing the di-jet amplitude in the Color-Glass-Condensate framework is a
formidable task.  The diagram shown in the upper right of
table~\ref{tab:pcounting} involves calculating the gluon propagator in the
full nonlinear classical background field.  In the dilute-dense limit this was
looked at
in~\cite{JalilianMarian:2004da,Baier:2005dv,Fukushima:2008ya,Iancu:2013dta} and
the resulting expression could not be written in a $k_T$-factorized form as the
end result involved correlators of four Wilson lines.  This quadrupole operator
obeys its own evolution equation and it cannot be factorized into a simple product of
dipoles. We point out, however, that its nonlinear Gaussian approximation \cite{Dominguez:2011wm} in terms of the dipole operator 
agrees numerically very well with the full result in JIMWLK evolution \cite{Dumitru:2011vk}.

In contrast, di-jet production has been extensively studied in collinear
factorization and at leading order the partons are produced strictly
back-to-back\footnote{The collinear factorized form for di-jet production can be obtained from the 
$k_\perp$-factorized form by substitution of the unintegrated gluon distribution with
the appropriate parton distributions, $\Phi(\kp)=xg(x)\; 4\pi^4\delta^{(2)}(\kp)$. 
From equation~\ref{eq:MRK} we obtain 
\begin{equation}
\frac{d^2N}{dy_p d^2\pp dy_q d^2\qp}=x_pg(x_p)\;x_qg(x_q)\;\frac{4\as^2 \Nc^2}{(\Nc^2-1)}\frac{\delta^{(2)}\left(\pp+\qp\right)}{\pp^2\qp^2}
\end{equation}
agreeing with equation~(60) of \cite{DelDuca:1995hf} when the
large-$y$ matrix element given in equation~(54) is used.
}.  
This is due to the fact that the internal momentum distribution
of the partons in the hadrons is neglected and conservation of momentum requires
the produced partons to have equal transverse momentum squared and opposite directions,
{\em i.e.}, $\sigma\sim\delta(\pp^2-\qp^2)\delta(\vert\phi_p-\phi_q\vert-\pi)$.
Any internal motion, for example due to Fermi motion or soft emissions, will
lead to a decorrelation of the jets. 

The soft radiation emitted between the produced gluons and their respective
hadrons in addition to the emission of rapidity ordered gluons between the two
tagged partons is encoded within the following $k_T$ factorized di-jet
cross-section
\begin{align}
\frac{d^2N}{d^2\pp d^2\qp dy_p dy_q} &=\frac{\as^2\Nc^2}{4\pi^8 (\Nc^2-1)}
\frac{\Sp}{\pp^2\qp^2}\\
&\times\int_{\kt{0},\kt{3}} \Phip(\kt{0})\Phit(\kt{3})\mathcal{G}(\kt{0}-\pp,\kt{3}+\qp,\Delta y_{pq})\,.\nonumber
\label{eq:BFKL}
\end{align}
In the context of BFKL resummation (where the above is rigorously defined) the
quantities $\Phi$ are the BFKL evolved impact factors and are known to
next-to-leading order\cite{Colferai:2010wu}.  In this work we will apply a {\em
hybrid} formalism that identifies the impact factor with the unintegrated gluon
distributions introduced earlier.  We employ in this work the LO BFKL Green's
function, the explict form of which can be found in \cite{DelDuca:1995hf} for example.  
Though its NLL form is known\cite{Colferai:2010wu} the corrections introduced to 
the observables considered here are small enough to be neglected.

The BFKL radiation can be {\em switched off} by taking the $\as\Delta y$ limit of the above Green's function and the resulting di-jet cross section becomes
\begin{equation}
\frac{d^2N}{dy_p d^2\pp dy_q d^2\qp}=\frac{\as^2 \Nc^2}{8\pi^{8}(\Nc^2-1)}\frac{\Sp}{\pp^2\qp^2}\int d^2\kp\Phi(\kp)\Phi\left(\kp+\pp+\qp\right)
\label{eq:MRK}
\end{equation}
The above expression is valid in the multi-Regge kinematics (MRK) corresponding to the leading $\Delta y\to\infty$ limit of the $2\to4$ amplitude.  Using the $2\to4$ matrix element in the eikonal approximation results in the quasi-multi-Regge kinematics (QMRK) whose precise form can be found in \cite{Leonidov:1999nc}.

\begin{figure*}
\centering
\includegraphics[height=6.cm]{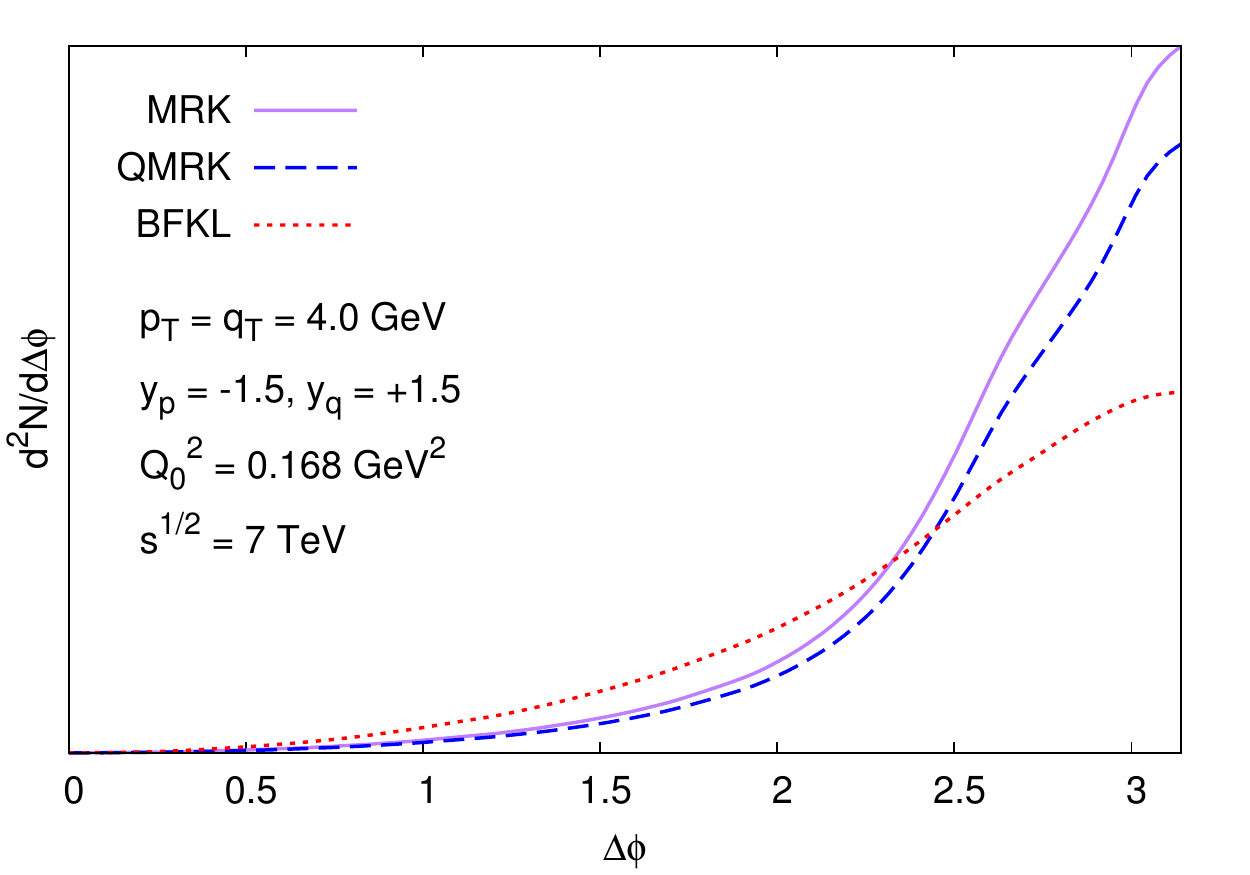}
\caption{Two-gluon production versus relative azimuthal separation for min. bias p+p collisions (details are labelled in the figure).  The three curves represent three approximations; Dashed blue: Eikonal approximation (parametrically small angle scattering of incident particles) to the $2-4$ matrix element without radiation between the tagged gluons ($\as\Delta y\to 0$), Solid purple: Leading term in $\Delta y\to\infty$ of the QMRK result, Dotted orange; Includes the full BFKL Green's function taking into account the emission of rapidity ordered gluons between the tagged final-state partons.}
\label{fig:jetqmrk}
\end{figure*}

\begin{figure*}[t]
\centering
\includegraphics[height=4cm]{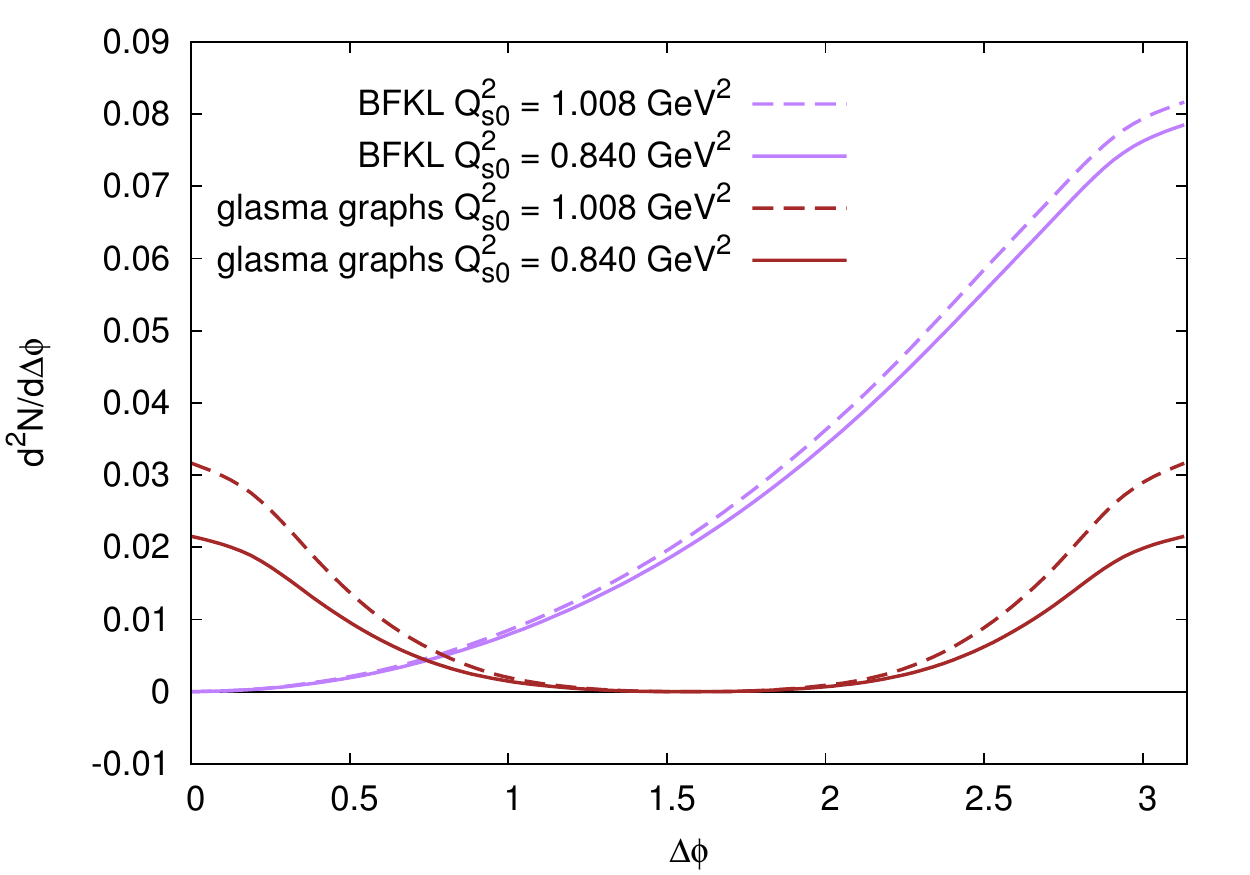}
\includegraphics[height=4cm]{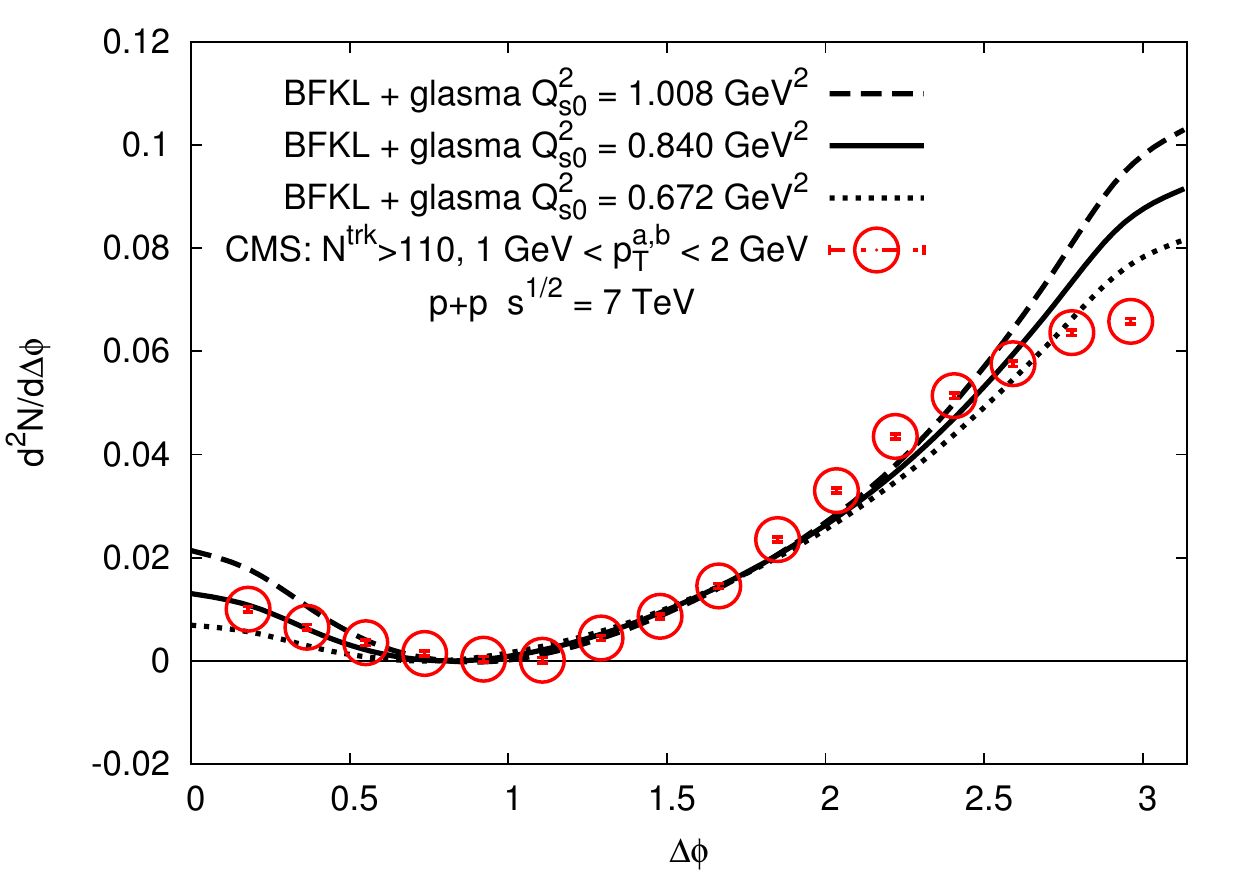}
\caption{Long range ($2\leq \Delta\eta \leq 4$) per-trigger yield of charged hadrons as a function of $\Delta\phi$ for \ppc collisions at $\sqrt{s}=7$ TeV.  Data points are from the CMS collaboration\cite{}. The curves show the results for $Q_0^2(x=10^{-2})=0.840$~GeV$^2$ and $Q_0^2(x=10^{-2})=1.008$~GeV$^2$. } 
\label{fig:pp110}
\end{figure*}

\begin{figure*}[t]
\centering
\includegraphics[width=\textwidth]{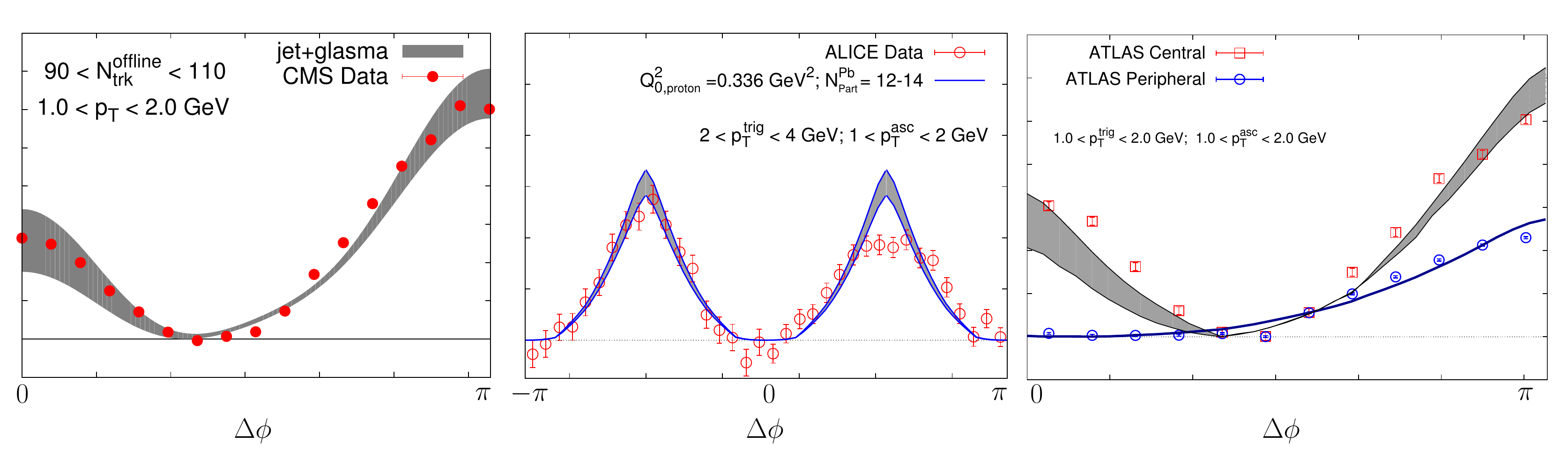}
\caption{Representative sample of \pPb data form the CMS, ALICE and ATLAS collaborations of the per-trigger yield along with calculations within the {\em glasma graph} framework. }
\label{fig:datapA}
\end{figure*}

Figure~\ref{fig:jetqmrk} demonstrates the results of the above discussion and
shows the relative importance of including rapidity ordered gluon emissions
even for moderate $\as\Delta y$ as encoded in the BFKL formalism.  Neglecting
these emission would spoil the agreement with the experimentally measured
away-side mini-jet.  The details of the full calculations will be discussed
further on, but The agreement between the BFKL evolution and the experimental
data can be seen the agreement of the away-side yield in \fig{fig:pp110} for
\ppc collisions.  Another example is the agreement between the blue curve in
the right plot of \fig{fig:datapA} that shows the mini-jet with BFKL evolution
in peripheral \pPb and the corresponding ATLAS measurement of the same.  The
mini-jet itself contains interesting QCD dynamics and may be a useful probe of
Pomeron exchange and BFKL evolution \cite{Dusling:2012cg} complementing studies
that look for the growth in di-jet cross section as proposed in
\cite{Mueller:1986ey}.  

\paragraph{Glasma Graphs}

We now review the mechanism behind the intrinsic collimation that was
predicted\cite{AD_rikenwkshp,Dumitru:2010iy} to exist in systems lacking
hydrodynamic flow or final state rescattering.  We will provide an overview of
the {\em glasma graphs} that are responsible for this intrinsic azimuthal
correlation and argue that the resulting signal is a quantum interference
effect sensitive to correlations present in the nuclear wave-function before
the collision occurs.

In order to set the stage and develop some notation we will begin with an
overview of the perturbative $\pp\gg\Qs$ calculation of single gluon production
and its resulting $k_\perp$-factorized expression.  The starting point
is the probability to produce
a gluon having rapidity $y$ and transverse momentum $\kp$,
\begin{equation}
\frac{d{\rm N}}{dy d^2\kp} =\frac{1}{2(2\pi)^3}\sum_{a,\lambda}\left<\vert M_{\lambda}^{a}\left({\bf k}\right)\vert^2\right>\,.
\label{eq:LSZ1}
\end{equation}
The classical contribution to the amplitude for the production of a single
gluon with four-momentum $k$ is $M_\lambda({\bf
k})=\epsilon^{\lambda}_{\mu}({\bf k})~k^2 A^{\mu}({\bf k})$ with
$\epsilon_\mu^\lambda$ representing the gluon polarization vector having
polarization states $\lambda$.  The above should be evaluated in the on-shell
$k^2\to 0$ limit.  The expectation value $\langle \cdots \rangle$ represents an
averaging over the color configurations of the two nuclei.  

At large transverse momentum the gauge fields can be expressed explicitly in terms of color
charge densities~\cite{Kovner:1995ts,Kovner:1995ja} of the projectile and
target nuclei, $\rhop$ and $\rhot$, 
\begin{equation}
p^2A^{\mu a}(p)=-if^{abc}g^3\int\frac{d^2\kp}{(2\pi)^2}L^\mu(\pp,\kp)\frac{\rhop^b(\kp)\rhot^c(\pp-\kp)}{\kp^2(\pp-\kp)^2}\,.
\label{eq:gf1}
\end{equation}
$L^\mu(\pp,\kp)$ is the effective Lipatov gluon emission
vertex\cite{Balitsky:1978ic}.
The resulting expectation value in \eq{eq:LSZ1} is a
product of two-point correlators over the color charge densities of the target and
projectile.  In the non-local MV model the momentum space correlator is
\begin{equation}
\langle \rho^a(\kt{1})\rho^{*b}(\kt{2})\rangle = (2\pi)^2\delta^{ab}\mu^2(\kt{1})\delta^{(2)}\left(\kt{1}-\kt{2}\right)\,.
\label{eq:rhorhokt}
\end{equation}
We should stress that in order to accommodate the small-$x$ evolution one needs
to generalize the local MV model of \eq{eq:mvmodel} to include a nonlocal
version such that $\mu^2\to\mu^2(\xp-\yp)$ and consequently in momentum-space a
function of the transverse momentum ({\em i.e. $\mu^2\to\mu^2(\vert\kp\vert)$}).
Further discussion of this subtlety can be found in \cite{Fujii:2006ab}.  Combing
equations~(\ref{eq:phitomu}), (\ref{eq:LSZ1}), (\ref{eq:gf1}), (\ref{eq:rhorhokt})
the following $k_T$-factorized form for the
single inclusive gluon production is obtained
\begin{equation}
\frac{dN}{dy_p d^2\pp}=\frac{\as\Nc}{\pi^4(\Nc^2-1)}\frac{\Sp}{\pp^2}\int\frac{d^2\kp}{(2\pi)^2}\Phip(\kp)\Phit(\pp-\kp)\,.
\label{eq:ktfac1}
\end{equation}
We should point out that the above factorized form is obtained in a
different limit than the usual-$k_T$ factorization derived from the scattering
of a dilute projectile off a quasi-classical target as examined in
\cite{Kovchegov:2001sc} for example.  Equation~\ref{eq:ktfac1} has been able to
successfully describe a wealth of single-inclusive data from both \ppc and
\dAu collisions~\cite{Tribedy:2010ab,ALbacete:2010ad,Tribedy:2011aa,Albacete:2012xq}.

We now consider the production of a pair of gluons with transverse momenta $\pp$ and $\qp$ and longitudinal rapidity $y_p$ and $y_q$ within the same framework.  The double-inclusive gluon distribution is 
\begin{equation}
\frac{d^2{\rm N}}{dy_p d^2\pp dy_q d^2\qp}=\frac{1}{4(2\pi)^6}\sum_{a,b,\lambda,\sigma}\left<\vert M_{\lambda\sigma}^{ab}\left({\bf p},{\bf q}\right)\vert^2\right>\,.
\label{eq:LSZ2}
\end{equation}
The {\em classical} contribution to the two-gluon matrix element is
$M^{ab}_{\lambda\sigma}({\bf p},{\bf q})=\epsilon^{\lambda}_{\mu}({\bf p})\epsilon^{\sigma}_{\nu}({\bf q})~p^2 A^{\mu, a}({\bf p})\;q^2 A^{\nu, b}({\bf q})$ and these gauge fields can be expanded in terms of the 
color charge densities when $\pp,\qp\gg \Qs$ as done in~\eq{eq:gf1}.  Putting all
the pieces together we obtain the following schematic form of the inclusive
two-gluon distribution 
\begin{align}
\label{eq:2gluon}
\frac{d^2{\rm N}}{dy_p d^2\pp dy_q d^2\qp}&\sim\int_{\substack{\kt{1},\kt{2},\\ \kt{3},\kt{4}}}\mathcal{T}(\pp,\qp,\kt{i})
\left< \rhop(\kt{1})\rhop^{*}(\kt{3})\rhop(\kt{2})\rhop^{*}(\kt{4})\right>\nonumber\\
&\times\left<\rhot(\pp-\kt{1})\rhot^{*}(\pp-\kt{3})\rhot(\qp-\kt{2})\rhot^{*}(\qp-\kt{4})\right>\nonumber
\end{align}
where color indices have been suppressed and a shorthand notation,
$\mathcal{T}$, has been used for the product of multiple Lipatov vertices, the
detailed form of which will not be needed for the qualitative discussion to follow. 

The basis of the {\em glasma graph} framework is the factorization of the
four-point functions into products of two point functions using the Gaussian
form of the color charge distribution such as the local MV model in \eq{eq:mvmodel}
or the non-local MV model in \eq{eq:rhorhokt} (used for the remainder of this
section) that accommodates small-$x$ evolution.  The resulting factorization
results in a total of nine diagrams, each carrying a unique dependence on the
azimuthal separation of the two produced gluons which we will discuss in turn.
 
An important piece of the above formalism, which remains to be discussed,
is the rapidity dependence of the above correlation.  At leading order in $\as$
({\em i.e.} working within the {\em local} MV model) the particle production is
boost-invariant and therefore independent of the rapidity separation of the
pairs and their respective separation from the leading hadrons.  Quantum
corrections modify this picture and lead to the breaking of boost
invariance.  The difficulty in the proper treatment of the above problem stems from
the fact that not only must quantum evolution be included between the measured
gluons and the leading nuclei but also {\em between} the two measured gluons.
A dense-dense factorization formalism resumming leading logarithms in $1/x$ corrections
for inclusive observables developed in
\cite{Gelis:2008rw,Gelis:2008ad,Gelis:2008sz} was used to compute the rapidity
dependence of gluon-pair production in\cite{Dusling:2009ni}.  However, this
factorization may break down when $\as\Delta y \gg 1$\cite{Iancu:2013uva}. 
Regardless, the framework used here serves as a useful starting point for addressing
the role of long-range rapidity correlations in the dense-dense limit.  

As noted above the glasma graphs are obtained by forming all possible contractions over the color sources.  One of the nine resulting graphs corresponds to the
single-inclusive spectra squared.  While it is leading in $\Nc$ it has no
dependence in azimuthal separation {\em within the approximations used here}.
However, event-by-event fluctuations or an impact parameter profile of the
nuclei would indeed generate azimuthal correlations within this framework (as
seen for example in \cite{Teaney:2002kn}).  We now focus on the remaining eight
diagrams and the azimuthal correlations they generate without
introducing a global geometry.

\begin{figure*}[t]
\centering
\includegraphics[height=6.cm]{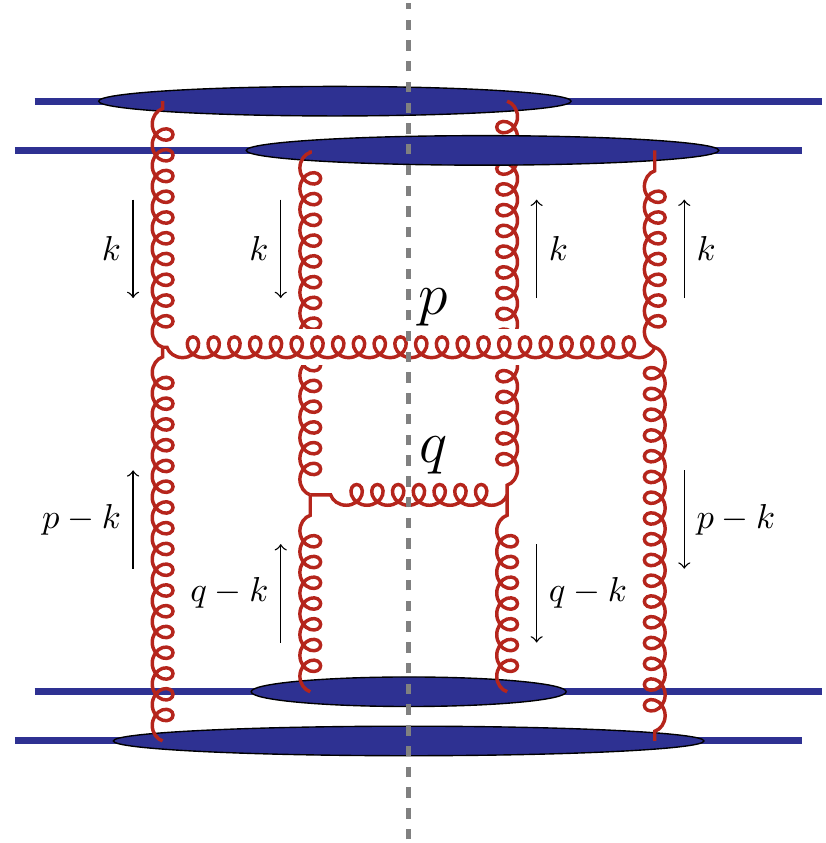}
\caption{Momentum flow demonstrating the generation of the near-side collimation.  The intrinsic parton momentum of either hadron is on the order $\Qs$ and therefore $\vert\pp-\kp\vert\sim \vert\qp-\kp\vert\sim \Qs$ where $\kp$ is a loop momentum constrained such that $\vert\kp\vert\sim\Qs$. }
\label{fig:glasmaflow}
\end{figure*}

Let us give as an example the expression from one such diagram responsible for the near-side ridge
\begin{equation}
\frac{d^2N_{\rm Fig.~\ref{fig:glasmaflow}}}{d^2\pp d^2\qp dy_p dy_q}= \frac{\as^2\Nc^2}{4\pi^{10}\zeta\;(\Nc^2-1)^3}\frac{S_\perp}{\pp^2\qp^2} \int_{\kp} \Phip^2(\kp)\Phit(\pp-\kp)\Phit(\qp-\kp)\nonumber\;,
\label{eq:glasma2}
\end{equation}
where the unintegrated gluon distributions are evaluated at (small) $x$ values
on the order $x\sim p_T/\sqrt{s}e^{\pm y_p}$ -- the precise prescription is
given in~\cite{Dusling:2012wy,Dusling:2013qoz}.  
The feynman diagram is shown in~\fig{fig:glasmaflow}.
There is a single loop momentum in this diagram $\kp$ that is integrated over.  The
transverse momentum flowing out of the nuclear projectile and target
(represented here by the blobs) is of the order of their respective saturation
momentum.  In this case we therefore must have $\vert\pp-\kp\vert\sim \Qs$ and
$\vert\qp-\kp\vert\sim \Qs$.  Furthermore the magnitude of the loop momentum
must also be near the saturation scale $\vert\kp\vert\sim \Qs$.  This leads to
the conditions that $q_\perp^2-2q_\perp \Qs\cos\phi_{kq}=0$ and
$p_\perp^2+p_T\Qs\cos\phi_{kp}=0$.  This pair of equations constrains the
produced gluons such that 1) $\phi_p=\phi_q$, collimation on the near side. 2)
$p_\perp\sim q_\perp$,  maximum correlation for similar transverse momentum and
3) $p_\perp\sim \Qs$ maximum correlation near a semi-hard scale.  All three of
these features are born out by the full numerical calculation and consistent
with the trends seen in the data.  

The full glasma graph calculation includes contributions from all eight graphs
-- the detailed expression can be found in~\cite{Dusling:2013qoz}.  All eight
diagrams provide a near- and/or away-side collimation and are necessary for a
quantitative description of the data.  Two diagrams of note are those having
delta function correlations, $d^2N/d\Delta\phi \sim \delta^{(2)}(\pp\pm\qp)$,
which was interpreted as HBT-like correlations in \cite{Kovchegov:2012aa}.  In
the dilute-dense limit (by dilute we mean $\phip(\kp)\sim\delta^{(2)}(\kp)$)
these two graphs are the only glasma graphs that remain and are therefore
crucial for our understanding of highly asymmetric collisions.  
The delta functions will be broadened by hadronization. For practical
considerations the delta function is replaced by a Gaussian, the precise form
of which can be found in~\cite{Dusling:2015rja}, originally fit to the \ppc
dihadron correlations before the availability of any data on \pPb.  The good
agreement with the subsequent \pPb data shows the robustness of this modeling
of non-perturbative hadronization dynamics.  We should stress that the
associated yield (the integral over $\phi_{pq}$ on the near side) is
insensitive to the functional form of the smearing function.

In order to convert the two-gluon correlations presented above into the
hadronic observables, an appropriate hadronization procedure must be used.  In
what follows we show results using the NLO KKP
parameterization\cite{Kniehl:2000fe} of fragmentation functions for gluon to
charged hadrons.  It has recently been found that the NLO KKP results are
troublesome at LHC energies\cite{dEnterria:2013aa} and suggested that the
gluon-to-hadron fragmentation functions were a probable source of this problem.
Extraction of new fragmentation functions by fits to the more recently available
single inclusive hadronic data would be highly valuable.

Figure~\ref{fig:pp110} shows a comparison of the di-hadron correlation compared
to the results for high multiplicity \ppc collisions.  The full numerical calculations of
all eight glasma graphs along with the away-side mini-jet contribution with 
BFKL evolution corroborate the qualitative picture shown in \fig{fig:ppsum}.
The centrality dependence is controlled by an appropriate choice of initial
saturation scale $Q_{s0}^2$ that fixes the initial condition in the rcBK
evolution equation.  Fits to deep-inelastic scattering constrain
$Q_{s0}^2=0.168$~GeV$^2$ and we take this value as representative of min. bias
\ppc collisions.  For convenience, we work with integer multiples of this
saturation scale.  For example, central \ppc ($\Ntrk\geq 110$) corresponds to
5-6 times this min. bias value.  

The overall strength of the glasma graph contribution is controlled by $\as$ which is
evaluated at one-loop running at the relevant momentum scale of the process.
In addition, a correction from non-perturbative dynamics (for example multiple
scattering) is taken into account through the multiplicative pre-factor
$1/\zeta$ that enhances the glasma graph contribution relative to
the jet contribution.  

\begin{figure*}[t]
\centering
\includegraphics[height=4.cm]{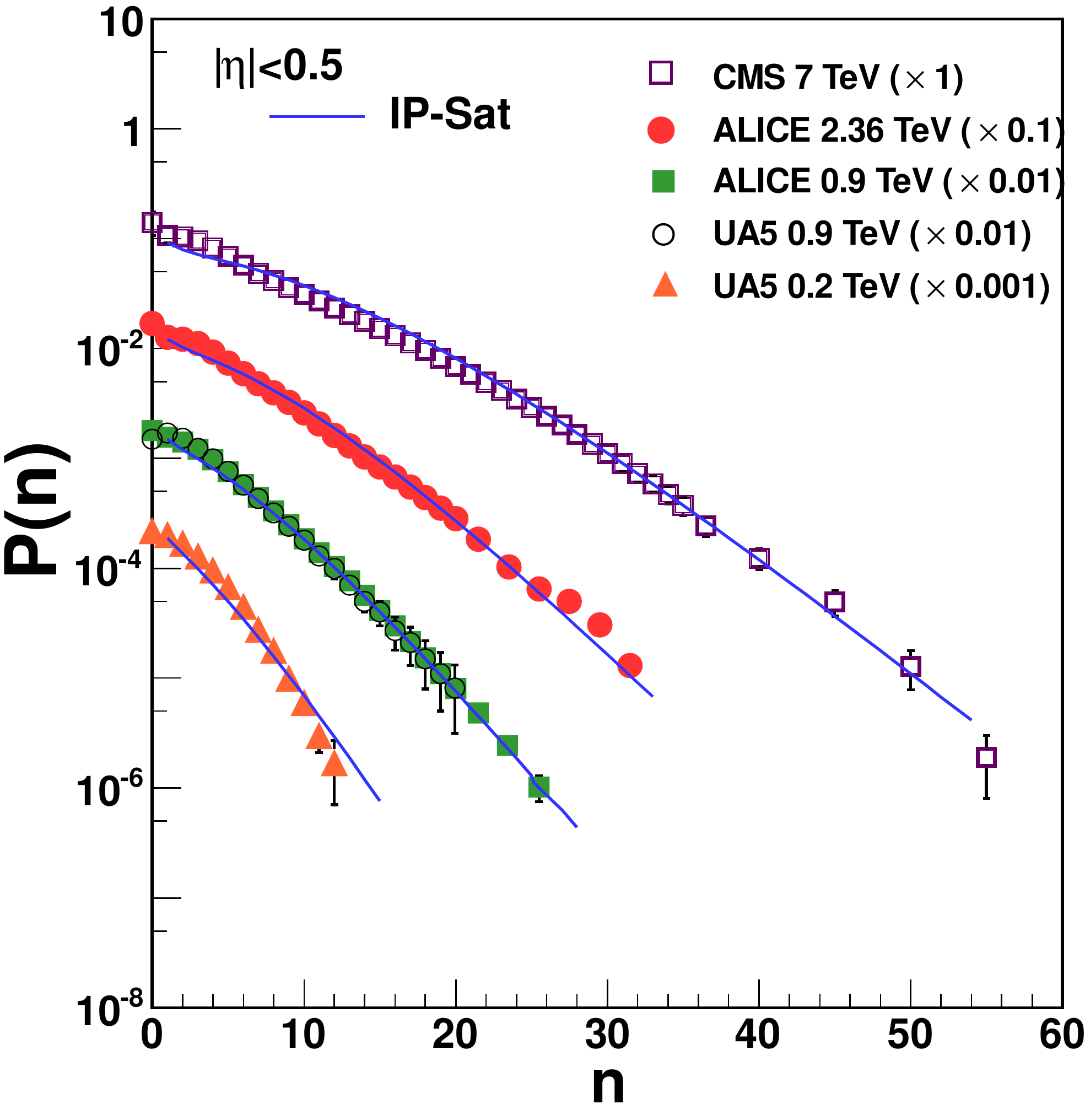}
\includegraphics[height=4.cm]{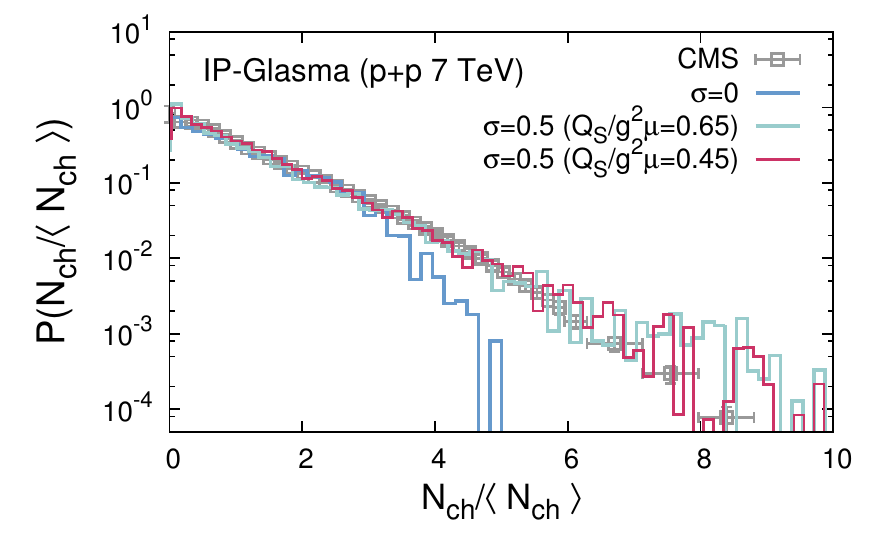}
\caption{Multiplicity distribution within $k_\perp$-factorization in the IP-Sat model\cite{Tribedy:2010ab} (left) and from classical Yang-Mills with intrinsic fluctuations of the proton scaturation scale\cite{McLerran:2015qxa} (right).}
\label{fig:nbd}
\end{figure*}

The parameter $\zeta$ is particularly sensitive to the multiplicity
distribution (see \fig{fig:nbd}) and independent fits corroborate the value of
$\zeta=1/6$ used in the glasma graph ridge analysis.  Lattice
calculations\cite{Lappi:2009xa,Schenke:2012fw} find that this constant can be
small, $\zeta\sim 0.2-1$, lending support that non-perturbative corrections
due to multiple-scattering enhance the signal. 

A comprehensive comparison of the glasma graph framework with all the available
\ppc, \pA and d-A data was presented in \cite{Dusling:2013qoz} and will not be
reproduced here.  Instead in \fig{fig:datapA} we show a representative set of
\pPb data from the CMS, ALICE and ATLAS collaborations along with the
corresponding glasma graph calculations.  

In summary, the glasma graph framework is able to account for many features
of the data on a qualitative and quantitative level.  These include 1)  the
long range nature of the correlations, 2) the nearly symmetric near- and
away-side ridge (the double ridge), 3) the strength of the correlation as a
function of transverse momentum, in particular the maximum correlation for
$p_\perp\sim q_\perp \sim Q_s$, 4) the non-trivial centrality dependence of
both \ppc and \pPb collisions.

Since the time of the above analysis more precise data on the experimental side have uncovered two shortcomings in the glasma graph description.  The first is the presence of a large and positive $v_3$ in central \pPb collisions.  The glasma graphs only contribute to the even harmonics and the peripheral jet diagram gives a negative $v_3$.  Whether modifications to the jet can accommodate a positive $v_3$ has yet to be fully explored.  Further, the interference diagram between the jet and glasma graph also appears to vanish\cite{Dusling:2014oha}.  Recent calculations within classical Yang-Mills provide one source of positive $v_3$ (see section~\ref{sec:cym}) but this has yet to confront the data.

A final shortcoming of the glasma graph description (at least within the framework
used here) is the negative $c_2\{4\}$ measured by CMS.  The glasma graph and
jet graph both give a positive contribution to $c_2\{4\}$.  The empirical
observation that $\sqrt[\leftroot{-2}\uproot{2}4]{-c_2\{4\}}\sim \sqrt[\leftroot{-2}\uproot{2}6]{c_2\{6\}}\sim \sqrt[\leftroot{-2}\uproot{2}8]{-c_2\{8\}}$
suggests that the correlation is driven by correlations among the disconnected
diagrams at the single-inclusive level.  We should stress that the multiple
particle correlations are dominated by soft momentum $\pp\lesssim\Qs$ where the
glasma graph framework no longer applies.  It would be very interesting to see
if and where $c_2\{4\}$ changes sign as the minimum $p_T$ used in the analysis
is increased and when a rapidity gap between the particles is introduced in order to 
suppress jet-like correlations.

\paragraph{Multi-gluon production in the CGC}

The framework laid out for two-gluon production in the previous section has
been extended to three gluons\cite{Dusling:2009ar} and generalized to
$n$-gluons\cite{Gelis:2009wh}.  

The results can be represented diagrammatically as {\em glasma graphs} in a
similar fashion as above and the subset of these diagrams that contribute for
$\pp\gg \Qs$ has been identified and the $n$-gluon distribution is
\begin{equation}
\left<\frac{d^nN}{dy_1 d^2\kt{1}\cdots dy_n d^2\kt{n}}\right>=(n-1)!\frac{\zeta(\Nc^2-                                        1)}{2\pi}\frac{\Qs^2\Sp}{\kt{1}^4\cdots\kt{n}^4}\left(\frac{\Nc(g^2\mu)^4}{2\pi^3g^2\zeta\Qs^2}\right)^n\,,
\end{equation}
where the sensitivity to the infrared dynamics is encoded in the constant
$\zeta$ introduced earlier.  The factorial cumulants $\left< d^nN\right>$ are
those of a negative binomial
\begin{equation}
P_n^{\rm NB}(\overline{n},k)=\frac{\Gamma(k+n)}{\Gamma(k)\Gamma(n+1)}\frac{\overline{n}^nk^k}{(\overline{n}+k)^{n+k}}
\end{equation}
where $\overline{n}$ is the average multiplicity and
$k=\zeta(\Nc^2-1)\Qs^2\Sp/(2\pi)$ is proportional to the number of flux tubes.
For $k=1$ the negative binomial is a Bose-Einstein distribution and for
$k\to\infty$ the distribution is Poissonian.  The left plot in \fig{fig:nbd}
shows the resulting multiplicity distribution in \ppc collisions in the IP-Sat
model \cite{Kowalski:2003hm} with the one free parameter $\zeta=0.155$ fit to data at 0.9 TeV and then
exhibited for other energies. 
 
The right plot in \fig{fig:nbd} shows the recent results of \cite{McLerran:2015qxa} for 
the multiplicity distribution in \ppc from classical Yang-Mills simulations. In this case
there is no free parameter $\zeta$ and the infrared behavior is instead determined by
the resulting non-linear dynamics of the gluon fields.  In this framework there are multiple sources of fluctuation.  First there is a geometric dependence from the saturation scale in the IP-Sat model which sets the initial field configurations.  Second, sub-nucleonic fluctuations from the sampling of different configurations of color charges according to the MV model give rise to the negative binomial distribution of gluon number fluctuations for fixed geometry.

However, convolution of these two sources alone are not sufficient to account for the broad
width of the multiplicity distribution.  In addition there must be intrinsic
fluctuations of $\Qs$ in the proton.  Modeling the distribution of saturation
scales of the proton as a Gaussian distribution provides a good description of
the data.  The key point for this discussion is that high multiplicity events
easily accommodate saturation scales 5--6 times the min. bias values as used to
explain the ridge data.

\subsubsection{Classical Yang-Mills dynamics}

The calculations within the {\em glasma graph} framework presented in the
previous section are strictly valid for $\pt\gtrsim\Qs$ and are infrared
divergent if extended to lower transverse momentum.  The divergences are
regulated by the non-perturbative dynamics of the gluon fields at scales of
order $\Qs$.  These dynamics can be captured by solving the classical
Yang-Mills equations numerically in the forward light-cone and are valid at any
transverse momentum -- only limited in the ultraviolet by the inverse lattice
spacing and in the infrared by the inverse nuclear size.  The perturbative,
high $\pt$, limit of the classical Yang-Mills simulations can be expressed
in terms of similar {\em glasma graphs} if a Gaussian distribution in the
sources is used.  Corrections due to small-$x$ evolution in numerical
simulations of quasi-classical gluon production have been done
\cite{Lappi:2011ju} but in the following section on classical Yang-Mills
dynamics a local Gaussian distribution of sources is used (which does not
accommodate small-$x$ evolution). 
 
Empirically, the near-side ridge persists to momenta well below the saturation
scale and it is therefore imperative to investigate whether the Glasma
flux-tube picture holds in this kinematic region.  It is also useful to
establish the size of the corrections due to multiple scattering at larger
transverse momentum.  In the remainder of this section we will review the
numerical results\cite{Schenke:2015aqa} of the classical Yang-Mills
simulations.  We will discuss how the classical Yang-Mills results complement
the {\em glasma graph} description and also provide possible resolutions to
some of the discrepancies with data alluded to in the previous section on
glasma graph phenomenology.  

In the CGC picture, the dynamics of a high energy collision is described by the solution of the classical Yang-Mills equations 
\begin{equation}
\left[D_\mu, F^{\mu\nu}\right]=J^\nu\;,
\end{equation}
in the forward light-cone of the collision.  In the above expression $J^\nu$ is an eikonal color current generated by a color neutral distribution of classical charges moving along the light-cone,
\begin{equation}
J^\nu=\delta^{\nu+}\rhop(x^-,\xp)+\delta^{\nu-}\rhot(x^+,\xp)\;.
\end{equation}
The sources are distributed according to a Gaussican distribution having variance
\begin{equation}
g^2\left<\rho^a(\xp)\rho^b(\yp)\right>=S_{\textrm{p/Pb}}(\bp)\delta^{ab}\delta^{(2)}(\xp-\yp)\;.
\end{equation}
The function $S_{\textrm{p/Pb}}(\bp)$ encodes information on the spatial
structure of color charge distributions of the projectile and target.  The
results shown here use a {\em constituent quark proton model} whereby the color
charge density is concentrated around the transverse positions of three
constituent quarks whose positions fluctuate event to event.  The use of a {\em
spherical proton} has a negligible effect on the results. \cite{Schenke:2015aqa}
The result is even less sensitive to the impact parameter dependence of the
color charge in the nucleus which is sampled from a Wood-Saxon distribution.

\begin{figure*}[t]
\centering
\includegraphics[height=4.cm]{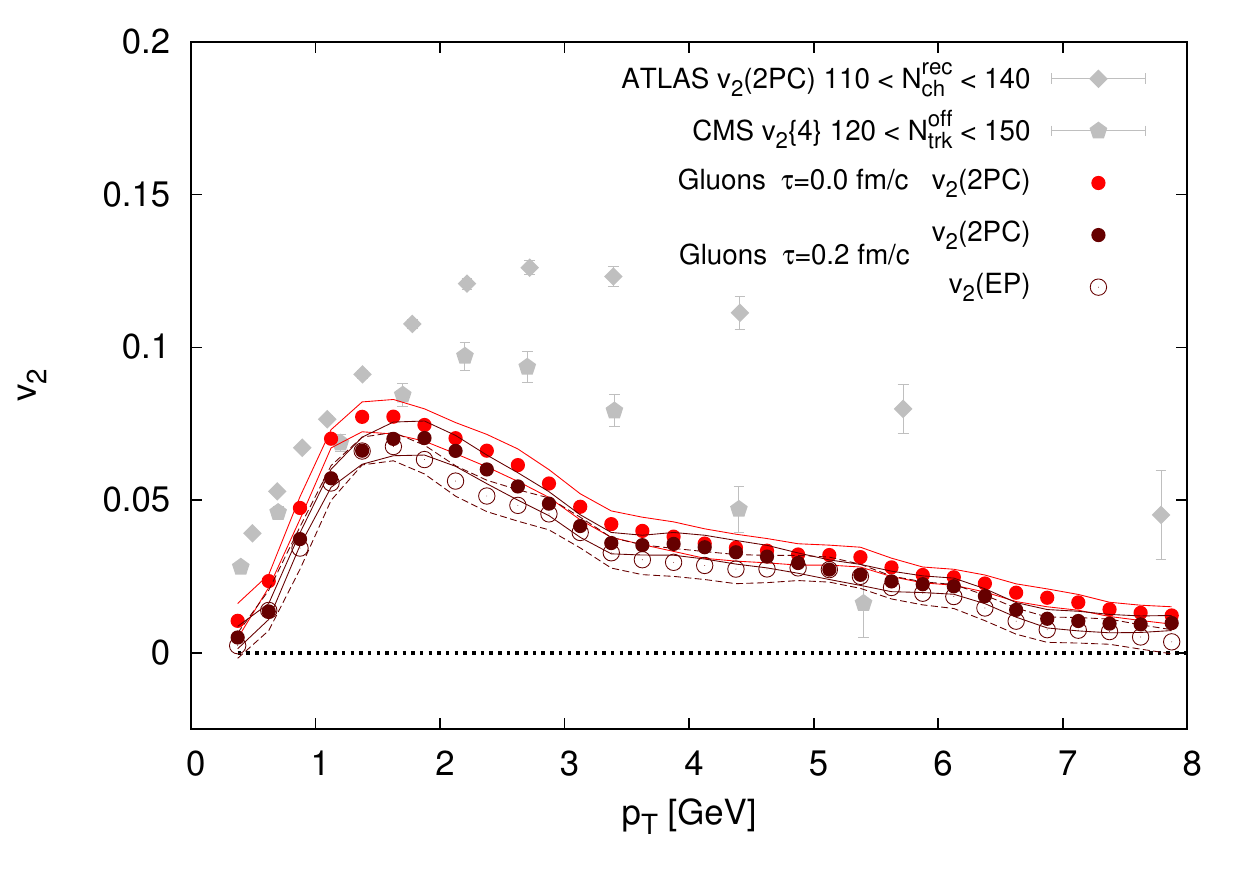}
\includegraphics[height=4.cm]{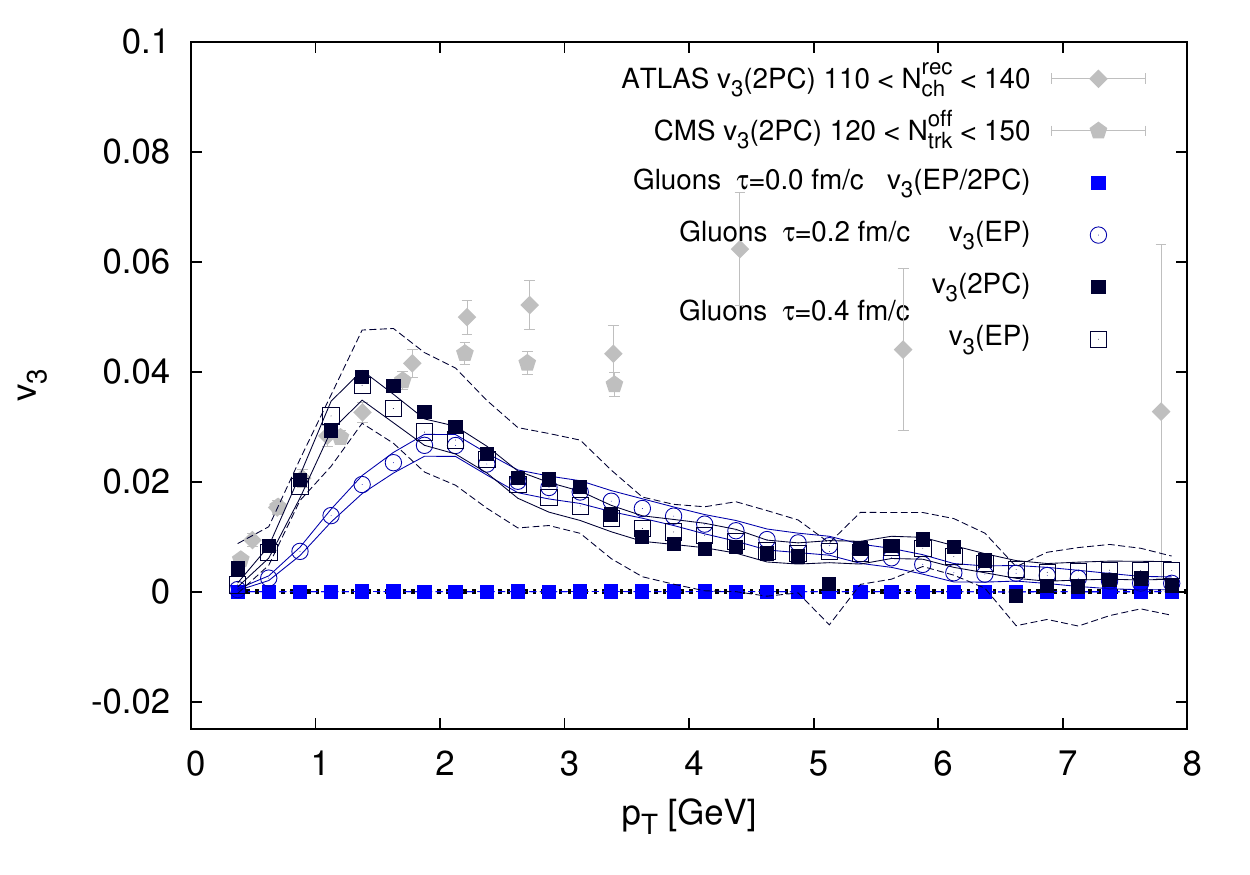}
\caption{
(Color online) Gluon $v_2(\pt)$ (left) and $v_3(\pt)$ (right) at times
$\tau=0,0.2,0.4$~fm/c in \pPb collisions at impact parameter $\bp=0$ in the
constituent quark proton model.  Open and closed symbols correspond to results
obtained using the event plane method and two particle correlations,
respectively.  Error bands include statistical errors only. Experimental
results by the ATLAS \cite{Aad:2014lta} and CMS collaboration
\cite{Chatrchyan:2013nka} for inclusive hadrons are also shown as a guideline
for comparison.} 
\label{fig:v23ecc}
\end{figure*}

Starting from field configurations derived from the color charges in a given
event the classical equations of motion are solved and the gluon distribution
extracted at a given proper-time by measuring equal-time correlation functions
of the gauge fields.  Using either two-particle correlations or the event plane
method $v_2(\pt)$ and $v_3(\pt)$ can be extracted from the gluon spectrum as show
in the left and right plot of \fig{fig:v23ecc}.  We now discuss the implications
of these results.

It is striking that a sizable $v_2(\pt)$ is present at the initial time
$\tau=0^+$ as it therefore cannot be attributed to a collective expansion of
the system.  This is in contrast to a hydrodynamic interpretation where the
elliptic flow is initially zero and gradually develops over the lifetime of the
collision.  Further evolution of the Yang-Mills simulation finds only a modest
decrease in the $v_2$ at later times indicating that the glasma flux-tube
picture is robust against gluon re-scattering (a similar conclusion was reached
by earlier lattice simulations\cite{Lappi:2009xa} as well). 
 
The observation that $v_2$ as extracted from the event-plane method and two
particle correlations are similar within errors suggests that the dominant
origin of the $v_2$ is the breaking of rotational symmetry of the single
particle spectrum on a per-event basis. To elaborate on this point, the
initial-state ({\em i.e.} $\tau=0^+$) classical Yang-Mills results can be
divided into two contributions.  One consisting of glasma graphs corresponding
to the connected diagrams shown in the previous section.  The second
contribution stems from seemingly disconnected graphs (disconnected on a
perturbative level) but connected when the two gluons scatter on a common
color field domain.  This is similar in spirit to the effect seen in \cite{Dumitru:2014dra}
(where the proton was treated as a dilute object) and will be discussed in
further detail in the upcoming section. 
 
The classical simulations fail to reproduce the magnitude of the $v_2(\pt)$ at
larger values of transverse momentum.  The situation should worsen when
hadronization is taken into account. Hadrons at a given $\pt$ fragment from
gluons of higher $\pt$ and therefore the $v_2$ shown in
\fig{fig:v23ecc} will be rescaled to lower $\pt$ when hadronization is
accounted for.  This demonstrates the importance of including small-$x$
evolution in the nuclear wavefunction absent in the classical Yang-Mills
simulations.  The classical Yang-Mills and glasma graph picture complement
each-other quite nicely; the latter responsible for the large $\pt \gtrsim \Qs$
ridge due to intrinsic ({\em i.e} non-factorizable) two-parton correlations
including small-$x$ evolution and the prior responsible for the softer ridge,
$\pt \lesssim \Qs$ generated from the event-by-event breaking of rotational
invariance.

We now turn to a discussion of the gluon $v_3(\pt)$ as shown in the right plot
of \fig{fig:v23ecc}.  The first observation is that the initial $v_3$ at
$\tau=0^+$ is zero.  This is a consequence of the initial gluon spectra being
symmetric under $\kp \to -\kp$ resulting in the odd Fouier harmonics being
identically zero at the initial time.  This is consistent with the vanishing of
odd harmonics in the glasma graph calculation.  At a proper-time of
$\tau=0.2$~fm/c a sizable $v_3$ has built up.  Further evolution to
$\tau=0.4$~fm/c, where the system is practically free-streaming, results in a
modest increase in the signal.  The agreement between the event plane method
and two particle correlations suggests that the event-by-event breaking of
translational invariance combined with coherent final state effects is
ultimately responsible for the $v_3$.  The most surprising finding is the lack
of correlation between the global initial state eccentricity, $\epsilon_3$, and
the $\pt$ integrated $v_3$ on a per-event basis; refuting the notion that
the source of $v_3$ is from the build up of a global energy-momentum flow via classical
Yang-Mills dynamics. 

Finally, we stress that the Yang Mills calculation \cite{Schenke:2015aqa} is not able to reproduce the
azimuthal anisotropy coefficients in \AAc collisions, because of the large amount of uncorrelated 
color field domains present in a single collision and the lack of hydrodynamic evolution.
This further confirms the standard interpretation of the $v_n$ in heavy ion collisions as 
strong final state effects.

To summarize, the azimuthal correlations generated via the classical Yang-Mills
simulations can be separated into three sources: 1) A genuine non-factorizable
two-particle correlation -- the glasma graph contribution  2) perturbatively
disconnected graphs connected by the event-by-event breaking of rotational
symmetry and 3) A contribution from final state interactions generated by the
Yang-Mills dynamics.  In the following section we review work that has focused
on the second contribution; the event-by-event breaking of rotational symmetry.

\subsection{Scattering from color-field domains}
\label{sec:cym}

The basic idea can be traced back to Kovner and Lublinsky
\cite{Kovner:2010xk,Kovner:2011pe} who argued that domains of color-electric
fields of size $1/\Qs$ will produce non-trivial angular correlations in the
single-particle distribution.  Following their arguments, this can be easily
seen by considering the scattering of two projectiles from a stationary (in the
lab frame) hadronic target.  The two projectiles may have different rapidities
but at high enough energies their wave functions are boost invariant.
Therefore the long-range nature of the correlations are a general consequence
of the high-energy limit.  If the two projectiles strike the target at the same
impact parameter both will see the same color configuration in the target and
be scattered in similar directions.  This provides the source of near-side
correlations.  If the projectiles under consideration are gluons, it is just as
likely that they can have {\em opposite charge} (as gluons transform in the
adjoint representation of SU($3$) ) and therefore scatter in opposite
directions.  For quarks in the fundamental representation this would not be the
case and the near-/away-side symmetry would be broken.  As the size of
correlated domains in the target are of order $1/\Qs^2$(target) we
parametrically expect the strength of the two particle correlation to go as
$\left(\Qs^2\Sp\right)^{-1}$ where $\Sp$ is the overlap area.

\begin{figure*}[t]
\centering
\includegraphics[height=4.5cm]{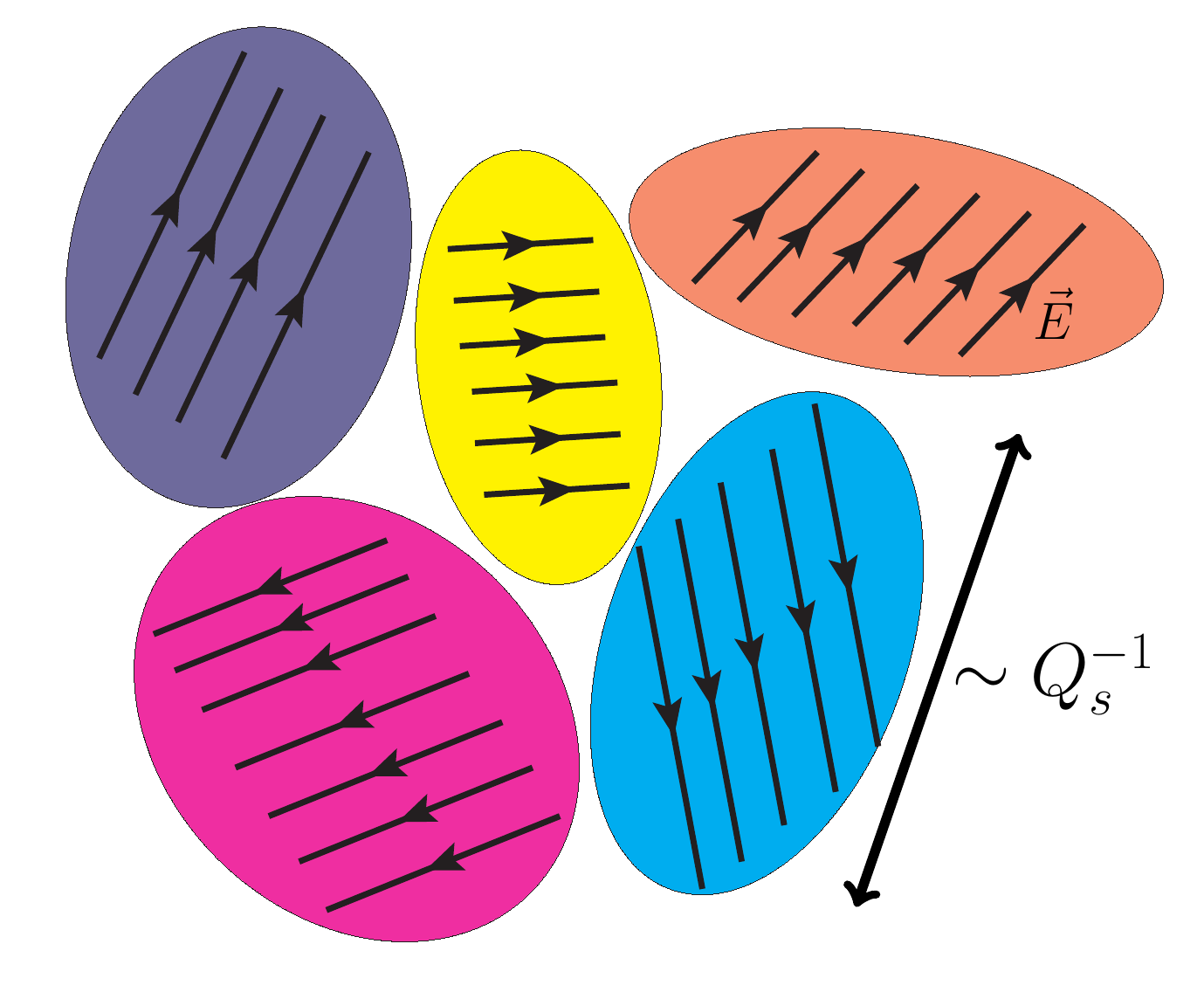}
\raisebox{.0cm}{\includegraphics[height=4.5cm]{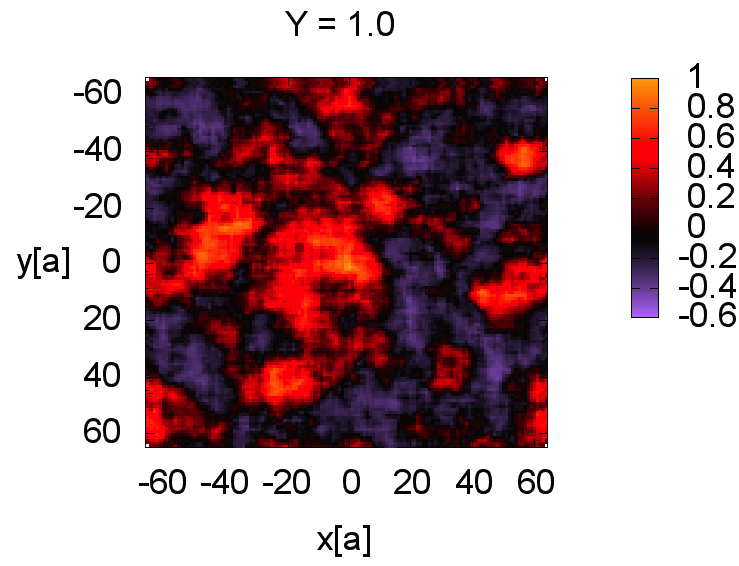}}
\caption{Left: Schematic illustration of localized domains of chromo-electric fields in the target.  Right: The dipole operator $1/\Nc\textrm{Tr}\;V(\xp)V^\dagger(\yp)$ after evolution by the JIMWLK equation demonstrating the persistence of local fluctuations of the Wilson line \cite{Dumitru:2011vk}.}
\end{figure*}

So far this is conceptually identical to what has been done in the previous
section on glasma graphs except considered in a different Lorentz frame.
Indeed the work of \cite{Kovner:2010xk} obtains the same glasma graphs
when using a Gaussian weight functional for the color charge densities.     

It was argued however that the Gaussian averaging procedure may underestimate
the strength of the correlation.  In the Gaussian approximation the two-gluon
production is written as a term that factorizes into a product of single
inclusive distributions and eight remaining connected diagrams (the glasma
graphs considered earlier), the latter of which are $\Nc^2$ suppressed, 
\begin{equation}
\left<\frac{d^2N}{dy_p d^2\pp dy_q d^2\qp}\right>_{\rm Gaussian}\sim \left<\frac{dN}{dy_p d^2\pp}\right>\left<\frac{dN}{dy_q d^2\qp}\right> + \frac{1}{\Nc^2}\left( \rm{8~glasma~graphs}\right)\,.
\end{equation}
However, in the Gaussian approximation this first term has no azimuthal
correlations -- it corresponds to the independent emission of two gluons.  An
analysis going beyond leading order in $\Nc$ has found that the four-point
function does not necessarily factorize into a product of two-point
functions\cite{Dumitru:2010mv} and these factorization breaking terms
contribute at leading order in $\Nc$ to correlated two-gluon production. 

A model for the generation of initial state azimuthal asymmetry from scattering from color domains was introduced in \cite{Dumitru:2014dra}.  It was shown that scattering of a dipole of size $\sim1/p_T$ from a target with fixed configuration of color electric field reproduces the $p_T$ dependence of $v_2$ and $v_4$ for a sufficiently polarized target.

Conventionally in the MV model the averaging over the field configurations of the target using a Guassian correlator,
\begin{equation}
\frac{g^2}{\Nc}\left<E^a_i({\bf b}_1)E_j^b({\bf b}_2)\right>=\frac{1}{(\Nc^2-1)}\delta^{ab}\delta_{ij}\Qs^2\Delta({\bf b}_1-{\bf b}_2)
\label{eq:EEMV}
\end{equation}
will result in isotropic particle production.
In order to take into account observables that are sensitive to the breaking of translation invariance on a per-event level a model is introduced such that the target configurations are constrained such that the electric fields point in a specific direction $\hat{a}$,
\begin{equation}
\frac{g^2}{\Nc}\left<E^a_i({\bf b}_1)E_j^b({\bf b}_2)\right>=\frac{1}{(\Nc^2-1)}\delta^{ab}\Qs^2\Delta({\bf b}_1-{\bf b}_2)\left(\delta_{ij}+2\mathcal{A}\left[\hat{a}_i\hat{a}_j-\frac{1}{2}\delta_{ij}\right]\right)
\label{eq:EEMVanis}
\end{equation}
The above expression takes into account that in the vicinity of ${\bf b}_1\sim{\bf b}_2$ there is a subclass of events that have a predefined orientation and the average over these subclasses are done {\em after} the observable has been computed. 

In this framework there exists a correlation even at the single particle level,
\begin{equation}
\left<\cos\;n(\phi_p-\phi_{\hat{a}}\;)\right>(\pp)=\int d\phi_p\left<\cos\;n(\phi_p-\phi_{\hat{a}}\;)\left<\frac{dN}{dy_p d^2\pp}\right>\right>_{\hat{a}}\;.
\end{equation}
and will correspondingly produce azimuthal correction in the disconnected terms in higher order cumulants. 

\begin{figure*}[t]
\centering
\includegraphics[width=6.cm]{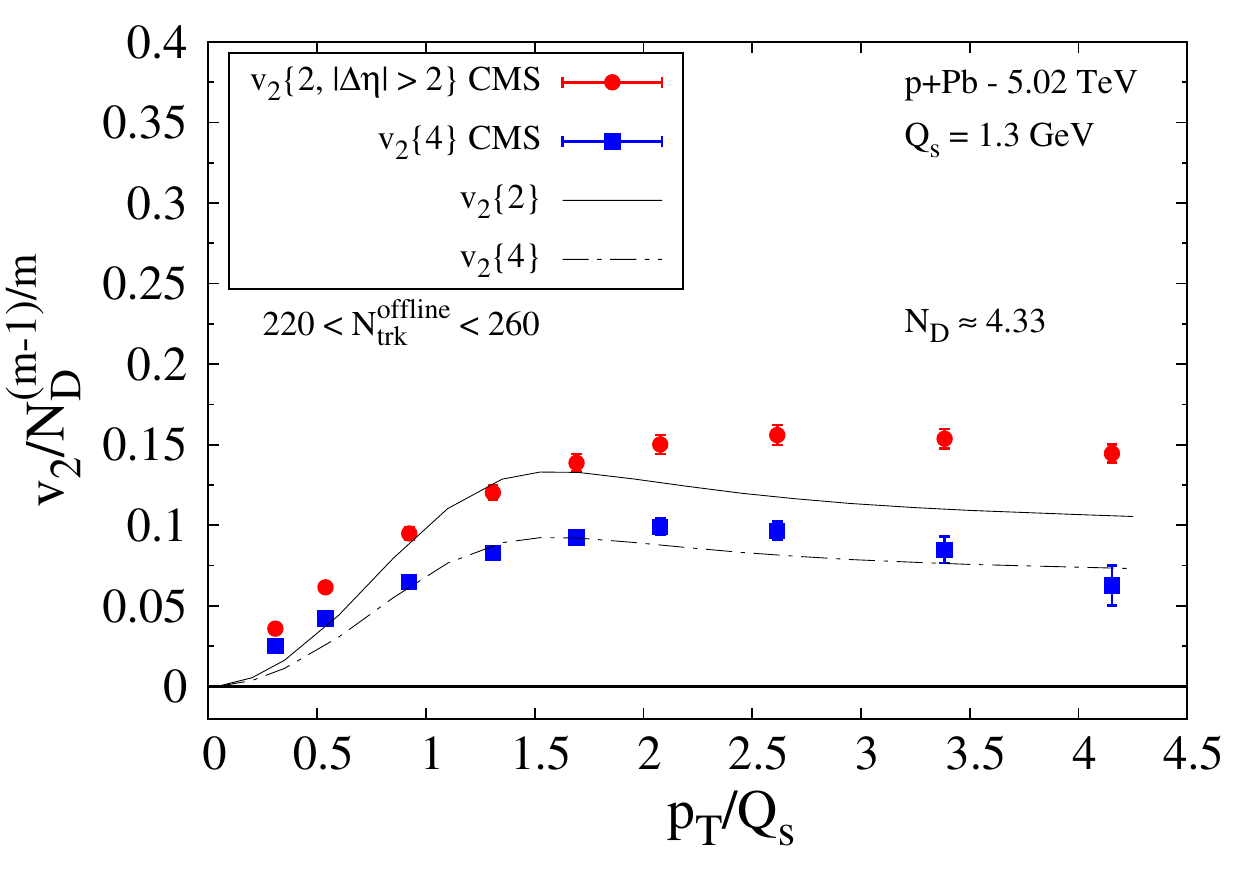}
\includegraphics[width=6.cm]{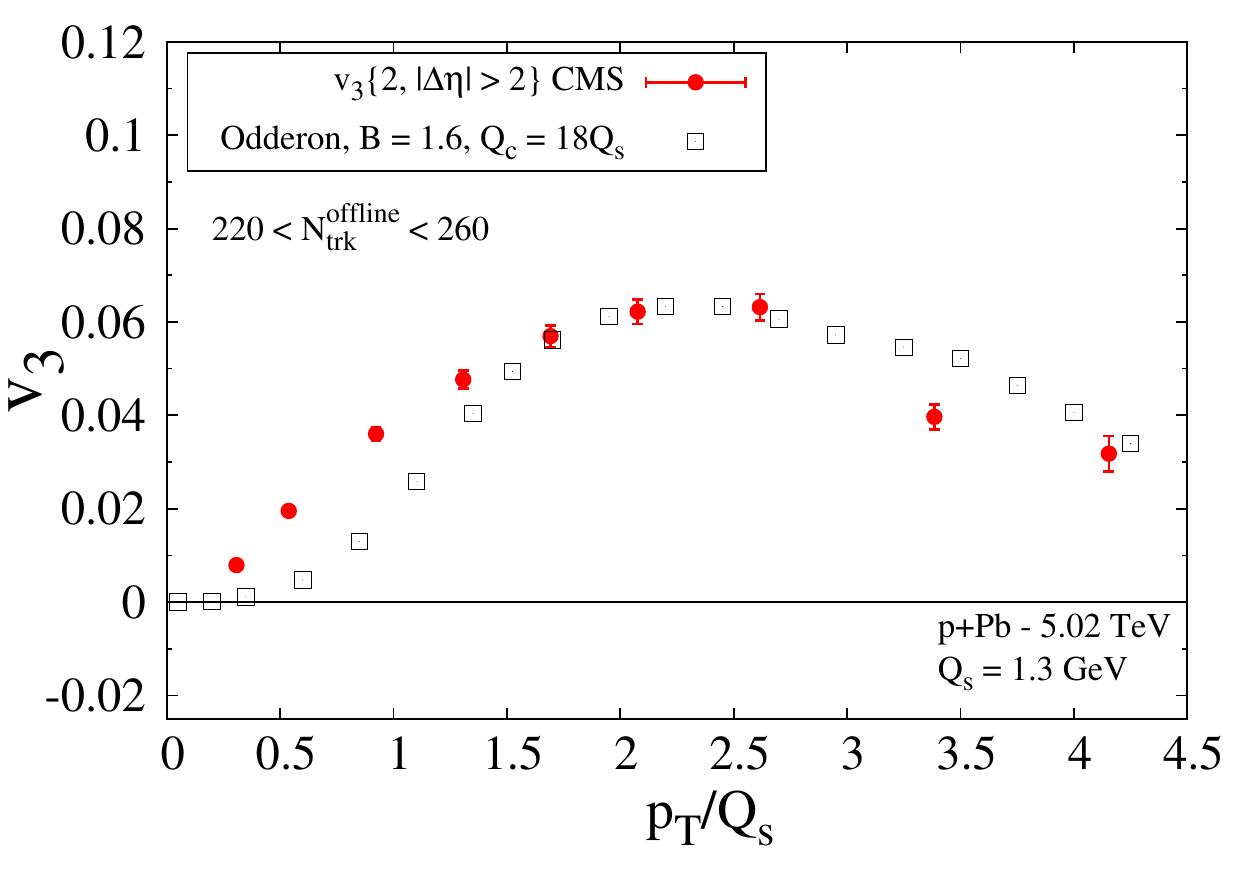}
\caption{Comparison of the color domain model of~\cite{Dumitru:2014dra} with $v_2(\pp)$ (left) and $v_3(\pp)$ (right) with high multiplicity p+Pb collisions.}
\end{figure*}

One of the strengths of the above model is that it is able to accommodate a
negative $c_2\{4\}$ \cite{Dumitru:2014yza} 
\begin{equation}
c_2\{4\}=-\frac{1}{N_D^3}\left(\mathcal{A}^4-\frac{1}{4(\Nc^2-1)^3}\right)
\end{equation}
Furthermore, the consideration of C-odd fluctations in the target, related to the odderon, are able to generate a $v_3$.

While the above color domain model appears to use only Gaussian fluctuations, ({\em i.e.} the only correlator used \eq{eq:EEMVanis} is a two-point function) the separation of averages is outside of the assumptions of the CGC framework.  Implicit in this separation of averages is that the angular fluctuations of the large-$x$ color sources, $rho$, evolve on a slower times scale.  First the fast modes are averaged over with the modified two point function \eq{eq:EEMVanis}, second the observable is computed, and only then is the averaging over all possible electric field directions, $\hat{a}$ performed as required by Gauge invariance.

A recent work \cite{Lappi:2015vha} has made the connection between the color domain model and non-gaussianites in the color glass condensate.  We direct the reader there for further discussion of the interpretation of the polarization factor $\mathcal{A}$ in the color domain model and its relation to the possible sources of non-Gaussianity in the CGC effective theory.

\subsection{Conclusions on initial state interactions in small systems}

In conclusion, the previous sections have demonstrated how glasma graphs and classical
Yang-Mills complement each other as two different approximations of the same
effective theory of high energy nuclear collisions able to address different
features of the data.  The connection to the scattering from localized domains
of color electric fields with finite polarization can be made by considering
non-Gaussianities of the color charge correlators.

We conclude this section by briefly discussing two other initial state
proposals.  While these have yet to address the data at the same quantitate
level as the hydrodynamic and CGC models discussed thus far, these proposals
contain interesting ideas which should be explored further.

The work of \cite{Levin:2011fb,Levin:2014kwa} demonstrates that angular
correlations can be generated by the exchange of two BFKL Pomeron ladders.  The
transverse momentum exchanged by the Pomeron is conjugate to the impact
parameter of the collision and as events with a fixed multiplicity correspond
to a finite impact parameter, an asymmetry is generated. Motivated by Gribov
Reggeon and Pomeron calculus it was shown that the correlation is long-range in
rapidity and that a double-ridge signal is generated.

A more recent work \cite{Gyulassy:2014cfa} argued that non-abelian beam jet
bremsstrahlung generates long range rapidity correlations having a hierarchy of
non-trivial azimuthal harmonics.  The anisotropies are generated from radiating
clusters that have accumulated a net transverse momentum kick on an
event-by-event basis.  In addition to explaining the ridge in small colliding
systems, it was argued that this mechanism may be able to explain the
approximate energy independence of azimuthal asymmetries observed in the RHIC
beam energy scan.

%\clearpage

%%% Local Variables: 
%%% mode: latex
%%% TeX-master: "../RidgeReview-ijmpe"
%%% End: 

\section{Future Directions}

While a lot of progress has been made towards understanding novel long-range 
correlation phenomena in small colliding systems, many questions remain open, requiring
future efforts by both experimental and theoretical communities.
We consider the following two big questions we feel should be addressed in order for progress to be made in the coming years:
1) Have we reached a consensus on the nature of long-range 
correlation phenomena in small (and/or large) systems?  2) If a strongly-coupled QGP is indeed formed, 
what fundamentally new knowledge do we gain from small systems?

To answer these questions, we discuss a few possible future directions in this section:
\begin{itemize}
\item Further scrutiny of hydrodynamic paradigm
\item Nature of the ridge in \ppc collisions
\item Jet-medium interactions in \ppc and \pPb collisions
\item Pre-equilibrium dynamics
\end{itemize}

\subsection{Further scrutiny of hydrodynamic paradigm}

The scrutiny of standard paradigms by identifying inconsistencies with
experimental observations has been the recipe for many major scientific
discoveries.  In this light, precision studies of the hydrodynamic
framework in small systems should be steadily pursued. The main theoretical
challenge for \ppc and \pA systems lies in the large uncertainties in
determining the initial-state geometry, due to the lack of knowledge of the
proton structure as well as its fluctuations, as discussed in Sec.~4.1.  To
make progress, identifying observables that are primarily sensitive to either
the initial-state geometry or final-state dynamics is the most promising
approach.

\begin{figure}[t]
\centering
\includegraphics[width=\linewidth]{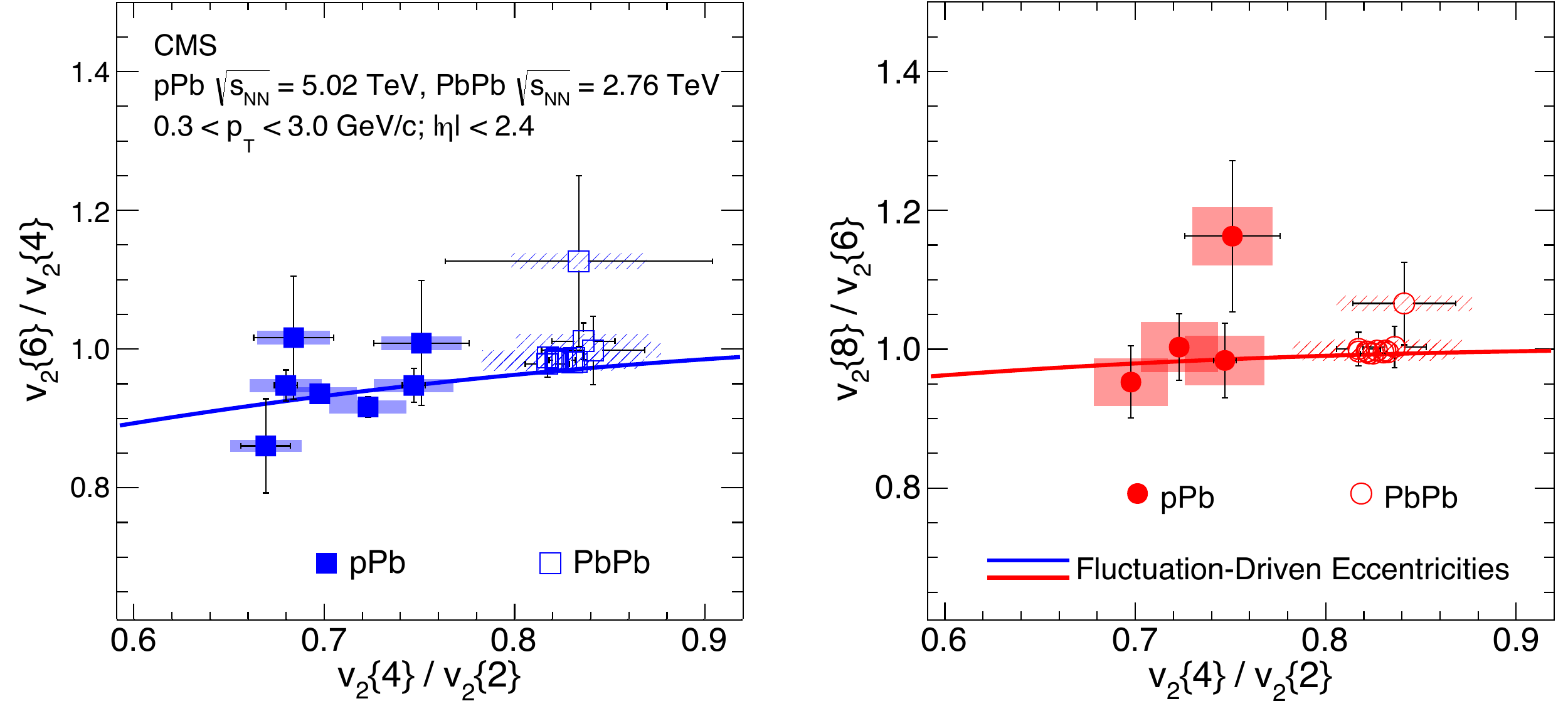}
  \caption{ \label{fig:ratios_v2n} 
  Cumulant $v_2$ ratios, $v_2\{6\}/v_2\{4\}$ (left) and $v_2\{8\}/v_2\{6\}$
(right), as a function of $v_2\{4\}/v_2\{2\}$ ratios for high-multiplicity
events in \pPb collisions at \rootsNN\ = 5.02\TeV and \PbPb collisions at
\rootsNN\ =    2.76\TeV~\cite{Khachatryan:2015waa}.  The solid curves are
predictions from a proposed universal behavior of fluctuation-driven
eccentricities.~\cite{Yan:2013laa} 
}
\end{figure}

It was argued in Ref.~\cite{Yan:2013laa}, that for purely fluctuation driven
initial-state anisotropies, such as in \ppc and \pA collisions,  the
initial-state eccentricity distribution is universal.  With the assumption of
$v_{2} \sim \epsilon_{2}$ motivated by hydrodynamics, the $v_2$ cumulants
measured from 2-, 4-, 6-, and 8-particle correlations are predicted to follow a
specific relation, as shown by the solid lines in Fig.~\ref{fig:ratios_v2n} for
ratios of $v_{2}\{6\}/v_{2}\{4\}$ and $v_{2}\{8\}/v_{2}\{6\}$ as a function of
$v_{2}\{4\}/v_{2}\{2\}$.  The recent experimental results in \pPb
collisions~\cite{Khachatryan:2015waa} (also shown in
Fig.~\ref{fig:ratios_v2n}), seem to favor the theoretical predictions but
clearly improvement in the experimental precision is needed before firm
conclusions can be drawn.  Additional information on the initial state
fluctuations can also be extracted by measuring event-by-event $v_2$
distribution in \pPb collisions, as was done in \PbPb
collisions.~\cite{Aad:2013xma}
 
Correlations among flow harmonics $v_n$ and the event planes ($\Psi_{n}$) are
another class of observable that may provide a valuable way of disentangling
the effects of initial-state geometry from final-state evolution within
hydrodynamical models.  Such correlations were studied in great detail by the
ATLAS collaboration in \PbPb collisions.~\cite{Aad:2014fla,Aad:2015lwa}  

Similar measurements should be pursued in \pPb collisions and compared to
hydrodynamic calculations.  For example, the correlation between $v_2$ and
$v_4$, or $\Psi_{2}$ and $\Psi_{4}$, is generated by the non-linear response of
hydrodynamic evolution, and one expects such a correlation to persist in small
systems as well.  The interpretation is different for $\Psi_{2}$ and $\Psi_{3}$.
Final-state hydrodynamic evolution does not induce a strong correlation between
$\Psi_{2}$ and $\Psi_{3}$ and indeed this correlation was observed to be small
in \PbPb collisions.~\cite{Aad:2014fla} However, it has been argued that a
sizable correlation between $\Psi_{2}$ and $\Psi_{3}$ may be induced by large
initial-state fluctuations in small colliding systems such as \pPb.~\cite{Yan:2015fva}  
Similar event plane correlations measurements in \pPb could
therefore provide valuable information on the initial-state geometry. 
 
Another observable that is mainly sensitive to the details of initial-state fluctuations
is the transverse momentum dependent event plane fluctuations (of the same order).
The measurement was carried out by the CMS collaboration in \pPb and \PbPb
collisions.~\cite{Khachatryan:2015oea} Hydrodynamic calculations have indicated that
this observable is sensitive to the granularity of the initial energy density fluctuations, 
instead of the $\eta/s$ of the medium.~\cite{Kozlov:2014fqa} Evidence of
significant rapidity-dependent event plane fluctuations in \pPb and \PbPb
collisions is also observed~\cite{Kozlov:2014fqa}, although it may be sensitive
to both initial- and final-state effects in the longitudinal
direction~\cite{Bozek:2015bna}.  Detailed studies of anisotropic flow
observables and their correlations among each other in small systems, could
provide crucial information on the initial-state in small systems if hydrodynamics
is the correct framework to model these systems.

\subsection{Collectivity in \ppc collisions}

While there has been a flood of new results on the long-range correlations in
\pPb collisions as summarized in Sec.~\ref{sec:exppA}, little progress has been
made in addressing the nature of the ridge in \ppc collisions since its first
observation in 2010.  Only recently have new measurements of the ridge in \ppc
become available~\cite{1384640} with theoretical analysis closely
following.~\cite{Dusling:2015rja}  A more recent work has attempted to extract the $v_2$
harmonics of long-range correlations by a careful fitting procedure between
peripherial and central event classes in 13\,TeV and
2.76\,TeV.~\cite{Aad:2015gqa}

As discussed in Sec.~\ref{sec:hydropA}, hydrodynamic descriptions of small
colliding systems are highly limited by the lack of knowledge of the initial conditions. 
Especially for \pA collisions, the initial-state eccentricity is highly sensitive to the
shape of the proton and its fluctuations on very short timescales.
Precision measurements of $v_n$ harmonics from long-range correlations in
\ppc collisions should provide new constraints on the proton shape, and thus
have the promise of improving the hydrodynamic modeling of \ppc and \pA systems.
Even more precise and independent measurements of the proton shape and 
its fluctuations would be possible with an electron ion Collider. \cite{Accardi:2012qut}

 \begin{figure}[t]
   \begin{center}
     \includegraphics[width=0.375\linewidth]{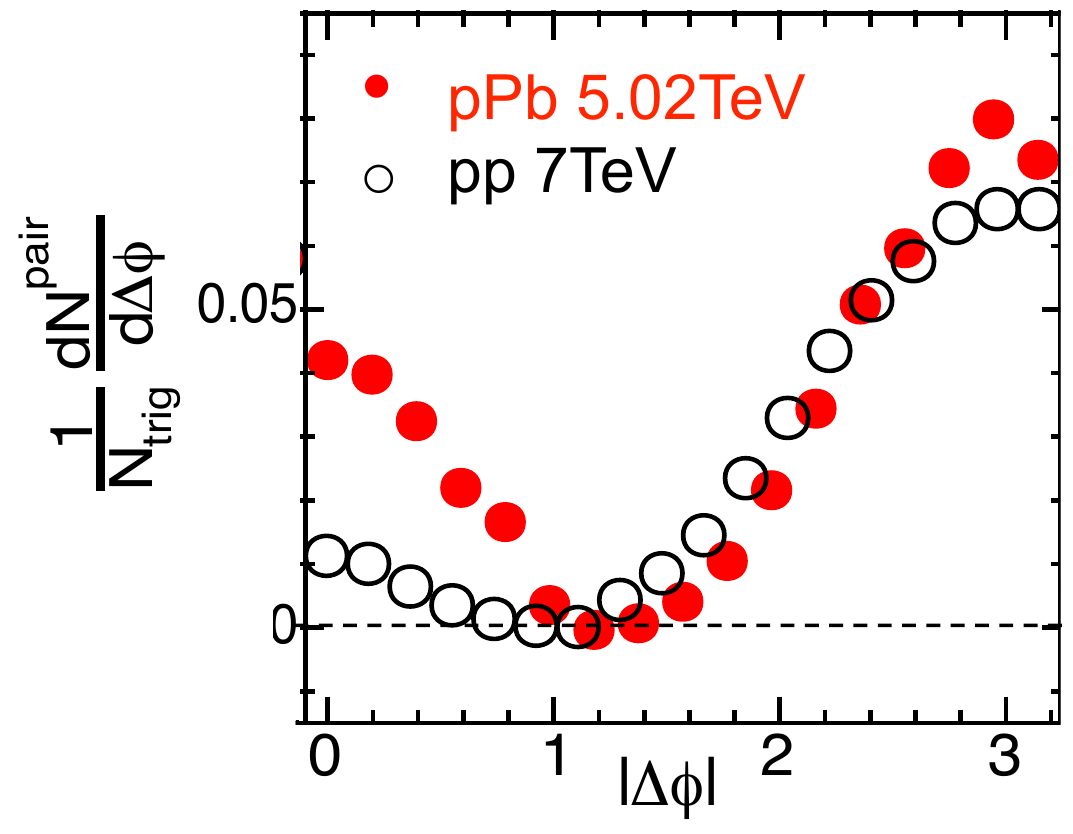}
     \hspace{0.35cm}
     \includegraphics[width=0.58\linewidth]{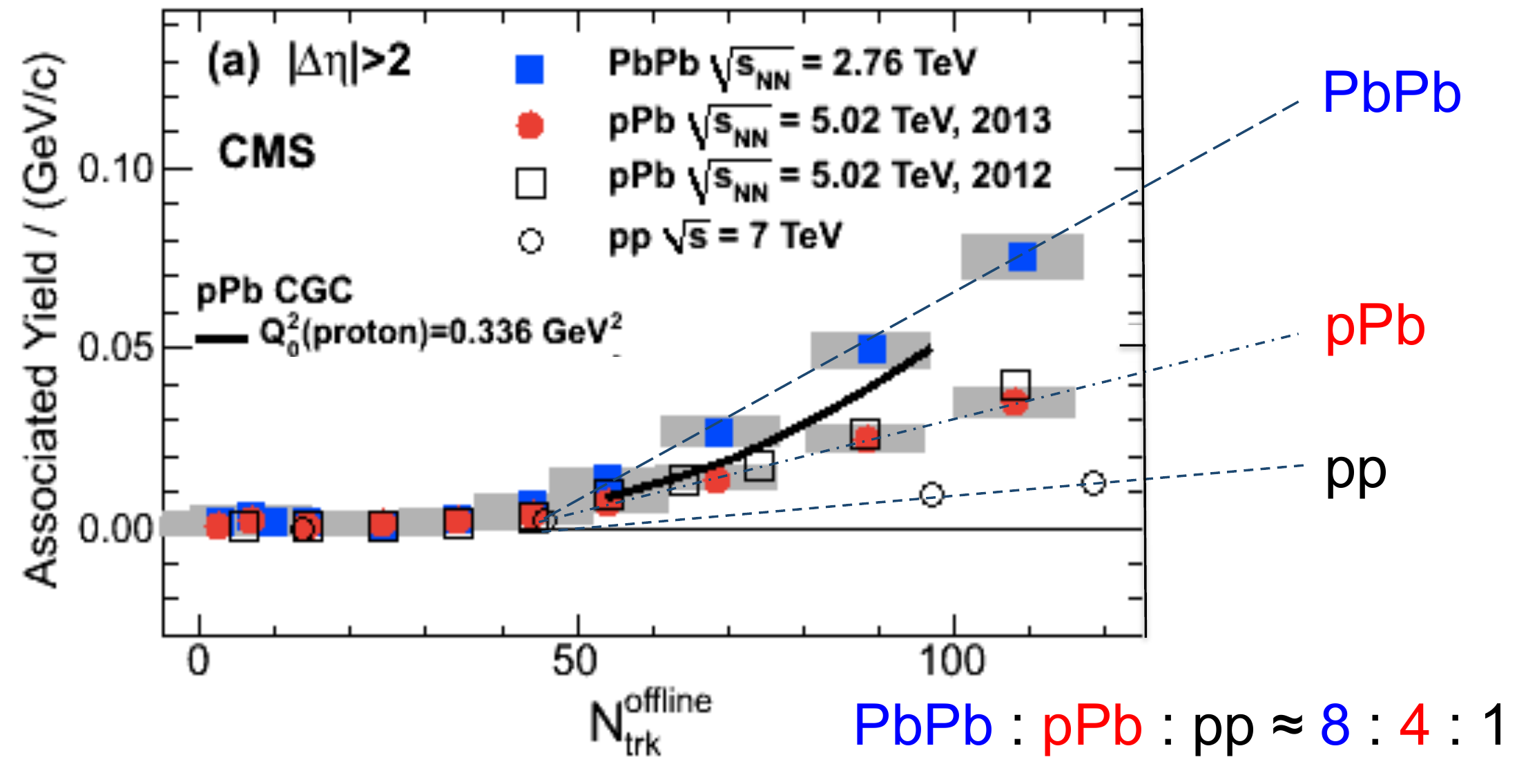}
     \caption{Left: The 1-D long-range azimuthal ($\Delta\phi$) correlation functions 
     in high-multiplicity \ppc collisions at \roots\ = 7\TeV\ and \pPb collisions at 
     \rootsNN\ = 5.02\TeV\ \cite{CMS:2012qk}. Right: long-range near-side associated yields
     for $1\GeVc<\pt<2$\GeVc\ as a function of multiplicity in \ppc, \pPb and \PbPb collisions~\cite{Chatrchyan:2013nka}.}
     \label{fig:ridgeyield_all}
   \end{center}
 \end{figure}

The $\Delta\phi$ correlation for hadron pairs with $1\GeVc<\pt<3$\GeVc\ in \ppc
and \pPb collisions are directly compared in Fig.~\ref{fig:ridgeyield_all}
(left) at a comparable multiplicity.  While the magnitudes of the away sides
are similar, the near-side ridge yield in \ppc is about four times smaller than in
\pPb.  Extracting $v_2$ from \ppc is difficult because of the larger relative
contribution from the away-side jet yield.  The near-side ridge yield as a
function of multiplicity in \ppc, \pPb and \PbPb collisions is shown in
Fig.~\ref{fig:ridgeyield_all} (right). Above around $N_{\rm trk} \sim 40$, the
ridge yield increases roughly linearly with multiplicity for all systems.  At a
given track multiplicity, the ridge yield in \ppc collisions is roughly 25\%
and 10\% of those observed in \PbPb and \pPb collisions, respectively.

Associated yields that are collective are expected to grow linearly
with event multiplicity, while short-range few-body correlations would be more or less
independent of multiplicity (except for the bias towards enhanced jet correlations
imposed by the selection of high multiplicity. However, this bias grows much more 
slowly than linearly with multiplicity). Therefore, pushing to sufficiently high multiplicities
(e.g., $N_{\rm trk} > 150$--$170$), the collective component of correlations would eventually
dominate. The application of the jet-yield subtraction procedure will be more reliable
in that regime.

To shed further light on the situation in \ppc collisions, one should aim to measure
$v_2$ with multi-particle correlations. $v_2$ from multi-particle cumulants 
is less susceptible to jet correlations and could provide further insights.
It is possible that a non-negligible amount of jet correlations may still be present 
in four-particle correlations for very-high-multiplicity \ppc events. In that case,
new methods of implementing an $\eta$ gap among four particles is worth pursuing
to suppress short-range jet correlations.

At the top LHC energy of \roots\ = 13--14\TeV\ for \ppc collisions, a data sample
with an integrated luminosity of 50~pb$^{-1}$ should be large enough to achieve the goals outlined above.
Cross sections of high-multiplicity \ppc events increase by more than a factor of 10 
from \roots\ = 7\TeV\  to \roots\ = 13\TeV.
However, it is necessary for these data to be delivered under a low pile-up condition
(e.g., an average pile-up of 1--2 at CMS and ATLAS) so that the experiments are able 
to trigger on high-multiplicity events from single \ppc interactions. A special run like
this lasting for a few days at the LHC in the future would yield very exciting physics.
Novel ideas of triggering on high-multiplicity events under high pile-up conditions
is another opportunity where progress can be made.

\begin{figure}[t]
  \begin{center}
    \includegraphics[width=0.9\linewidth]{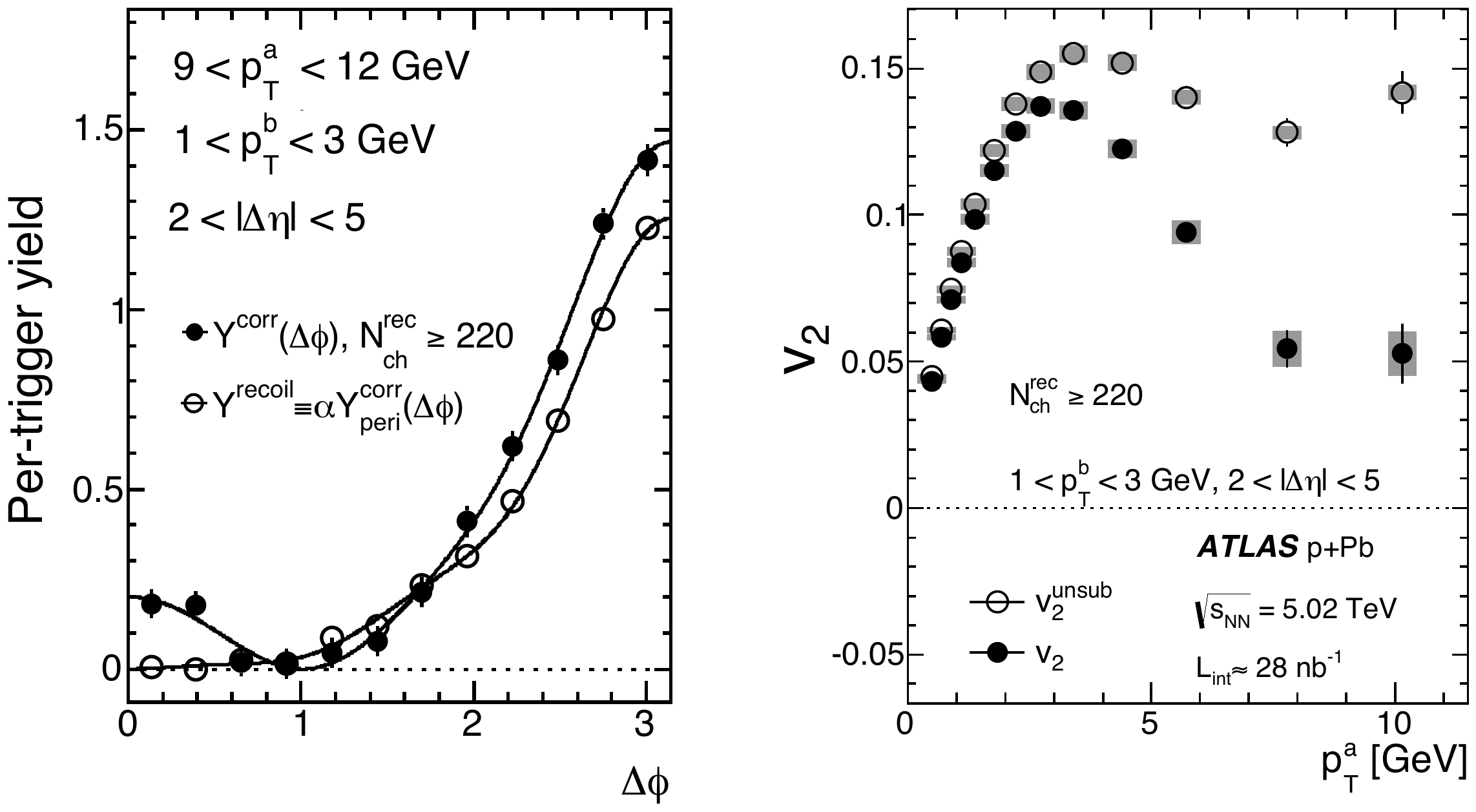}
    \caption{ Left: The 1-D long-range azimuthal ($\Delta\phi$) correlation functions 
    for high-\pt\ (9--12\GeVc) trigger particles in pPb collisions at \rootsNN\ = 5.02\TeV.
    Right: The $v_2$ values from long-range correlations as a function of \pt\ in 
    high-multiplicity pPb collisions, up to very-high-\pt\ region, with (closed)
    and without (open) correcting for back-to-back jet correlations~\cite{Aad:2014lta}. 
    }
    \label{fig:v2_highpt_pPb}
  \end{center}
\end{figure}

\subsection{Jet quenching in small systems}

If the observed ridge-like correlations are related to strong final-state 
rescatterings inside the medium, interactions between high-\pt\ partons and the medium
should also be present, leading to the ``jet-quenching'' phenomenon first 
observed in heavy-ion collisions at RHIC. 

Due to a much smaller system size, the average path length of a parton
traversing through a \ppc or \pA system is much shorter.  For this reason one
might argue that little jet quenching should be expected.  However, at
similar multiplicities, a smaller system represents a higher energy density or
temperature state.  The parton energy loss depends on both the path length,
$L$, and the transport coefficient $\hat{q}$, the mean transverse momentum
squared accumulated by a hard parton per unit length.  While the average $L$ is
reduced in small systems, $\hat{q}$ increases with $T^{3}$.  A jet might lose
just as much energy in a smaller but denser \pPb system as in a larger but more
dilute \PbPb system.  While early model calculations predicted a large jet
quenching \cite{Zakharov:2013lga} one would like to see modeling with the more
realistic space-time evolution predicted by hydrodynamic calculations. 
 
\begin{figure}[t]
  \begin{center}
    \includegraphics[width=\linewidth]{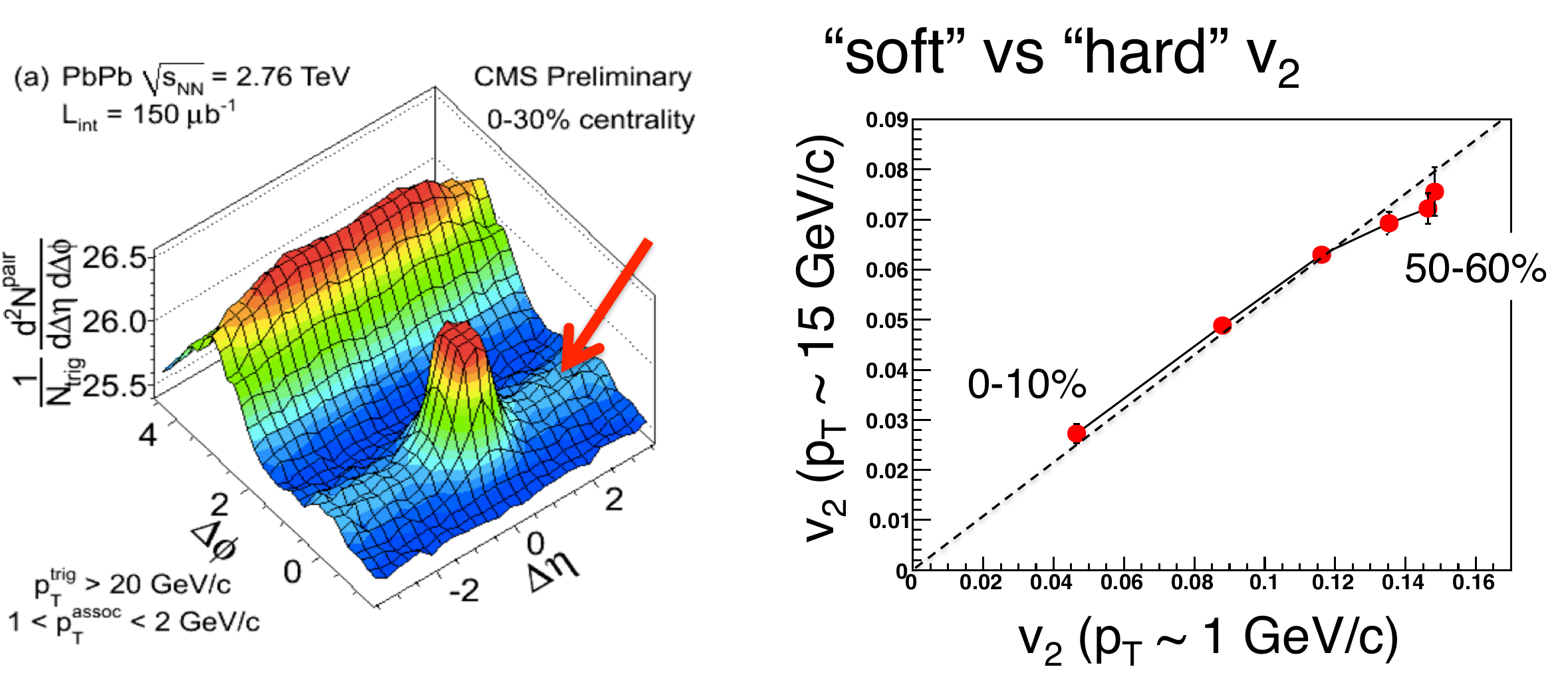}
    \caption{ Left: The 2-D $\Delta\eta$--$\Delta\phi$ correlation function for 
    high-\pt\ ($>20$\GeVc) trigger particles in \PbPb collisions at \rootsNN\ = 2.76\TeV.
    Right: The $v_2$ values at high \pt\ ($\sim 15$\GeVc) versus low \pt\ ($\sim 1$\GeVc)
    for different centralities in \PbPb collisions~\cite{Chatrchyan:2012xq}.}
    \label{fig:v2_highpt_PbPb}
  \end{center}
\end{figure}

Initial results of high-\pt\ jets or hadrons
in \pPb collisions at the LHC indeed suggest no significant modification of 
high-\pt\ parton production~\cite{Chatrchyan:2014hqa,ALICE:2012mj}. However, 
what has been explored so far mainly concerns well-reconstructed jets at very 
high-\pt\ (e.g., $>120$\GeVc)~\cite{Chatrchyan:2014hqa}, or lower-\pt\
hadrons but only in minimum bias pPb events~\cite{ALICE:2012mj}. Studies of the
nuclear modification factor as a function of centrality in \pPb collisions is
highly challenging due to multiplicity selection biases, which remain 
inconclusive~\cite{Adam:2014qja}. Many studies have also indicated
non-trivial correlations between the production of hard processes and underlying 
event activity, which is used for centrality determination
in \pPb collisions~\cite{Adam:2014qja,Alvioli:2014eda,Armesto:2015kwa}.
So far, no clear path forward is laid out for the study of jet quenching
in \ppc and \pA collisions.

Looking for azimuthal anisotropies ($v_2$) of high-\pt\ particles 
can avoid the multiplicity biases and provide us a hint of the (non-)existence of
jet quenching. The fact that the near-side ridge yield
persists to the high-\pt\ region suggests finite $v_2$ values of high-\pt\ particles,
as shown in Fig.~\ref{fig:v2_highpt_pPb} by the ATLAS collaboration~\cite{Aad:2014lta}.
Assuming back-to-back jet correlations are not significantly modified, 
the ATLAS collaboration extracted $v_2$ up to high \pt\ in high-multiplicity \pPb collisions,
where a sizable $v_2$ of about 5\% is observed at $\pt \sim 10$\GeVc, shown
in Fig.~\ref{fig:v2_highpt_pPb}. 

In \PbPb collisions, a long-range near-side correlation structure 
for a trigger particle with $\pt > 20$\GeVc\ is clearly visible
(Fig.~\ref{fig:v2_highpt_PbPb}, left)~\cite{Chatrchyan:2012xq}.
Furthermore, the $v_2$ values obtained at high \pt\ and low \pt\ from
different centrality ranges of \PbPb collisions are found to be strongly correlated 
(Fig.~\ref{fig:v2_highpt_PbPb}, right), indicating a common 
origin, i.e., that both are related to the initial-state geometry. It will be
very interesting to perform the same study in \ppc and \pPb collisions to gain
some insight into the nature of long-range correlations at high \pt. Moreover,
performing multi-particle $v_2$ measurements 
in the high-\pt\ region in \ppc and \pPb would help minimize the influence of 
short-range jet correlations and help clarify the picture of high-\pt\ azimuthal 
anisotropy.

\subsection{Pre-equilibrium dynamics}

The strong anisotropic flow of final-state particles in heavy-ion collisions
has been regarded as evidence for thermalization realized during the very early
stages of the collision.  However, understanding the detailed mechanism of the
fast thermalization process has been a big challenge in the
field~\cite{Strickland:2013uga}. In most hydrodynamic models, a QGP
thermalization or formation time, $\tau_{0}$, is usually assumed, after which
the hydrodynamic evolution is switched on. This additional freedom allows for
the tuning of models to match experimental data.  Often, physics in the
pre-equilibrium stage as well as its transition to a strongly-coupled QGP state
is ignored, even though it is likely to have relevant effects on final-state
observables (for example, its influence on HBT radii has been argued to be
important~\cite{Pratt:2008qv}). 

\begin{figure}[t]
\centering
\includegraphics[width=0.45\linewidth]{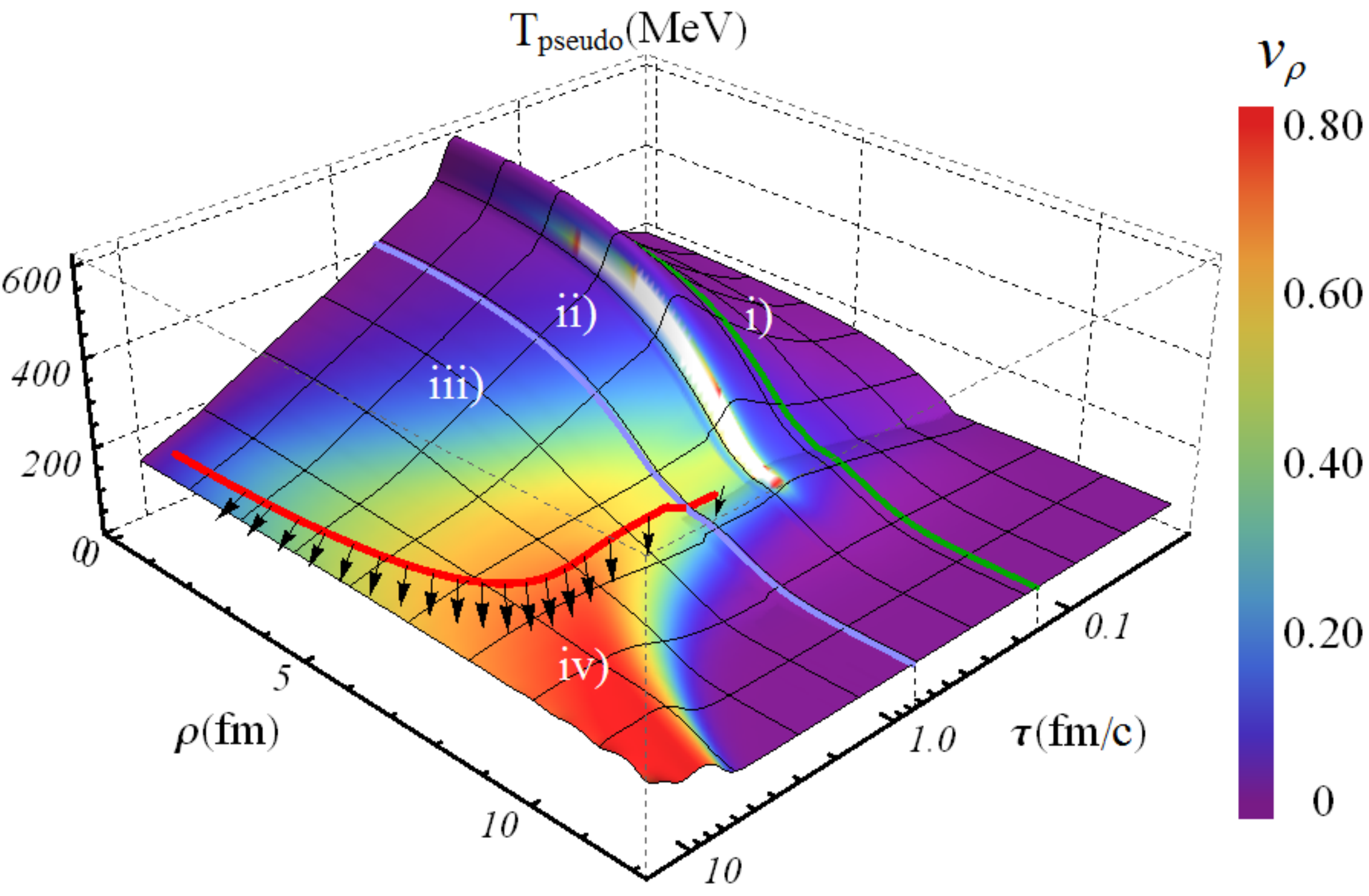}
\hspace{0.5cm}
\includegraphics[width=0.48\linewidth]{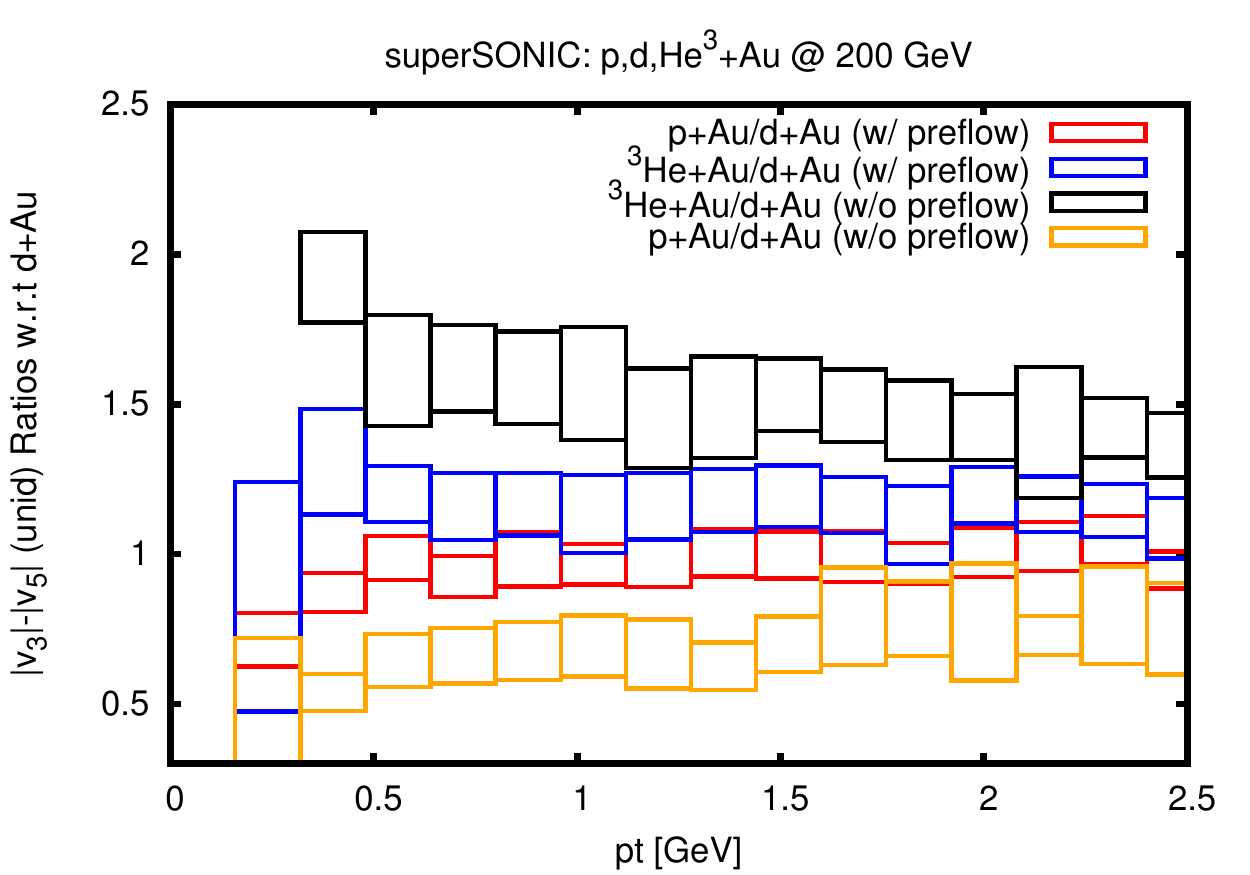}
  \caption{ \label{fig:preflow} 
  Left: The radial velocity $v_{\rho}$ extracted as a function of $\tau$ (time) and $\rho$ (radius) 
  for a representative simulation from Ref.~\cite{vanderSchee:2013pia} with pre-equilibrium dynamics.  
  Right: Ratios of $v_3$ between $^{3}$He-Au, \pAu and \dAu as a function of \pt\ at \rootsNN\ = 200\GeV\
  from the superSONIC model with and without pre-equilibrium flow~\cite{Romatschke:2015gxa}.
   }
\end{figure}

The IP-Glasma model~\cite{Schenke:2012wb,Schenke:2012fw} includes some
pre-equilibrium dynamics via the Yang Mills equations from time zero.  However,
a switch to hydrodynamics at a time $\tau_0$ is still necessary, because the
2+1 dimensional Yang Mills simulation does not contain the necessary dynamics,
which possibly includes the development of Weibel instabilities, to isotropize
the system to the point where a matching with hydrodynamics can be
completed.  However, the build up of flow within the first $\sim 0.4\,{\rm fm}/c$ is
comparable in the Yang Mills and hydrodynamic picture, making the final state
observables independent of the switching time $\tau_0$. \cite{Gale:2012rq}

Progress has also been made in the superSONIC model, where pre-equilibrium dynamics 
and its transition to the hydrodynamic regime is treated in a coherent way without any 
explicit switching time~\cite{vanderSchee:2013pia,Romatschke:2015gxa}.
Fig.~\ref{fig:preflow} (left) shows the development of the radial flow velocity profile 
as a function of time and radius going from the pre-equilibrium stage to the viscous 
hydrodynamic regime~\cite{vanderSchee:2013pia}. Most importantly, simulation results from the superSONIC model
indicate that flow developed during the pre-equilibrium stage has a significant 
contribution for small colliding systems since the total lifetime of a small system
is much shorter. This can be clearly seen in Fig.~\ref{fig:preflow} (right) for 
the ratios of $v_3$ between $^{3}$He-Au, \pAu and \dAu as a function of \pt\ at \rootsNN\ = 200\GeV,
compared with and without pre-equilibrium flow~\cite{Romatschke:2015gxa}. A lot of work
still lies ahead before any solid conclusion can be drawn but high-multiplicity, 
small colliding systems promise to open a new window for probing the
pre-equilibrium dynamics in heavy-ion collisions.

\section{Summary}

In summary, the observation of long-range correlations in high-multiplicity
\ppc and \pPb collisions has opened up new opportunities for investigating novel
high-density QCD phenomena in small colliding systems.  Experimental results
from RHIC and the LHC over the past several years have provided crucial insights
and imposed stringent constraints on possible theoretical interpretations.

We reviewed the theoretical progress in the hydrodynamic modeling of small
colliding systems.  The current status is that hydrodynamic models are able to
describe all the features of the experimental data at a largely quantitative
level. However, a large sensitivity to the unknown initial state in small
systems remains.  Without further constraints on initial state models it will
be difficult to distill information on the medium produced in small systems and
scrutinize the hydrodynamic paradigm.  Clearly hydrodynamics is being pushed
to the edge of its validity in these small colliding systems and further
theoretical work is needed to asses higher order corrections. 
 
Calculations based on the Color Glass Condensate effective theory can also
describe many characteristic features of the experimental data.  If one can
conclusively show that the correlations are generated by glasma graphs, valuable
information about multi-gluon correlations in the nuclear wave-function can be
obtained;  multi-particle correlations can serve as a sensitive probe of
saturation dynamics.

Ideally these frameworks should be merged.  We have a wealth of information
from deep inelastic scattering that should be incorporated into our modeling of
the initial state.  How the initial gluon fields decohere, isotropize and
possibly thermalize to a system that can be described hydrodynamically is an
open question.  We have argued above that classical Yang-Mills dynamics
contains many of the missing pieces of the glasma graph framework.
Furthermore, the recently proposed color domain models can accommodate a
negative $c_2\{4\}$, and is highly interesting from a theoretical point of view
as it could provide a measure of non-Gaussianities in the hadronic wavefunction.

The strongest conclusion that can be made at this time is that the discovery
potential from small colliding systems is immense.  The detailed experimental
information that has and will continue to come from these experiments will
allow the community to test theoretical proposals at an unprecedented level of
accuracy.   Studying collectivity in high-energy proton-proton and
proton-nucleus collisions provide us with access to a rich variety of emergent
(and possibly yet undiscovered) QCD phenomena.

%%% Local Variables: 
%%% mode: latex
%%% TeX-master: "../RidgeReview-ijmpe"
%%% End: 

\section*{Acknowledgements}
We thank Jurgen Schukraft for providing valuable feedback on an early version
of this manuscript.  BPS is supported under DOE Contract No. DE-SC0012704. BPS
acknowledges a DOE Office of Science Early Career Award. WL acknowledges
funding from a DOE Office of Science Early Career Award (Contract No.
DE-SC0012185), from the Welch Foundation (Grant No. C-1845) and from an Alfred
P. Sloan Research Fellowship (No. FR-2015-65911).

\bibliographystyle{ws-ijmpe}
\bibliography{RidgeReview-ijmpe}

\end{document}